\documentclass[letterpaper,twocolumn,10pt]{article}
\usepackage{usenix,epsfig,endnotes}

\usepackage{epsfig,endnotes}
\usepackage{subfiles}
\usepackage{times}
\usepackage{tabularx}
\usepackage{graphicx}
\usepackage{epstopdf}
\usepackage{caption}
\usepackage{capt-of}
\usepackage{subcaption}
\usepackage{xcolor}
\usepackage{booktabs}
\usepackage{multirow}
\usepackage{amssymb}
\usepackage{amsmath}
\usepackage{mathptmx}
\usepackage[blocks]{authblk}
\usepackage[small,compact]{titlesec}
\usepackage{footnote}
\usepackage{cases}
\usepackage{listings}
\usepackage{algorithm}
\usepackage{algorithmic}
\usepackage[pdftex,colorlinks,urlcolor=blue,linkcolor=blue,citecolor=blue,hyperfootnotes=false]{hyperref}
\usepackage{authblk}

\usepackage{framed}

\let\OldS\S
\renewcommand{\S}{\OldS{}}

\usepackage[skip=6pt]{caption}

\def\A{\mathcal{A}}
\def\E{\mathbf{E}}
\def\x{\mathbf{x}}

\newcommand\observation[1]{
        \noindent {\paragraph{\textit{Summary:}} \textit{#1}}
}

\def\compactify{\itemsep=2pt \topsep=2pt \partopsep=1pt \parsep=1pt \leftmargin=1.2em}
\let\latexusecounter=\usecounter

\newcommand{\squishlist}{
   \begin{list}{$\bullet$}
    { \setlength{\itemsep}{0pt}      \setlength{\parsep}{3pt}
      \setlength{\topsep}{3pt}       \setlength{\partopsep}{0pt}
      \setlength{\leftmargin}{1em} \setlength{\labelwidth}{1em}
      \setlength{\labelsep}{0.5em} } }

\newcommand{\squishlisttwo}{
   \begin{list}{$\bullet$}
    { \setlength{\itemsep}{0pt}    \setlength{\parsep}{0pt}
      \setlength{\topsep}{0pt}     \setlength{\partopsep}{0pt}
      \setlength{\leftmargin}{2em} \setlength{\labelwidth}{1.5em}
      \setlength{\labelsep}{0.5em} } }

\newcommand{\squishend}{
    \end{list}  }
\usepackage[font={small}]{caption}
\usepackage[labelfont=bf]{caption}

\makeatletter
\let\@oldmaketitle\@maketitle
\renewcommand{\@maketitle}{\@oldmaketitle
  \vspace*{9em}}
\makeatother

\title{\textsc{AdaRes}: Adaptive Resource Management for Virtual Machines}
\author[ ]{\and}
\author[w]{Ignacio Cano}
\author[w]{Lequn Chen}
\author[p]{Pedro Fonseca}
\author[w]{Tianqi Chen}
\author[ ]{\and}
\author[n]{Chern Cheah}
\author[n]{Karan Gupta}
\author[n]{Ramesh Chandra}
\author[w]{Arvind Krishnamurthy}
\affil[w]{\normalsize{Paul G. Allen School of Computer Science \& Engineering, University of Washington}}
\affil[ ]{\small{\texttt{\{icano,lqchen,tqchen,arvind\}@cs.washington.edu}}}
\affil[p]{\normalsize{Department of Computer Science, Purdue University}}
\affil[ ]{\small{\texttt{pfonseca@purdue.edu}}}
\affil[n]{\normalsize{Nutanix Inc.}}
\affil[ ]{\small{\texttt{\{chern,karan.gupta,ramesh.chandra\}@nutanix.com}}}

\begin{document}
\setlength{\abovedisplayskip}{0pt}
\setlength{\belowdisplayskip}{0pt}

\maketitle

\begin{abstract}

Virtual execution environments allow for consolidation of multiple
applications onto the same physical server, thereby enabling more
efficient use of server resources.  However, users often statically configure
the resources of virtual machines through guesswork, resulting in either
insufficient resource allocations that hinder VM performance, 
or excessive allocations that waste precious data center resources.
In this paper, we first characterize real-world resource allocation and utilization of VMs through the analysis of an extensive dataset, consisting of more than 250k VMs from over 3.6k private enterprise clusters.
Our large-scale analysis confirms that VMs are often misconfigured, either overprovisioned or underprovisioned, and that this problem is pervasive across a wide range of private clusters. 
We then propose \textsc{AdaRes}, an adaptive system that dynamically adjusts VM resources using machine learning techniques. In particular, \textsc{AdaRes} leverages the \textit{contextual bandits} framework to effectively manage the adaptations. Our system exploits easily collectible data, at the cluster, node, and VM levels, to make more sensible allocation decisions, and uses \textit{transfer learning} to safely explore the configurations space and speed up training. Our empirical evaluation shows that \textsc{AdaRes} can significantly improve system utilization without sacrificing performance. For instance, when compared to threshold and prediction-based baselines, it achieves more predictable VM-level performance and also reduces the amount of virtual CPUs and memory provisioned by up to 35\% and 60\% respectively for synthetic workloads on real clusters.

\end{abstract}

\section{Introduction}


Virtual execution environments are widely used in industry as they provide a high degree of flexibility and allow efficient use of cluster resources. An application that might otherwise require a dedicated server to run, can be deployed as a virtual machine (VM) and executed together with other VMs on the same physical hardware, thus enabling more efficient use of resources~\cite{Vasic:2012:DAR:2150976.2151021}.

There are however many hurdles in achieving both high system efficiency and optimal VM performance. 
For example, users typically allocate resources to VMs based on guesswork, which hardly matches the actual resource needs of the applications. 
Even more, the application workload for a VM typically changes over time~\cite{Barker:2010:EEL:1730836.1730842, Farley:2012:MYM:2391229.2391249,Delimitrou:2016:HRP:2872362.2872365,Iosup:2011:PVP:2007336.2007402}, rendering static resource allocation settings inappropriate. 

Incorrect resource allocations can result in a variety of problems. VMs that are not provided enough resources could experience significant application level penalties, such as trashing or swapping.
Further, VMs that underutilize their resources could affect the overall system efficiency, whereas VMs that starve resources could potentially damage other VMs, which could have otherwise benefited from those extra resources~\cite{vSphere6, MemMgmtESXServer, MemOverCommitment, vmware-best_practices}. This motivates the need for a system that adaptively changes the amount of system resources allocated to each VM in a cluster.

In this paper, we first perform a large-scale measurement study of clusters to characterize the resource needs for VMs in the real-world. 
We gather an extensive dataset by instrumenting more than 3.6k enterprise clusters running a commercial computation and storage virtualization product developed by Nutanix.\footnote{For more details refer to \url{http://www.nutanix.com}.} Our analysis allows us to quantify the extent to which user-configured resource allocations are incorrect and the overall impact on cluster efficiency.
Among our main findings, we observe VM instances with significant amounts of overprovisioning as well as some underprovisioning. Further, we find significant variation across time and VMs within a cluster, which renders static resource allocations ineffective.

Unlike most existing traces~\cite{Cortez:2017:RCU:3132747.3132772, Reiss:2012:HDC:2391229.2391236,clusterdata:Wilkes2011,Mishra:2010:TCC:1773394.1773400}, our data refers to privately managed, enterprise clusters that are provisioned and operated independently by 2k+ different companies. Such environments have received little attention despite representing an important virtualization environment that is extensively used by companies~\cite{rightscale_cloudstate}.
Furthermore, the traces we collect contain a richer set of metrics (e.g., VM memory usage, effective I/O operations, etc.) than most other traces, enabling a more thorough analysis of the resource allocation problem.

Based on our findings, we design and build \textsc{AdaRes}, an adaptive system that automatically optimizes VM resource allocations in real clusters. \textsc{AdaRes} uses the multi-armed bandit framework with context information~\cite{NIPS2007_3178}, also known as contextual bandits, to dynamically tune the VMs resource settings, namely virtual CPUs (vCPUs) and memory. By design, the contextual bandits framework allows a cluster manager to adapt to the VM workload characteristics 
through online learning, and represents a natural half-way point between supervised learning and reinforcement learning~\cite{Agarwal14tamingthe,Li:2010:CAP:1772690.1772758,pmlr-v15-beygelzimer11a,NIPS2007_3178}. 

A key challenge in leveraging contextual bandits in our setting is the ``unsafe'' exploration that is required for learning something useful. In other words, we need to be careful of the changes we perform to the VMs as we do not want to (permanently) impair them. To address this challenge, we build a cluster simulator from data collected by running different benchmarks in experimental clusters. We then pre-train (or warm-up) our model(s) offline using the simulator, and transfer the knowledge gained in the simulated environment to the real clusters in order to conduct safer configuration changes as well as speeding up training~\cite{Pan:2010:STL:1850483.1850545, Garcia:2015:CSS:2789272.2886795}, which translates into up to 2$\times$ resource savings when compared to models learned from scratch.
We also leverage the cluster's instrumentation by providing our model a full picture of the cluster, node and VM states, so that it can make more informed decisions. 

Summarizing, our main contributions are:

\squishlist
\item We present a large-scale study of VM resource allocations and usage within thousands of enterprise clusters, which enables us to characterize the  overprovisioning, underprovisioning, and variation in resource utilization over time that occurs in this context.
\item We propose, design, and build \textsc{AdaRes}, an adaptive system capable of tuning VM resources to increase overall system efficiency
that is compatible with existing cluster schedulers. 
\item We propose a contextual bandit-based approach to drive the resource adjustments, and we instantiate our model with an appropriate formulation that results in better resource allocations in real clusters, with resource savings up to 35\% (CPU) and 60\% (memory) in synthetic workloads executed on real clusters, when compared to threshold and other ML-based baselines.
\squishend

\section{Resource Utilization Measurements of Enterprise Clusters}
\label{sec:motivation}


This section presents our measurement study on resource allocation and utilization of enterprise clusters with virtual execution environments. Our study characterizes the VM resource allocation problem in the context of enterprise clusters and motivates the need for \textsc{AdaRes}.


\subsection{Measurement Methodology}

We perform our measurements on enterprise clusters running a commercial virtual execution platform developed by Nutanix.
Nutanix cluster manager transparently allocates and migrates VMs based on user configured resource settings and cluster-level utilization metrics. In addition, the platform provides transparent access to highly available virtual storage (virtual disks) located within each cluster node.  

Our dataset was collected from sensors deployed on the cluster nodes that record data regarding a broad class of metrics, such as the resources utilized by a VM (e.g., CPU and memory) and cost of various operations (e.g., average I/O latency).
Our dataset consists of a subset of the clusters that push diagnostic information to a centralized data collection service and refers to the period from April 23$^{rd}$ to May 20$^{th}$, 2018.
Table~\ref{tab:dataset} shows an overview of the virtual execution environments that we study, containing more than 250k VM traces. 

\begin{table}[ht!]
\small
\begin{center}
\begin{tabular}{cc} 
\toprule
  \textbf{Statistic} & 
  \textbf{Value}
\\ 
\midrule
{\# of Companies} & 2,003 \\
{\# of Clusters} & 3,669 \\
{\# of Nodes} & 17,633 \\
{\# of VMs} &  252,941 \\
\bottomrule
\end{tabular}
\end{center}
\vspace{-0.3cm}
\caption{Dataset Overview}
\label{tab:dataset}
\end{table}


\subsection{Private Cluster Configurations}
\label{sec:cluster_configs}

To better understand the configuration patterns of enterprise clusters, we perform an analysis of configurations at cluster, node, and VM levels.

\begin{figure}[h!]
  \centering
  \begin{subfigure}{0.48\columnwidth}
    \includegraphics[width=1\linewidth]{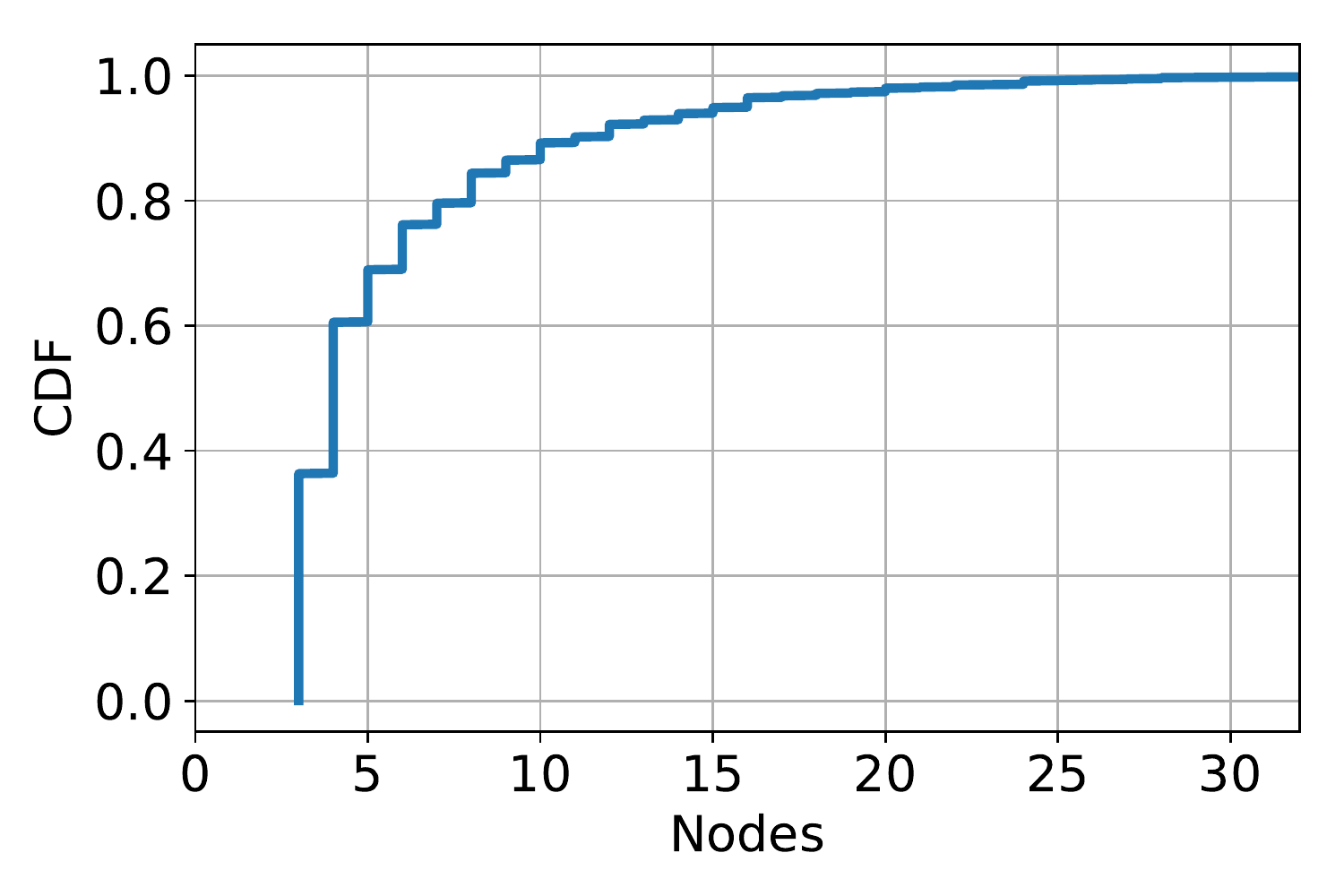}
    \caption{Nodes per Cluster}
    \label{fig:nodes_per_cluster}
  \end{subfigure}
  \begin{subfigure}{0.48\columnwidth}
    \includegraphics[width=1\linewidth]{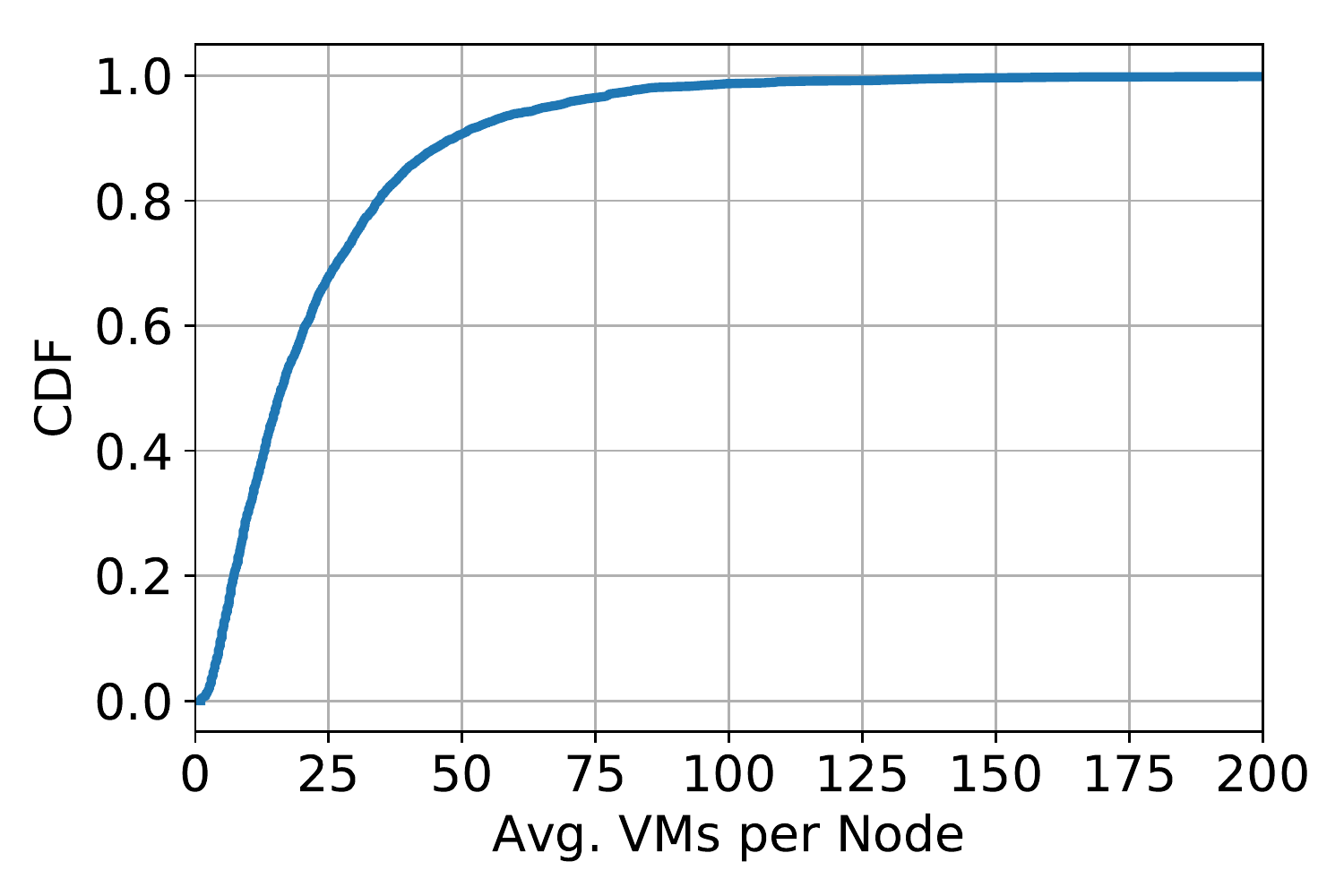}
    \caption{Consolidation Factor}
    \label{fig:avg_nr_vms_per_node}
  \end{subfigure}
  \caption{Cluster-level Configuration}
  \label{fig:cluster_sizing}
\end{figure}

\paragraph{Cluster-level Configuration}
Figure~\ref{fig:cluster_sizing} shows the distribution of nodes per cluster (\ref{fig:nodes_per_cluster}) and the consolidation factor, i.e., the average number of VMs per node, (\ref{fig:avg_nr_vms_per_node}).
From Figure~\ref{fig:nodes_per_cluster}, we observe that 60\% of the clusters have 4 nodes or less, and 30\% have between 5 and 10 nodes. In general, the clusters have a modest number of nodes. We find that under these environments, when users need additional nodes, companies tend to expand their computational resources by adding clusters, as opposed to adding nodes to existing clusters.
There are three main reasons for this: (1) smaller clusters provide better fault isolation, (2) most of the analyzed clusters are deployed on premise, in remote office/branch office (RoBo) configurations, and (3) some companies prefer to create clusters for each line of business.
Figure~\ref{fig:avg_nr_vms_per_node} shows that 50\% of the clusters have, on average, at most 16 VMs per node, and that 20\% have more than 35 VMs per node, up to $\sim$200 VMs per node.



\begin{figure}[h!]
  \centering
  \begin{subfigure}{0.48\columnwidth}
    \includegraphics[width=1\linewidth]{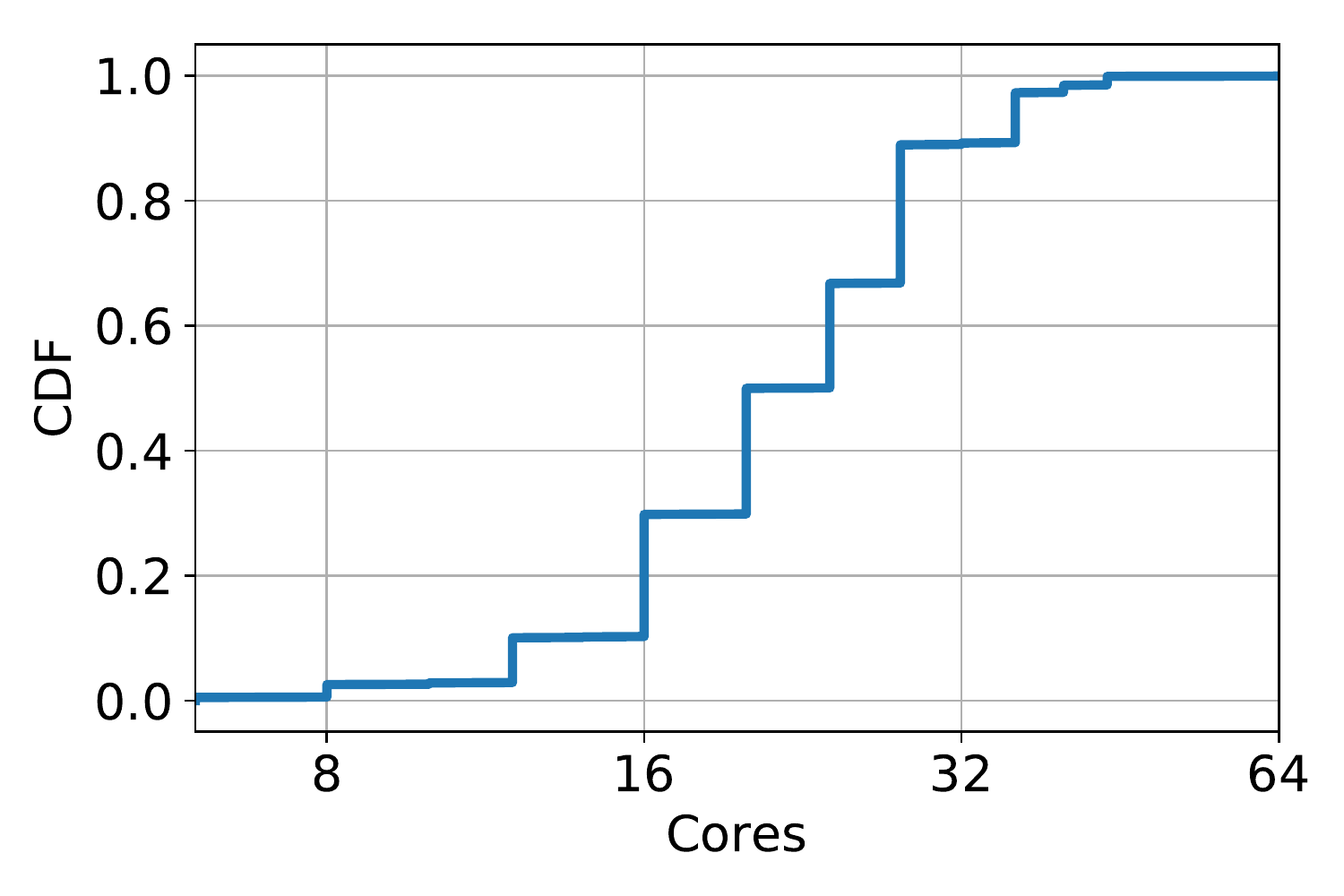}
    \caption{CPU}
    \label{fig:nodes_cores}
  \end{subfigure}
  \begin{subfigure}{0.48\columnwidth}
    \includegraphics[width=1\linewidth]{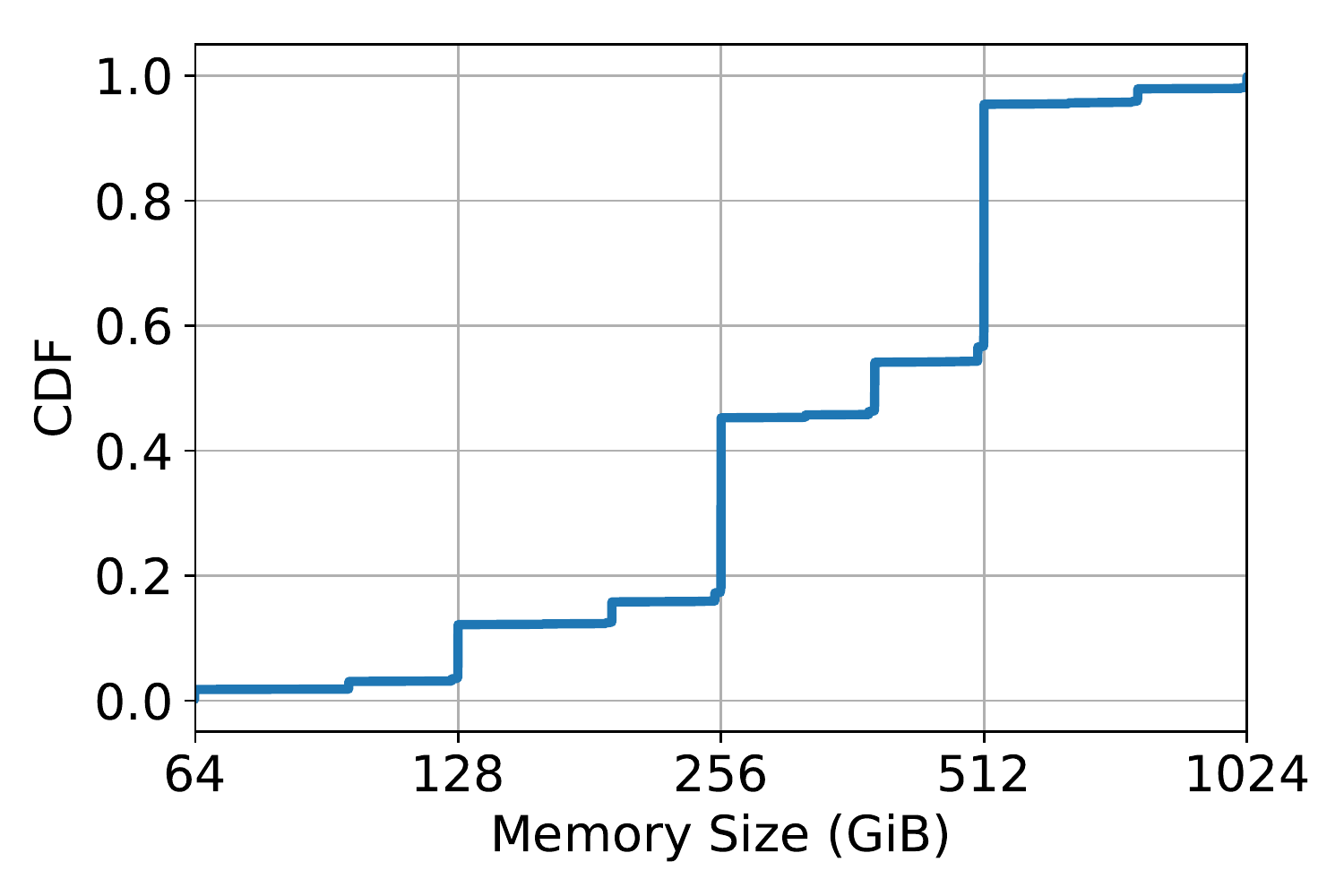}
    \caption{Memory}
    \label{fig:nodes_memory}
  \end{subfigure}
  \caption{Node-level Configuration}
  \label{fig:node_sizing}
\end{figure}

\paragraph{Node-level Configuration}
Enterprise clusters often have powerful nodes, as shown in Figure~\ref{fig:node_sizing}. We observe that 50\% of the nodes have more than 24 physical cores and 384 GiB of RAM, and 10\% have at least 36 cores and more than 512 GiB of RAM.

\begin{figure}[h!]
  \centering
  \begin{subfigure}{0.48\columnwidth}
    \includegraphics[width=1\linewidth]{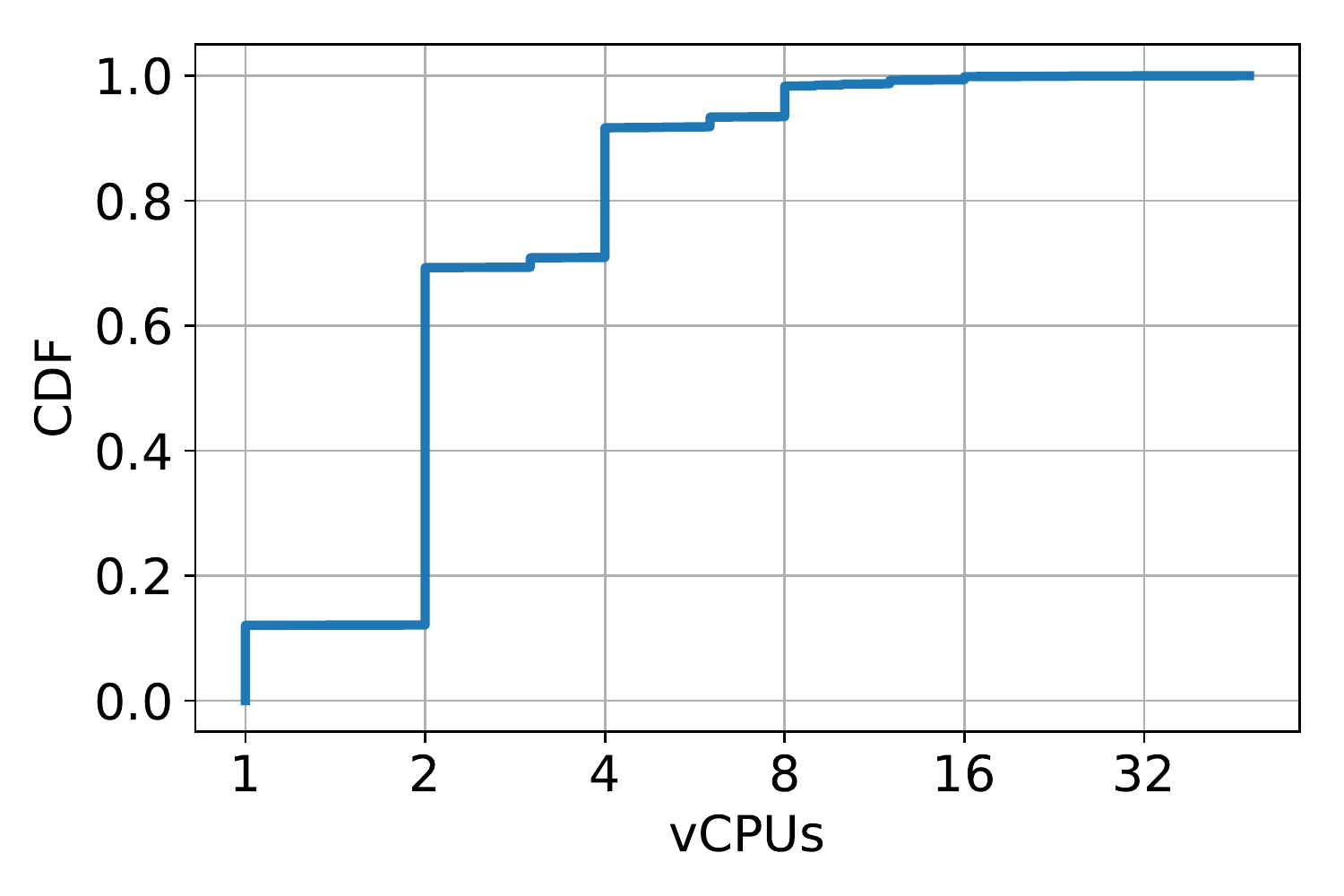}
    \caption{CPU}
    \label{fig:vm_vcpus}
  \end{subfigure}
  \begin{subfigure}{0.48\columnwidth}
    \includegraphics[width=1\linewidth]{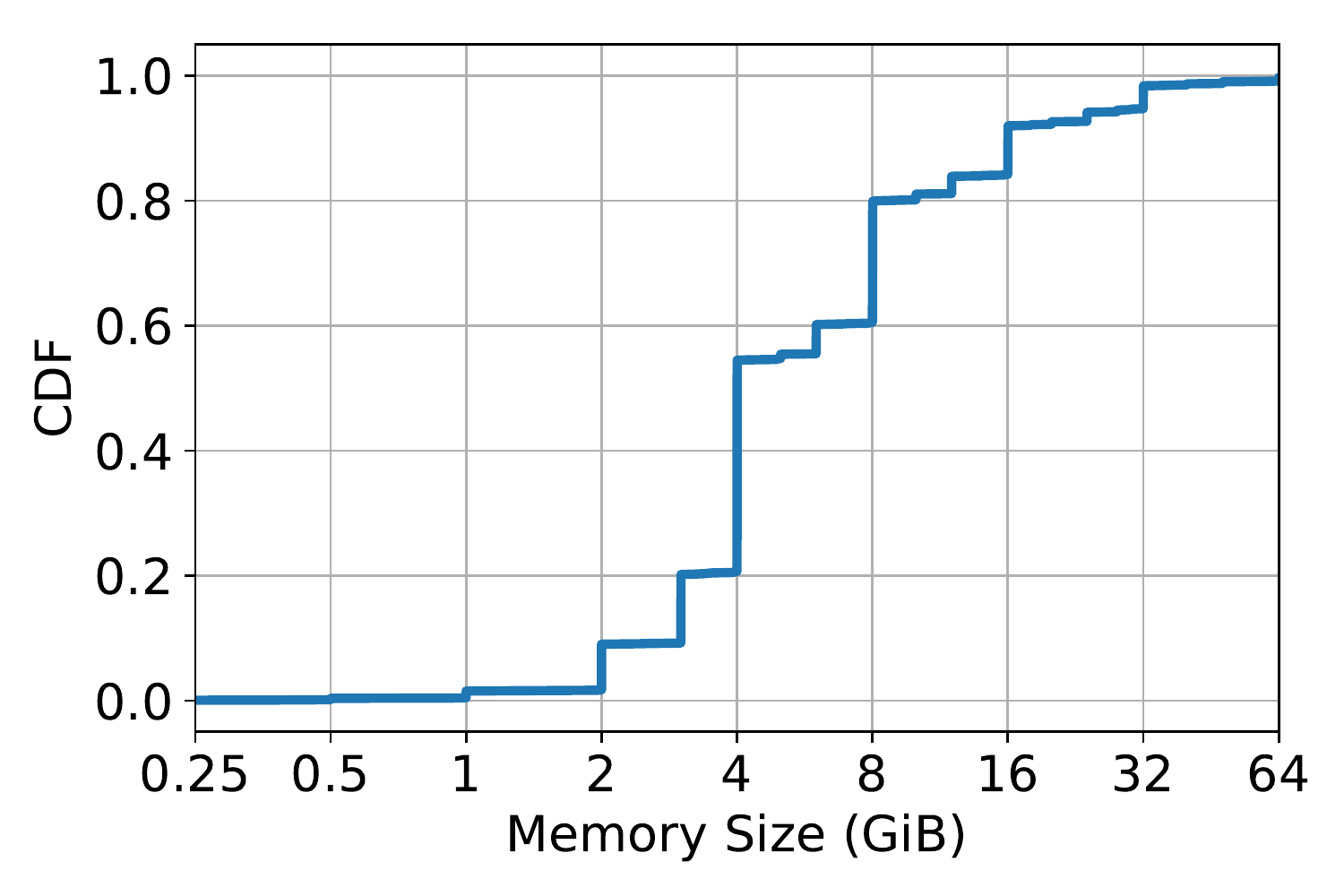}
    \caption{Memory}
    \label{fig:vms_memory}
  \end{subfigure}
  \caption{VM-level Configuration}
  \label{fig:vm_sizing}
 \end{figure}

\paragraph{VM-level Configuration}
Figure~\ref{fig:vm_sizing} provides an analysis of the VM sizes in terms of virtual CPUs (vCPUs) and allocated memory.
Our dataset shows that approximately half of the VMs are configured with 2 vCPUs, whereas 20\% are configured with 4 vCPUs.  
Regarding memory, around 35\% of the VMs are deployed with 4 GiB of RAM, and 20\% with 8 GiB. In both resources, we note a ``human'' sizing pattern of using powers of 2.

We also observe a correlation between the size of the clusters and the number of VMs per node: small clusters have on average the lowest VM density because such clusters typically run a small number of applications supporting limited workloads. 
In contrast, larger clusters typically support a broad mix of workloads, with some supporting applications such as Virtual Desktop Infrastructure (VDI), which typically deploy a large number of VMs for each connected user. 
Further, we note that many medium-sized VMs (i.e., VMs with 2-4 vCPUs) are typically used to deploy server applications such as SQLServer, MS Exchange, etc.

\observation{Enterprise clusters are often small-sized single-tenant clusters, with powerful nodes, that support the workload requirements of small and medium-sized businesses. 
}

\subsection{Problem Characterization}


This section provides an analysis on the utilization of the clusters. Our analysis relies on several key metrics that we collect and are representative of the VMs resource usage. For each metric, we record, on each  cluster node, the average measurement over a 5-minute interval at the VM-level.  
This data enable us to calculate the mean, maximum, and the 95$^{th}$ percentile (P95) of a series of 5-minute measurements for any given metric.

\begin{figure}[h]
  \centering
  \begin{subfigure}{0.48\columnwidth}
    \includegraphics[width=1\linewidth]{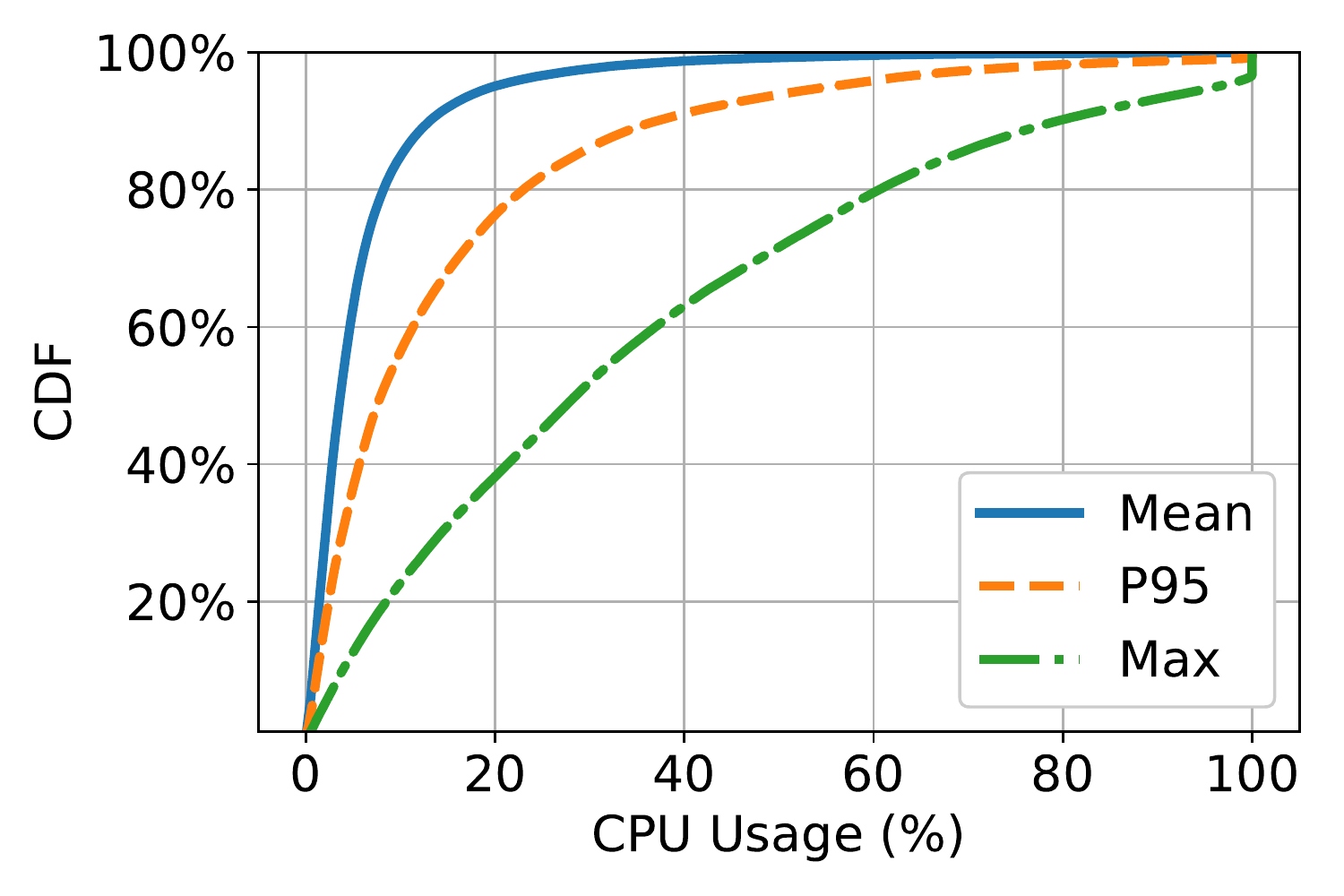}
    \caption{CPU}
    \label{fig:vm_cpu_usage}
  \end{subfigure}
  \begin{subfigure}{0.48\columnwidth}
    \includegraphics[width=1\linewidth]{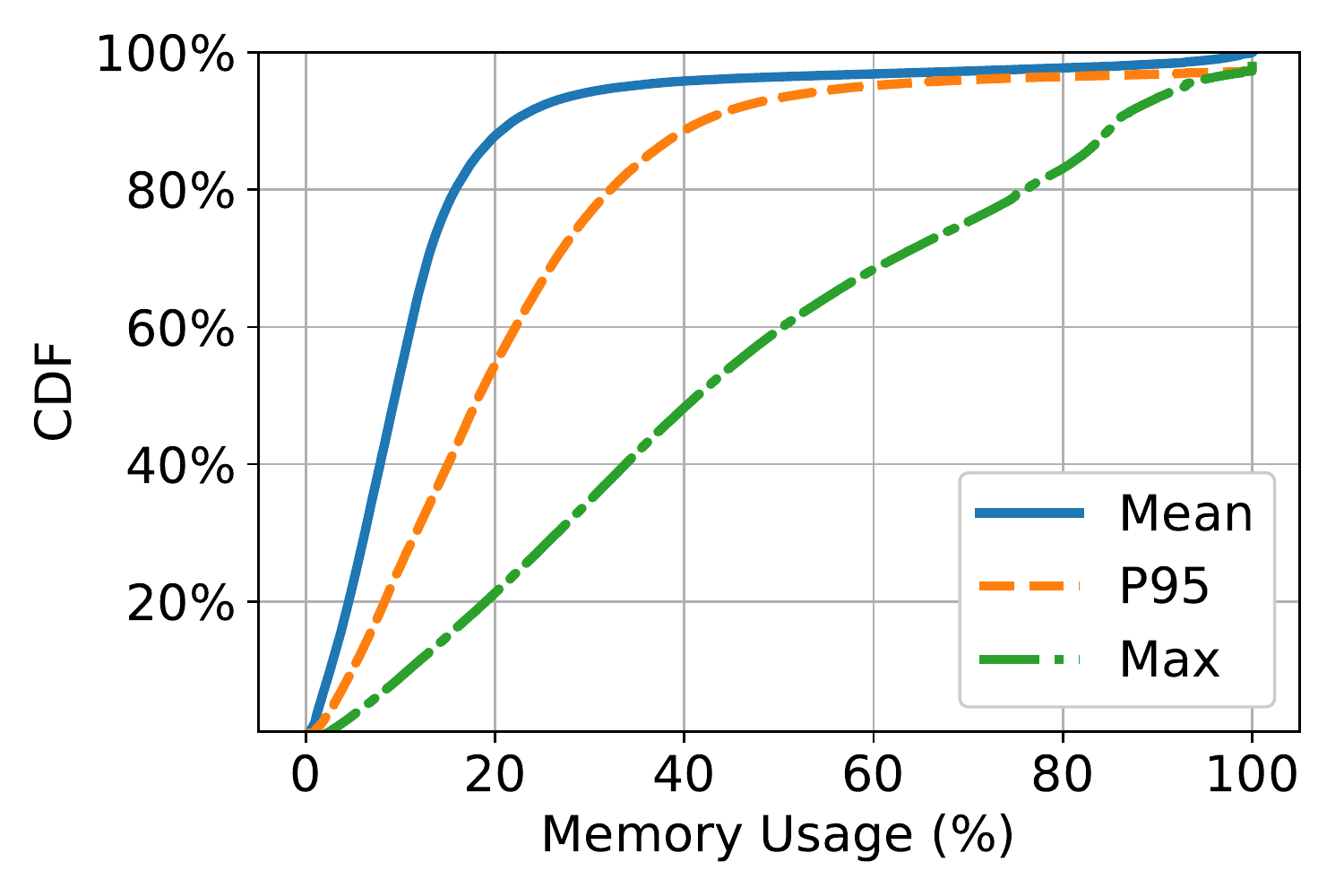}
    \caption{Memory}
    \label{fig:vm_mem_usage}
  \end{subfigure}
  \caption{VM Resource Usage}
\end{figure}




\begin{figure*}[t!]
  \centering
  \begin{subfigure}{0.32\textwidth}
    \includegraphics[width=1\linewidth]{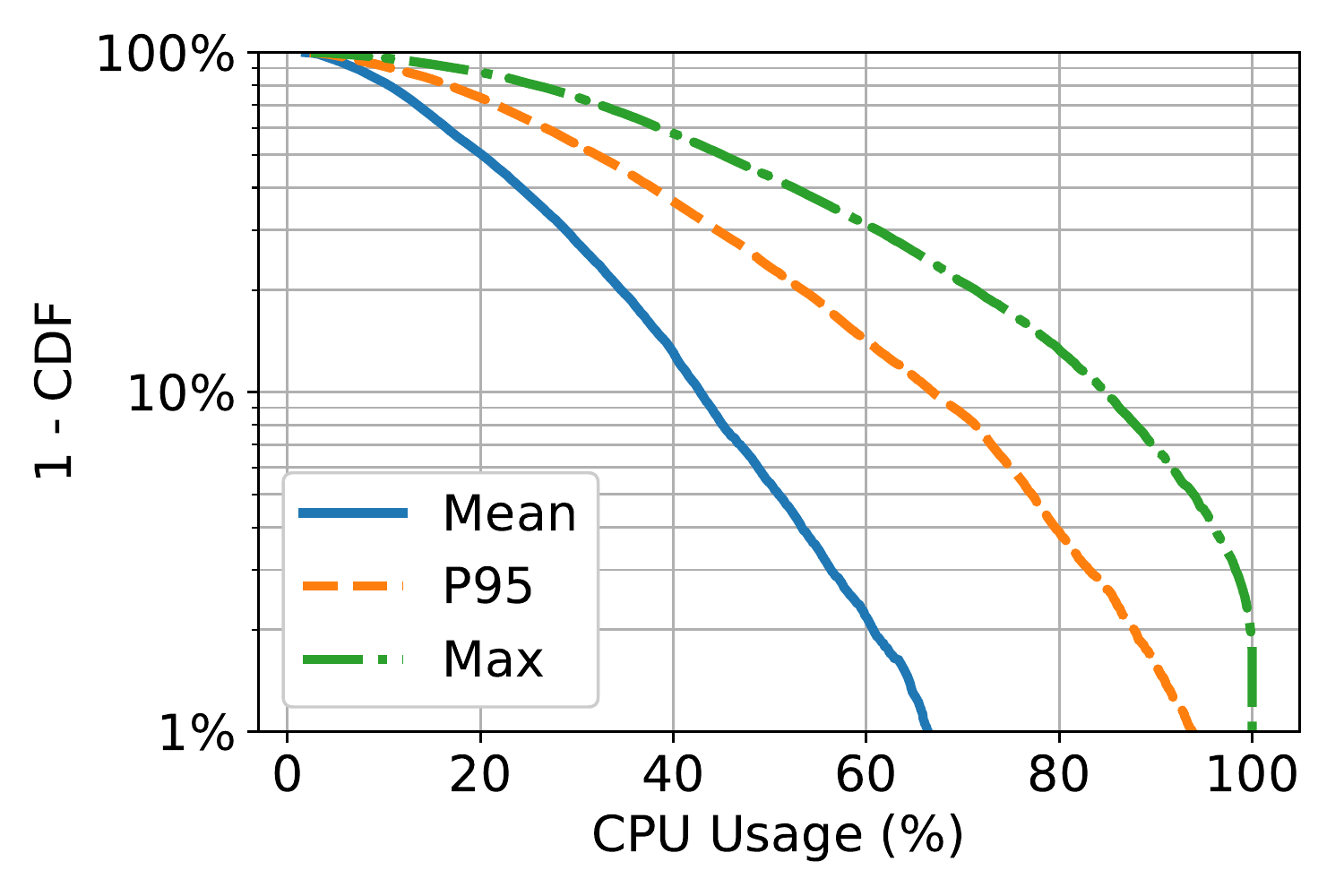}
    \caption{CPU}
    \label{fig:nodes_cpu_usage}
  \end{subfigure}
  \begin{subfigure}{0.32\textwidth}
    \includegraphics[width=1\linewidth]{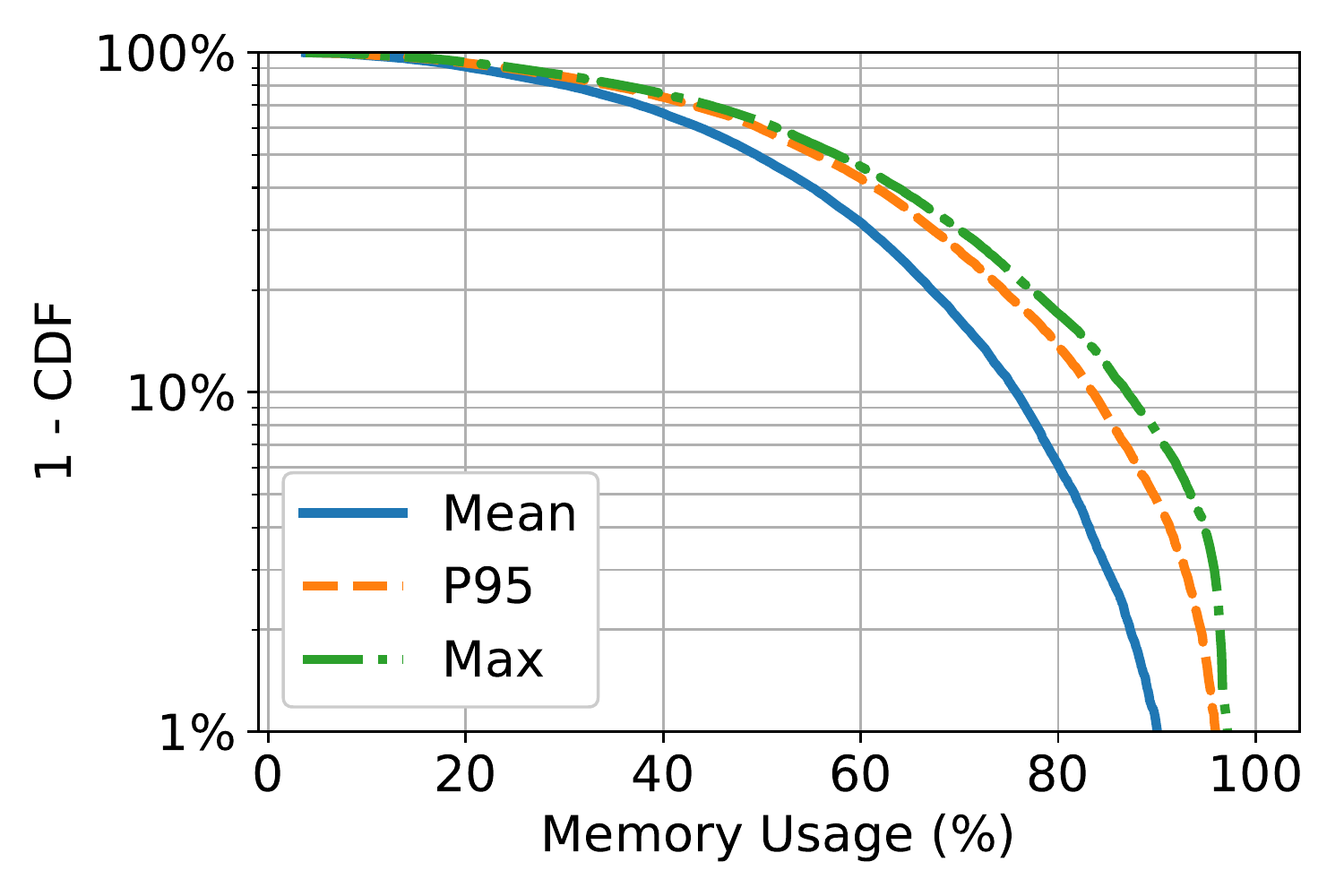}
    \caption{Memory}
    \label{fig:nodes_memory_usage}
  \end{subfigure}
  \begin{subfigure}{0.32\textwidth}
    \includegraphics[width=1\linewidth]{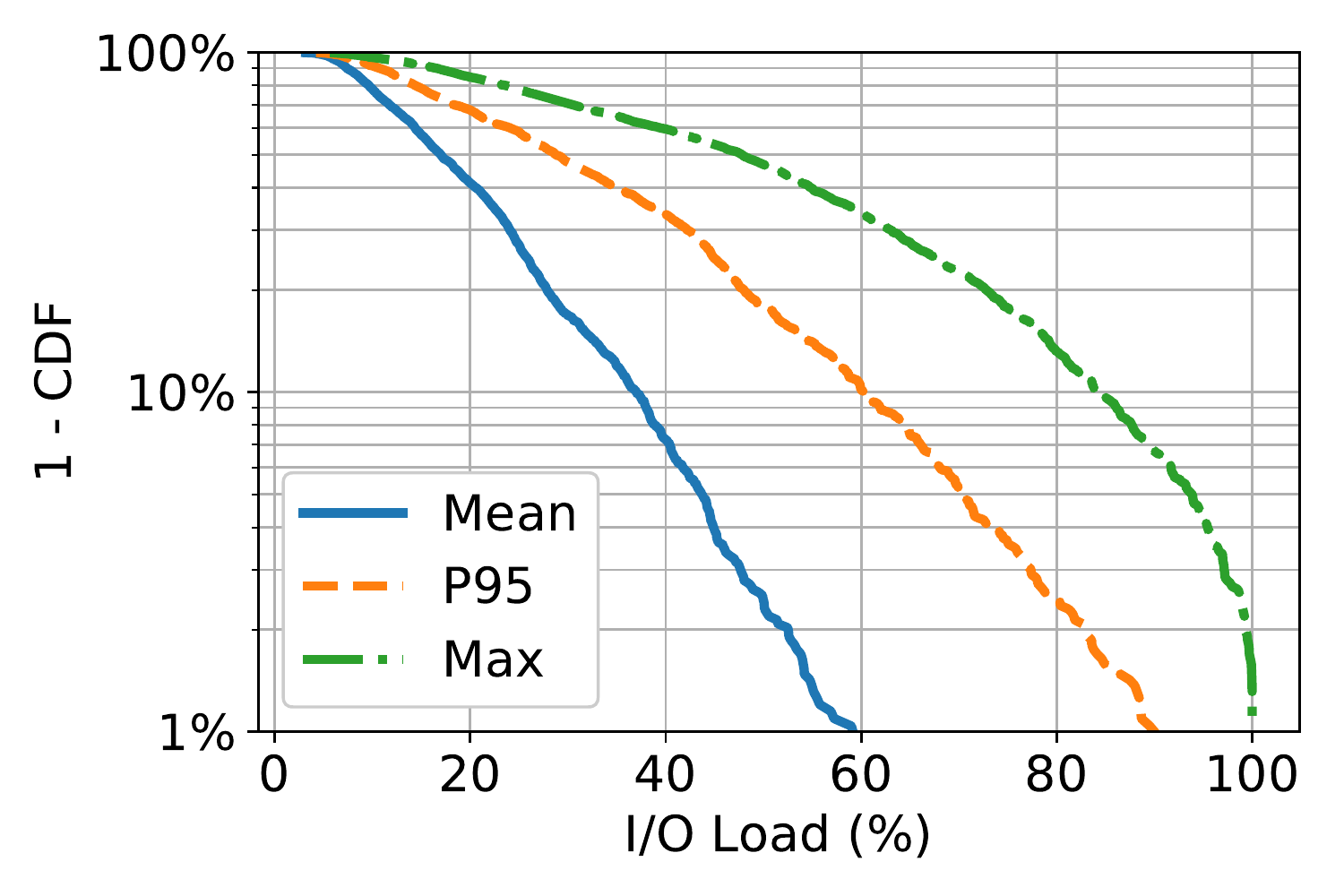}
    \caption{I/O}
    \label{fig:nodes_io_load}
  \end{subfigure}      
  \caption{Node Resource Usage}
  \label{fig:node_usage}
\end{figure*}


Figures~\ref{fig:vm_cpu_usage} and \ref{fig:vm_mem_usage} present the cumulative distribution function (CDF) of the mean, P95 and maximum VM resource usage for CPU and memory.
These results show that many of the VMs are \textit{overprovisioned} with respect to both CPU and memory. In particular, 90\% of the VMs have P95 CPU and memory usages lower than 40\%. Further, 80\% of the VMs have a maximum resource usage that is lower than 60\% (CPU) and 80\% (memory) throughout their lifetime; in other words, 40\% and 20\% of the allocated resources are \emph{never} used by 80\% of the VMs. By analyzing the dataset we calculate that the global resources allocated but never used correspond to 26\% (CPU) and 27\% (memory) of the total allocated resources by all VMs.\footnote{Intuitively, the areas to the right of the maximum line in Figure~\ref{fig:vm_cpu_usage} and \ref{fig:vm_mem_usage} represent the global wasted resources that are never used, but our numbers additionally take into consideration the different absolute sizes of the VMs.} Such allocated but sparsely used resources are the result of two main factors: (a) manual VM resource allocation, and (b) users inability to accurately predict the resource demands of their workloads.

We observe a similar trend at the node level, i.e., many nodes have low average utilization but experience high peak resource usage. We show the
complementary cumulative distribution functions (CCDF or 1-CDF) of
node-level usage in Figure~\ref{fig:node_usage}. Note that CCDFs are useful for highlighting the tails of distributions. Besides CPU and memory usage, we also analyze the compute processing load of the storage controller on each node and use it as a proxy of the node's I/O load. In general, we see that node usage is higher than VM-level usage, especially memory utilization, due to oversubscription, where around 10\% of nodes have, on average, more than 80\% memory usage, but still, many nodes are underutilized. 



\begin{figure}[h]
  \centering
  \begin{subfigure}{0.48\columnwidth}
    \includegraphics[width=1\linewidth]{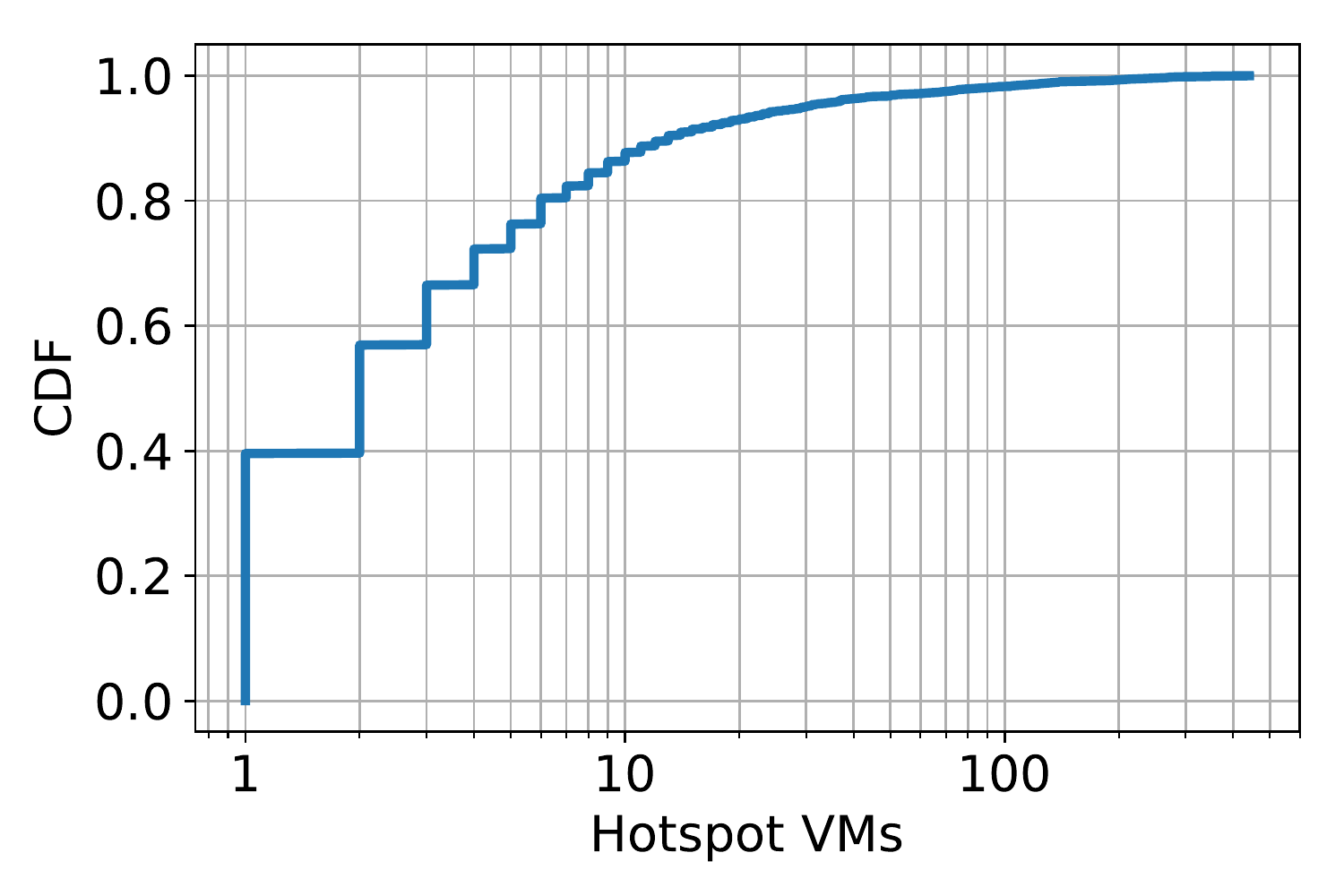}
    \label{fig:hotspot_vms_per_cluster}
    \caption{Distribution of Hotspot VMs per Cluster}
  \label{fig:hotspots}
  \end{subfigure}
  \begin{subfigure}{0.48\columnwidth}
    \includegraphics[width=1\linewidth]{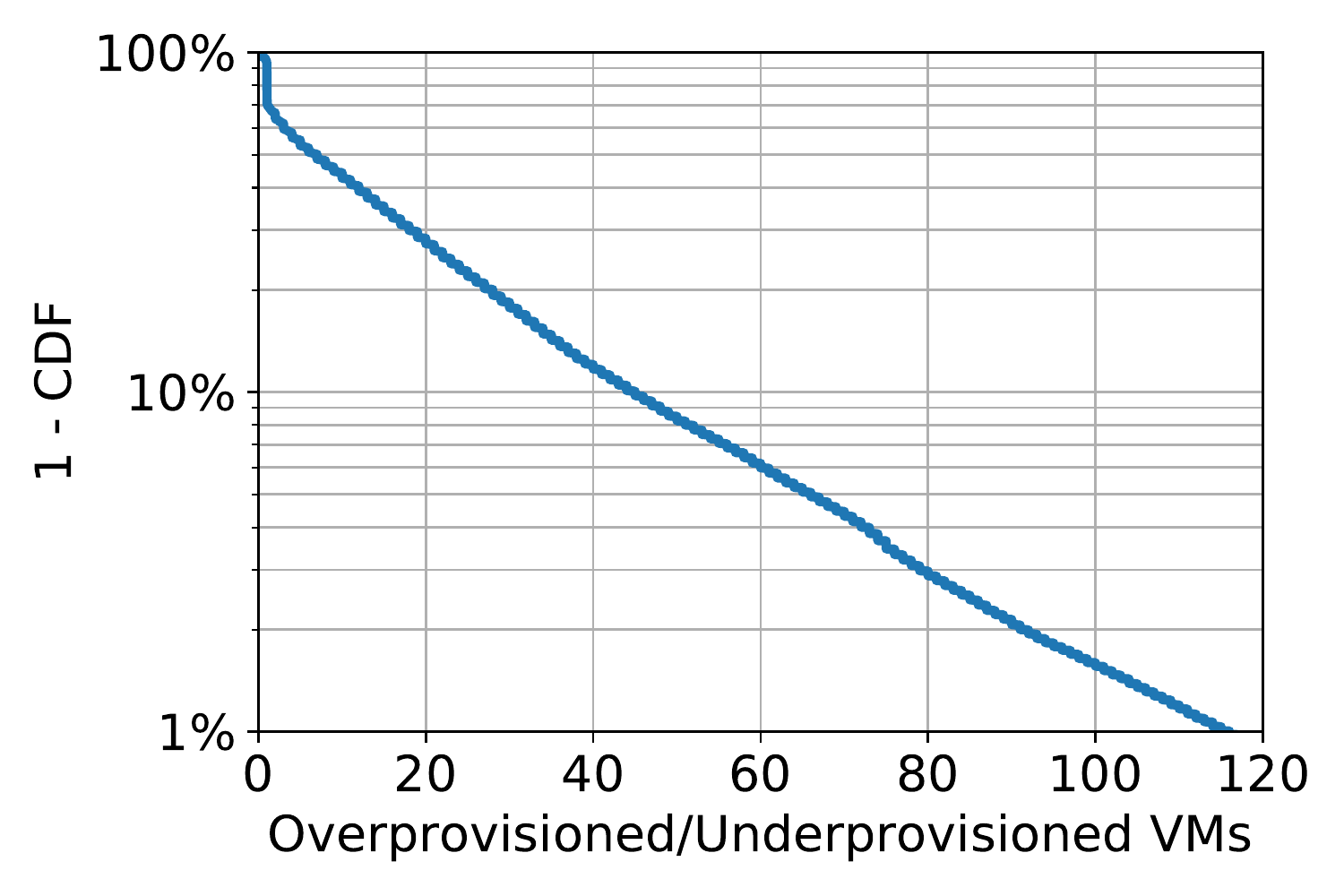}
    \label{fig:over_under_ratio_cluster}
  \caption{Ratio of Over/Underprovisioned VMs per Cluster}
  \label{fig:over_underprov_ratios}
  \end{subfigure}
  \caption{Hotspots and Over/Underprovisioned VMs Ratio}
\end{figure}

Although average utilization is generally low, our data still reveals that many VMs are underprovisioned. Figure~\ref{fig:hotspots} shows the distribution of hotspot VMs per cluster. We consider a VM to be a hotspot if its 95$^{th}$ percentile metric utilization is greater than 75\%. 
We observe that 40\% of the clusters with hotspot VMs have at most 2 underprovisioned VMs, whereas 
10\% of the clusters with underprovisioned VMs contain at least 10 hotspot VMs.
From the total clusters in the dataset, 45\% contain either CPU-hotspot VMs, memory-hotspot VMs, or both. Thus, our data suggests that underprovisioning is not limited to few, possibly incorrectly managed, clusters; instead, our data reveals that the hotspot problem impairs a large fraction of clusters.

\observation{Most VMs in today's enterprise clusters are not sized appropriately, with many VMs either overprovisioned or underprovisioned. This motivates the need for developing an automated system to determine VM resource allocations as opposed to relying on user-provided configurations.
}


\subsection{Opportunities and Challenges for Adaptive Resource Allocation}

Figure~\ref{fig:over_underprov_ratios} shows the distribution of the ratio of overprovisioned divided by underprovisioned VMs (when such underprovisioned VMs exist) per cluster, at a given point in time. 
We consider a VM to be overprovisioned if its 95$^{th}$ percentile metric utilization is less than 25\%. Recall that underprovisioned (or hotspot) VMs are those with a P95 utilization greater than 75\%. 
In general, when there are hotspots, there are also VMs with overprovisioned resources at the same time. For example, we observe that 50\% of the clusters with underprovisioned VMs have at least a 7:1 overprovisioned/underprovisioned VMs ratio.

\begin{figure}[h!]
  \centering
  \begin{subfigure}{0.47\columnwidth}
    \includegraphics[width=1\linewidth]{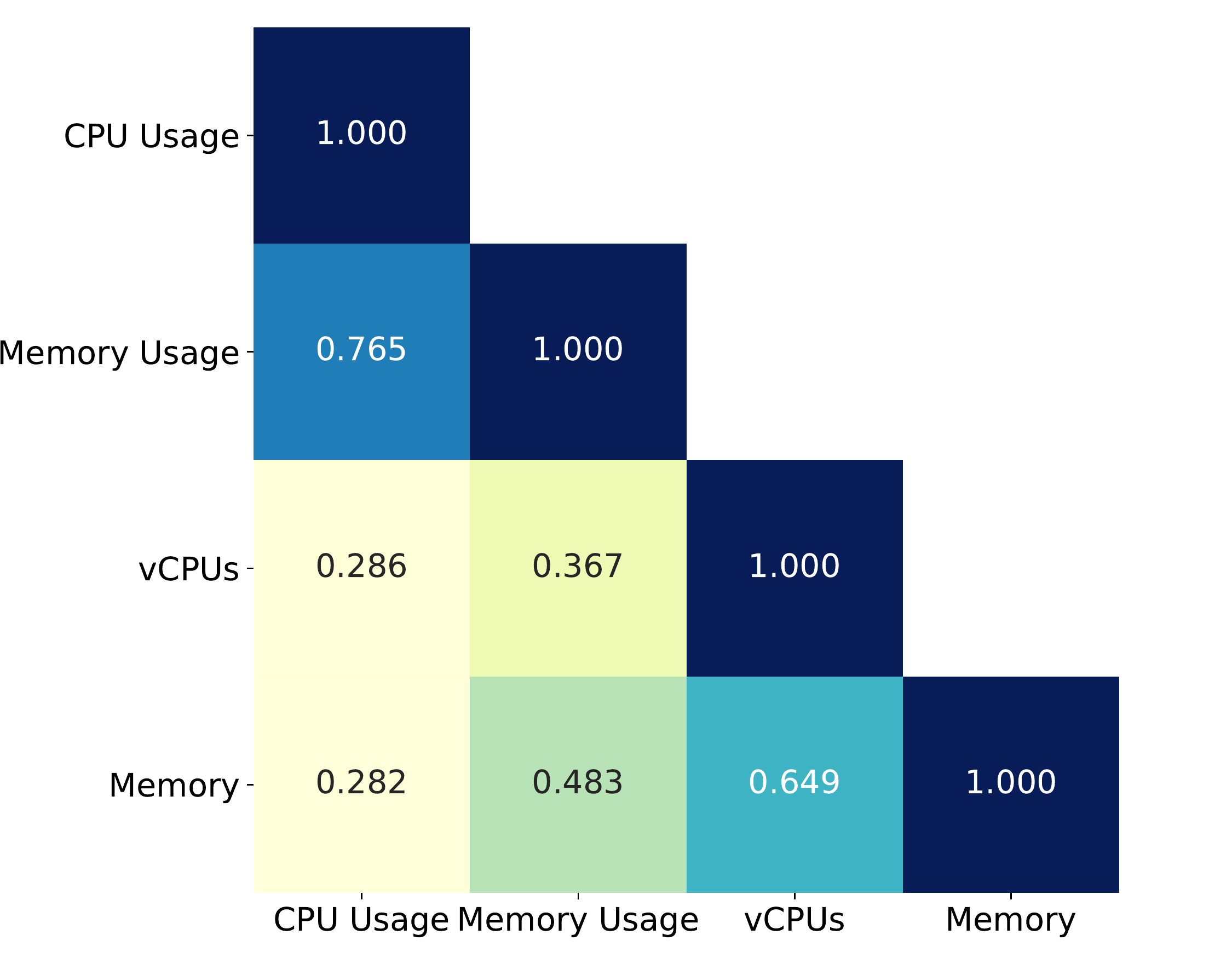}
    \caption{VM-level}
    \label{fig:vm_correlations}
  \end{subfigure}
  \begin{subfigure}{0.49\columnwidth}
    \includegraphics[width=1\linewidth]{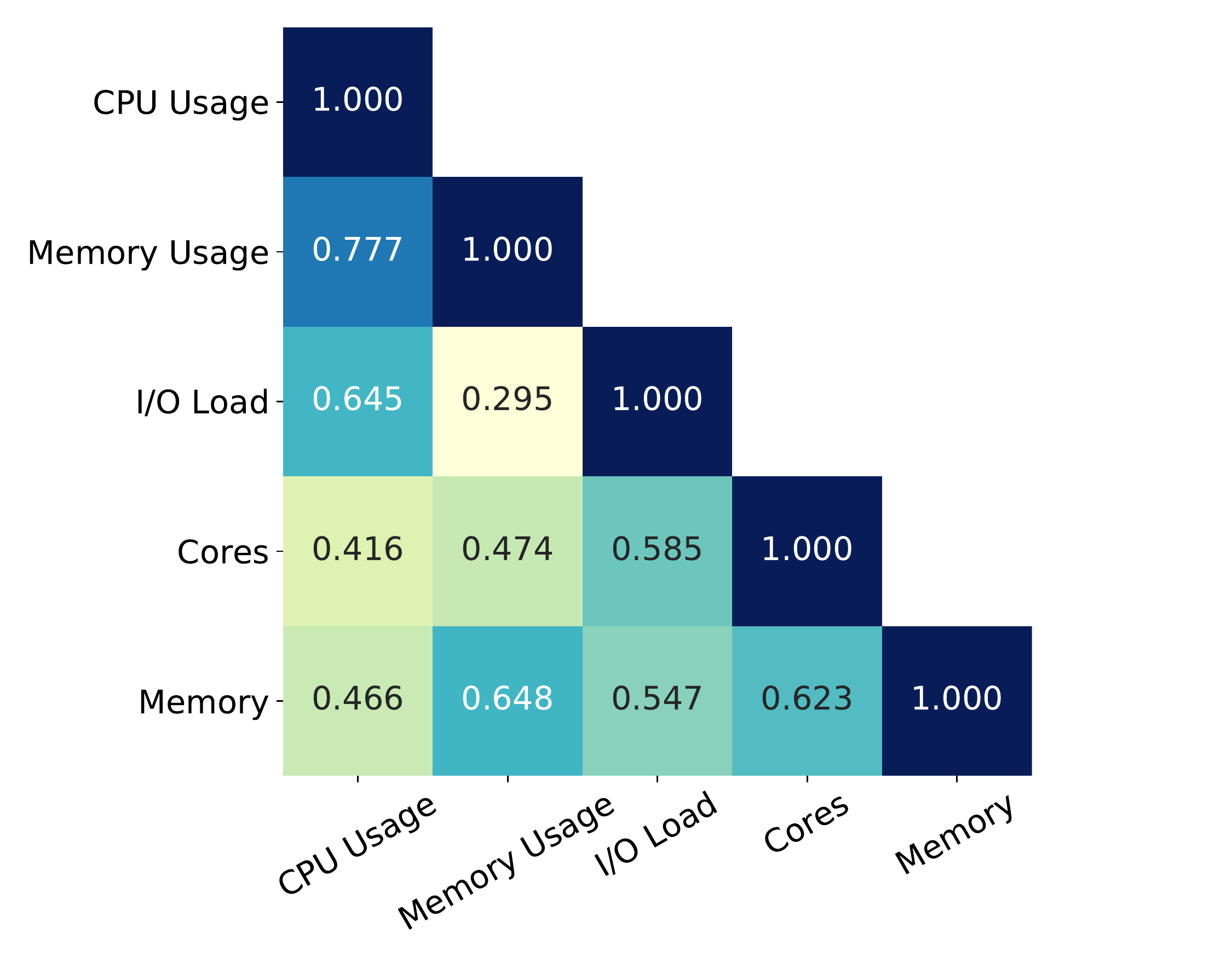}
    \caption{Node-level}
    \label{fig:node_correlations}
  \end{subfigure}
  \caption{Provisioning and Utilization Metrics Correlations}
  \label{fig:correlations}
\end{figure}

We also correlate the VM/node provisioning and utilization metrics using Spearman's correlation~\cite{spearman}, which assesses monotonic relationships between variables (linear or not). We use P95 values of each VM for this analysis. 
We show the results in Figure~\ref{fig:correlations} as a heat map, which intuitively can be interpreted as follows. If metric \textit{x} tends to increase when \textit{y} increases, the correlation coefficient is positive. If \textit{x} tends to decrease when \textit{y} increases, the correlation is negative. A zero correlation indicates that there is no tendency for \textit{x} to either increase or decrease when \textit{y} increases.
A perfect correlation of $\pm1$ occurs when each of the variables is a perfect monotone function of the other.
We observe that CPU and memory usage have a strong positive (but not perfect) correlation, which seems to indicate that the compute-heavy workloads in our dataset are also memory-intensive, but VM-specific tuning is still necessary to determine how much memory should be provided to a VM to go with the amount of CPU resources allocated to it. Further, the node-level I/O usage is not that strongly correlated with memory and CPU usage, indicating that there is an opportunity to co-locate VMs that are just I/O intensive with VMs that are memory or CPU-intensive.

Next we examine the variation in resource utilization across time. The purpose of this analysis is to quantify the need for reallocating resources across VMs within a cluster and to examine the implications of static thresholds.

\begin{figure}[h!]
  \centering
  \begin{subfigure}{0.48\columnwidth}
    \includegraphics[width=1\linewidth]{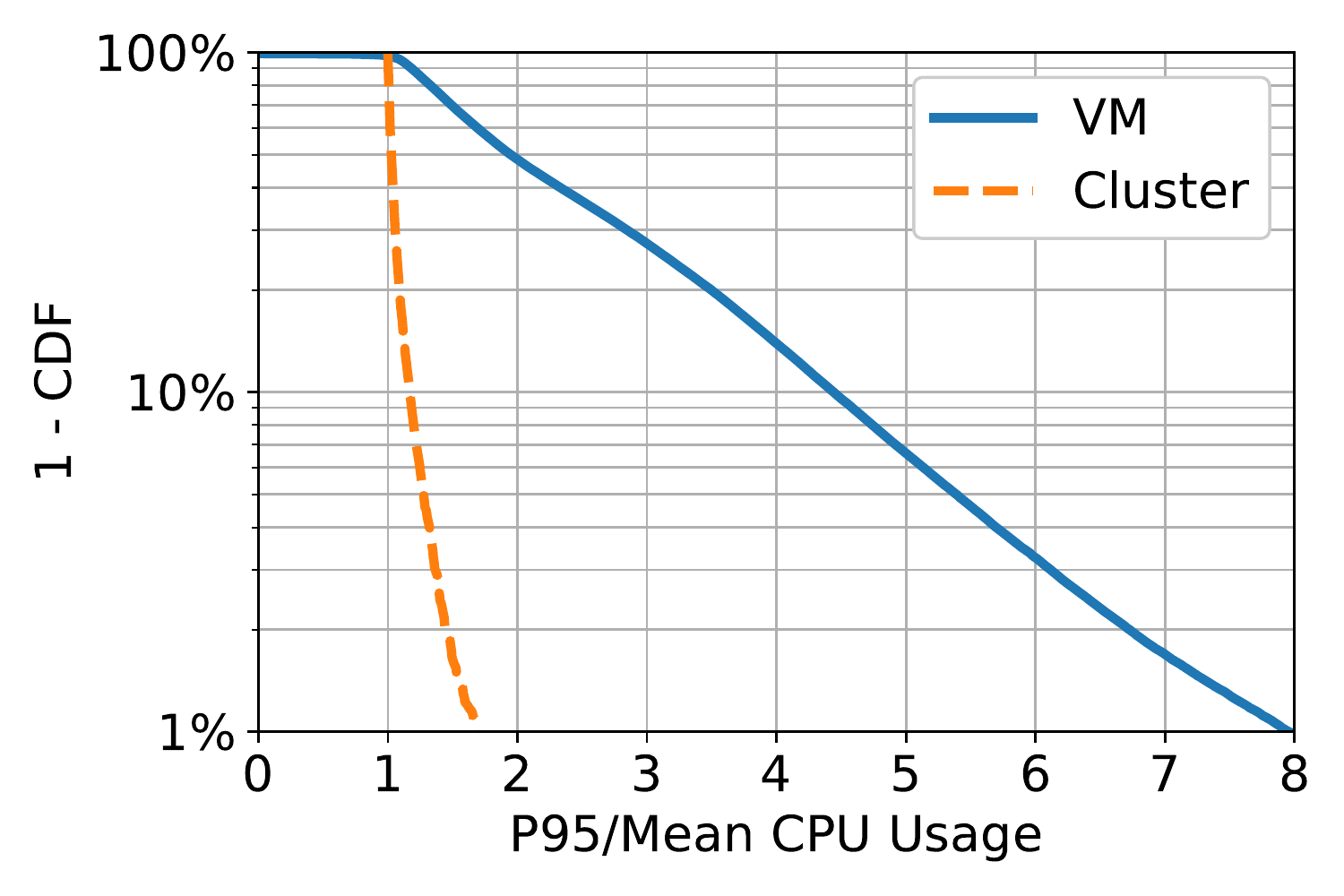}
    \caption{CPU Usage}
    \label{fig:p95_mean_cpu}
  \end{subfigure}
  \begin{subfigure}{0.48\columnwidth}
    \includegraphics[width=1\linewidth]{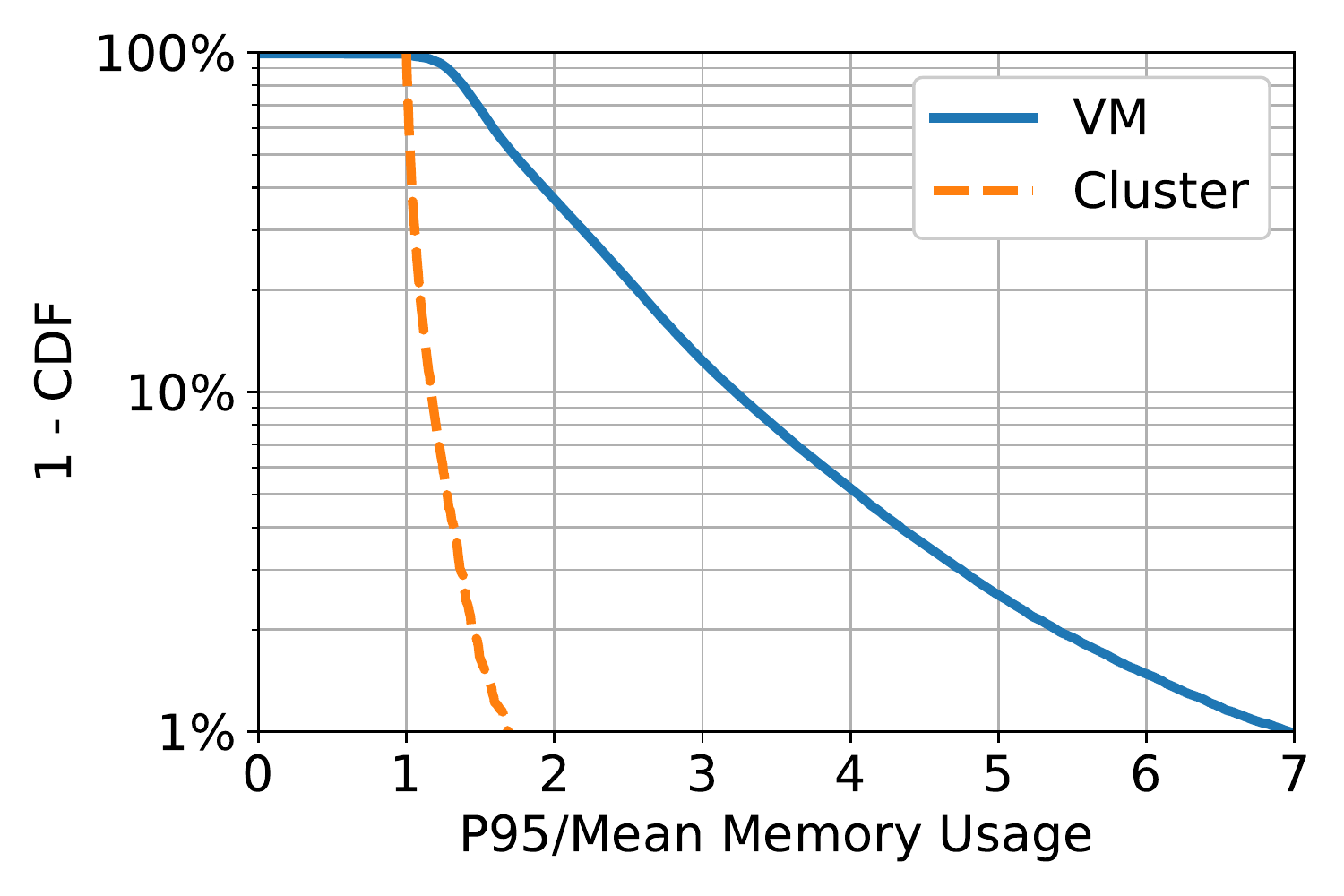}
    \caption{Memory Usage}
    \label{fig:p95_mean_memory}
  \end{subfigure}
  \caption{P95/Mean Usage Ratios}
  \label{fig:p95_mean}
\end{figure}

Figure~\ref{fig:p95_mean} shows the CCDF of the 95$^{th}$ percentile divided by the mean of CPU (\ref{fig:p95_mean_cpu}) and memory (\ref{fig:p95_mean_memory}) usages for both VMs and clusters. 
We notice that $\sim$45\% of the VMs have a P95 at least 2x bigger than the mean, for both metrics, which indicates that there is significant variation across time for many VMs.
However, at a cluster-level, the variation of CPU and memory usage over time is insignificant, indicating that usage spikes are not highly correlated across VMs.

\observation{Many clusters have both underprovisioned and overprovisioned VMs.  In fact, there is significant disparity between the utilization levels of VMs in a cluster, regardless of the resource type. 
This disparity, in turn, provides an opportunity to reallocate resources from VMs that are overprovisioned onto VMs that are underprovisioned, potentially solving both problems. 
However, such a mechanism would have to address two important challenges: (1) it can only reallocate resources between VMs running at a given time, and (2) it has to continuously adapt to the current load given the large temporal variations in VM resource usage.
}

\section{Design}
\label{sec:design}
This section describes the design of \textsc{AdaRes}, a system that changes the physical resources allocated to VMs based on current workload and other attributes of the virtual execution environment. Our system crucially relies on the contextual bandits framework and other techniques to guide the resource adjustment. This section starts with a high-level description of the goals that determined the design of \textsc{AdaRes}, an overview of the system, and a description of its core components. This section complements the system description with background information to assist readers unfamiliar with the contextual bandits framework.

\subsection{\textsc{AdaRes} Goals}
\label{sec:goals}

\textsc{AdaRes} is designed to identify the appropriate resource allocation settings for VMs in enterprise clusters. The goal is to improve cluster execution efficiency by allocating the optimal amount of resources to each VM but without compromising VM performance; that is, the resources allocated to a VM should be just adequate for it to operate without experiencing a slowdown. Thus, \textsc{AdaRes} reduces the resources allocated to overprovisioned VMs and increases resources allocated to underprovisioned ones.  

Note that the VM assignment problem is orthogonal and is out of the scope of this paper, i.e., \textsc{AdaRes} does not determine the optimal node to which a VM is assigned or migrated to; instead, it relies on existing tools, such as VMware's vShpere/vMotion~\cite{vmotion}, to address this challenge. Nevertheless, by optimally setting the resource allocation, \textsc{AdaRes} allows such tools to both pack more VMs into clusters as well as migrate VMs to the appropriate nodes that have sufficient resources to host them~\cite{Ruprecht:2018:VLM}.

We design our system with the goal of achieving the following properties:

\squishlist
    \item \emph{Highly adaptive:} The system should work in a diverse set of operating conditions and identify optimal operating points for a diverse set of cluster, node, and VM configurations.  It should continuously adapt VM configurations in response to changes in workloads. Our choice of contextual bandits is primarily driven by its ability to learn and adapt to such settings.
    \item \emph{Safe allocations:} A key challenge with using bandits in our setting is that the adaptive controller might require a significant amount of unsafe exploration to distill a decent model of cluster behavior. We seek to build a system that can transfer the knowledge gained from simulations and thereby safely streamline the model distillation process in real clusters.
    \item \emph{Modular and configurable:} Our system should provide a configurable framework that can integrate a variety of measurement sensors and operate using configurable prediction models.  Further, we desire a framework that can integrate system management policies defined by the cluster operator (e.g., ensuring that VMs never exceed a certain amount of  utilization for a given resource). Moreover, the approach should be general enough to be able to work with many hypervisors and virtualization environments.
    \squishend

\subsection{\textsc{AdaRes} Components}
\label{sec:adares_overview}

\begin{figure}[tb!]
  \centering
  \includegraphics[width=1\columnwidth]{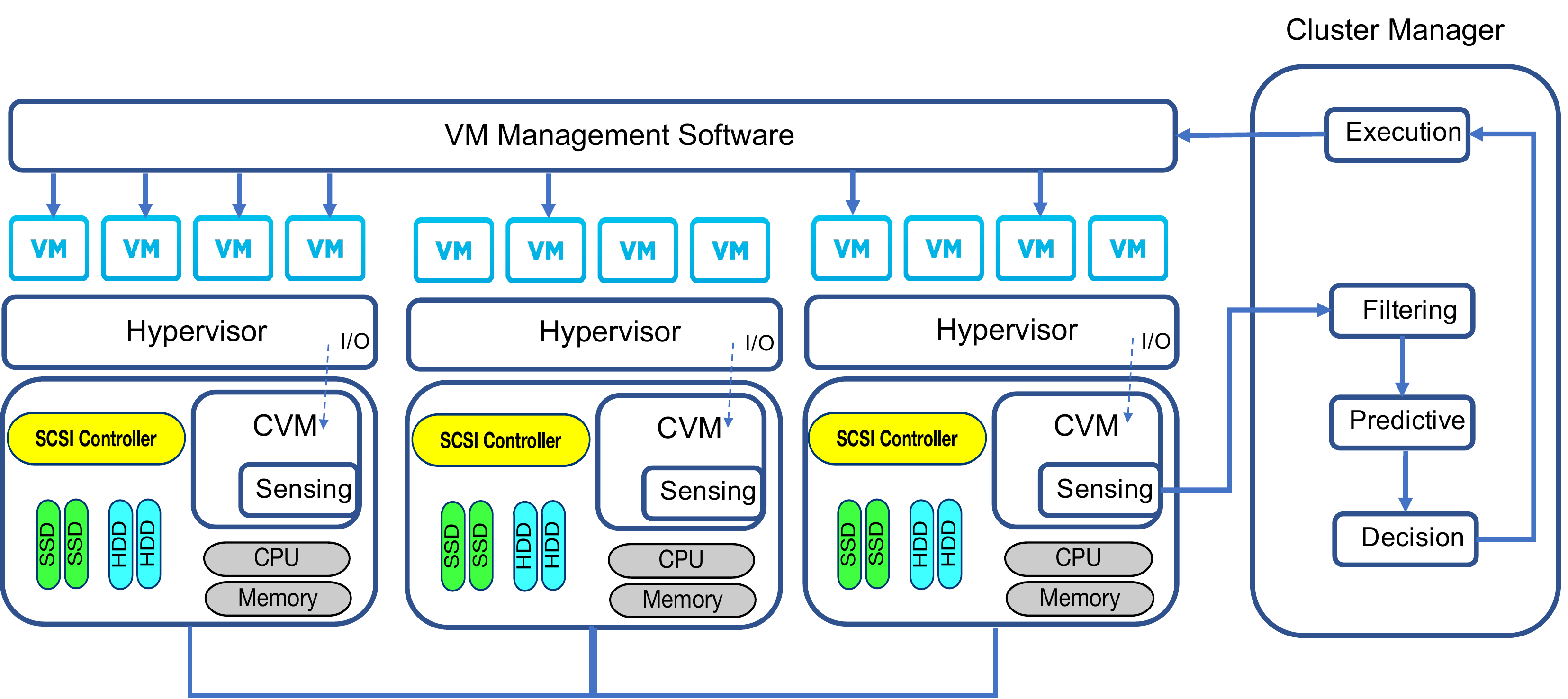}
  \caption{\textsc{AdaRes} Architecture}
  \label{fig:adares}
\end{figure}

This section provides an overview of our system and introduce its core components. Figure~\ref{fig:adares} shows a high level overview of its architecture. 
\textsc{AdaRes} is composed of five core components to optimize VM configurations: the Sensing Service component is deployed on each node in the cluster, whereas the remaining components are executed within the cluster manager node.

\subsubsection{Sensing Service (SS)}  
The Sensing Service is in charge of collecting telemetry data.
The current version collects data at cluster, node, and VM-level. It utilizes sensors on each of the nodes in the cluster to continuously collect information regarding the utilization levels of resources as well as some key performance metrics of the VMs. For instance, it collects information on the CPU and memory utilization of VMs and the number of IOPS performed by each VM, as well as performance metrics such as CPU ready times, virtual memory swap rates, and the latency of I/O operations.  These sensors are typically deployed on the controller VMs (CVMs) running on each node, which not only have access to VM-level metrics (e.g., CPU or memory utilization), but also interpose on I/O operations performed by the VMs on the virtual disks exported by the cluster software.

\subsubsection{Filtering Service (FS)} 
\label{sec:fs}
The Filtering Service component serves as a pre-processing step running on the cluster manager node and is designed to limit the number of VM configuration changes made by the system at a given time. It enables the operator to filter the collected telemetry data based on different strategies. For instance, the FS can filter VMs with CPU usage greater than a certain threshold, randomly select a percentage of the total VMs in the cluster, etc. 
The output of this service is typically a subset of VMs that will be tuned in a given round of the contextual bandit algorithm. As such, the FS component functions as a throttling mechanism, as it can control the rate at which changes are made. This is especially important for highly loaded systems, where changing a large number of VMs at the same time could be counterproductive. It is worth noting that although this component aims to filter inputs in order to avoid unnecessary computations, \textsc{AdaRes} can be configured to also discard outputs---as we shall see in the Decision Service.

\subsubsection{Predictive Service (PS)}  
The Predictive Service along with the Decision Service encompass the core contextual bandit logic in \textsc{AdaRes}. At a high-level, a machine learning (ML) model identifies the appropriate \emph{arms} or \emph{actions} (e.g., scale up/down a VM's memory allocation), given the current \emph{context} or \emph{state} of the VMs in the cluster (e.g., utilization level and other metrics). The actions are chosen based on some expected \emph{reward}, i.e., the effect of taking the actions on the VM performance metrics. We discuss these concepts in greater detail in~\OldS\ref{sec:cb}.

PS exports two methods as part of its interface: (a) \emph{predict}, which outputs the recommended actions for the selected VMs based on the ML model trained to maximize the expected reward, and (b) \emph{learn}, which supports updating the ML model in an online fashion, in order to fold in the actual observed rewards as a consequence of pulling arms (or taking actions). 


\subsubsection{Decision Service (DS)} 
The Decision Service component makes the final decision regarding changes to resource allocations. PS gives hints to DS (e.g., with high confidence PS can recommend to scale down the vCPUs of a particular VM), but it is up to the DS service to follow PS's advice. DS can be seen as a component that leverages the ML-based predictions, but additionally, folds in two other considerations when determining the actual decisions performed by the cluster manager: (a) \emph{exploring} the configuration space to discover the rewards associated with a diverse set of actions, and (b) \emph{leveraging domain knowledge} to make more sensible decisions given the application domain.

For the latter consideration, DS enables users to configure different rules, such as min-max (hard) bounds of utilization and resources, as well as update levels of resources per VM (or group of VMs). 
For example, a user could set a configuration to ensure that VDI VMs can only have between 1 and 4 vCPUs, and 2-8 GiB of memory, and that the system must \emph{always} scale up the vCPUs of those VMs if their CPU usage is more than 90\%. 
Further, on every scaling operation the user can configure, for example, to limit the number of updates of vCPUs to $\pm~1$ and memory to $\pm~40\%$. This feature allows \textsc{AdaRes} to be more cautious or aggressive in accordance with the workload resource tolerance.

\subsubsection{Execution Service (ES)} 
The decisions made by DS are handed to this service, which triggers the operations. Our current prototype supports integration with VMware vSphere,\footnote{\url{https://www.vmware.com/products/vsphere.html}} which acts as the VM Management Software layer. Therefore, we also use VMware ESXi as the nodes' hypervisor.
In order to perform provisioning changes on-the-fly, the underlying guest OS kernel needs support for hot addition/removal of CPUs and memory. However, VMware vSphere only provides native support for hot addition of both resources but not removal. We therefore use other vSphere APIs, in particular, the ability to execute programs directly on guests using the VMware tools installed on the VMs, to perform the adaptations.
Finally, this component also keeps track of the execution progress and notifies the main controller of any failures during the process.

\subsection{Bandit-based Approach}
\label{sec:cb}

We now describe how \textsc{AdaRes} uses contextual bandits for the VM resource allocation problem.  We begin by describing the abstract contextual bandits framework and the rationale behind this choice. We then outline how we apply it to our problem setting. Importantly, this section identifies the challenges in using contextual bandits and how \textsc{AdaRes} addresses them.

\subsubsection{Background}
\label{sec:cb_model}

In the multi-armed bandit (MAB) problem with contextual information, an agent collects rewards for actions taken over a sequence of rounds. In each round, the agent chooses the action to take based on: (a) \emph{context} (or features) of the current round, and (b) \emph{feedback} (or rewards) obtained in the previous rounds. In any given round, the agent observes \emph{only} the reward for the chosen action, thus the feedback is said to be \emph{incomplete}~\cite{Agarwal14tamingthe}.

More formally, the learning agent proceeds in a sequence of discrete trials, $t = 1, 2, 3 ..$.
At each trial $t$, the agent observes the context $\x_t$, and selects an action, $a_t \in \A_t$, where $\A_t$ is the set of all actions available at time $t$. The agent then receives a reward, $r_{t,a_t} \in [0,1]$, and improves its action-selection strategy with the tuple $(\x_t, a_t, r_{t,a_t})$~\cite{Li:2010:CAP:1772690.1772758}.

The total reward for the agent after $T$ trials is defined as $\sum_{t=1}^{T} r_{t,a_t}$. Similarly, the optimal expected $T$-trial reward is defined as $\E [\sum_{t=1}^{T} r_{t,a_t^*}]$, where $a_t^*$ is the action with the maximum expected reward at trial $t$. 
The goal of the agent is to maximize the expected reward, or, equivalently, minimize the \emph{regret} with respect to the optimal action-selection strategy. The regret of the agent after $T$ trials is formally defined as follows:

\begin{equation}
R(T) =  \E [\sum_{t=1}^{T} r_{t,a_t^*}] - \E [\sum_{t=1}^{T} r_{t,a_t}]
\label{eq:regret}
\end{equation}


A fundamental challenge in bandit problems is the need for balancing exploration and exploitation. In order to minimize the regret in Equation~\ref{eq:regret}, the agent \emph{exploits} its past experience and chooses the action that appears to be the best. However, that action might be suboptimal due to the agent's insufficient knowledge. Instead, the agent may need to \emph{explore} by selecting seemingly suboptimal actions in order to gather more knowledge about them~\cite{Li:2010:CAP:1772690.1772758}.
Common applications of contextual bandits include, but are not limited to, personalized news recommendations, clinical trials, and mobile health interventions~\cite{ads2interventions, pmlr-v15-beygelzimer11a}.

\subsubsection{Why Contextual Bandits?}


Contextual bandits can be considered a hybrid between supervised learning and reinforcement learning. The construction of context using features comes from supervised learning, while exploration, necessary for good performance, is inherited from reinforcement learning~\cite{pmlr-v15-beygelzimer11a}.

Training a model offline using any supervised learning algorithm would not work in our case because VM workloads change frequently and many incoming VMs do not have historical records at all. Such approach would require re-training the model with a high frequency to try to keep up with workload changes and its unclear how often this process would be required to attain acceptable results. 
Instead, using an online learning algorithm is more suitable because it automatically and dynamically adapts to new patterns as new data becomes available. 
One can think of an online model trained to predict workload characteristics of VMs. For example, given a new VM context, a model would predict its maximum CPU usage in the next hour, and if it is above certain target threshold, then the system would scale its vCPUs up. However, even if we had a perfectly accurate predictive model, we would not have an easy way to properly fold the result of taking the action into the model, as the prediction task is decoupled from the result of the action. Furthermore, the action taken would have affected the actual max CPU usage of the VM during the hour, complicating the learning process.

We therefore need an online formulation where the learning task itself could estimate the result (i.e., reward) of taking an action, given side information (i.e., VM context). As we do not know what \emph{would have happened} had we taken a different action, our model should take different actions so as to refine its estimates. The two main models that encompass the above characteristics are contextual bandits and reinforcement learning. 
Reinforcement learning (RL)~\cite{Sutton:1998:IRL:551283, Sutton:1984:TCA:911176, Sutton88learningto} is oftentimes seen as an extension of the contextual bandit setting. One difference is that the reinforcement learning agent can take many actions until it observes a reward. For example, in a chess game, the player makes many moves but the reward is only revealed at the end of the game (win, loss, draw). This sparsity makes the problem harder to learn and gives rise to the so-called credit assignment problem, i.e., which actions along the way actually helped the player win? In our setting, however, we do not have to deal with sparse rewards; after we scale a VM, we can sense its performance metrics with our Sensing Service and get an idea of how much the scaling action affected the VM. 
Further, although recent successes in deep reinforcement learning~\cite{mnih-dqn-2015,mnih-atari-2013} have made it quite popular among practitioners, most RL algorithms lack theoretical guarantees. On the contrary, there are many contextual bandit algorithms with strong theoretical guarantees that ensure convergence to an optimal solution~\cite{Li:2010:CAP:1772690.1772758, Agarwal14tamingthe, pmlr-v15-beygelzimer11a}, and they typically have a faster ramp up than their RL counterparts. 

\subsubsection{Context-Actions-Reward}
\label{sec:formulation}


In order to apply contextual bandits to manage VM resources, we need to define the set of features that represent the contexts $\x$, the set of possible actions $\A$, and the reward function. Crucially, all this setup depends on how the rest of the system is structured, as in what can be measured and how the performance of an application VM can be quantified.

\paragraph{Context}

We represent the context of VMs by cluster, node and VM-level features, as well as temporal information.
The context attributes include the various measurements collected by the Sensing Service, e.g., the resource allocations made to VMs, current and historical resource utilization levels (at  VM, node, and cluster  granularities), summary statistics of those (e.g., max, min, average, and P95 utilization), performance metrics that characterize VM behavior (e.g., latency, IOPS, swap rates, CPU ready time, etc.), overcommitment factors of the node and cluster where the VM is running, and others. 

We consider is worth noting that the ability to feed side information into the agent, allows the agent to do context-dependent adaptations, and makes the whole contextual MAB framework well-suited for our setting.
The intuition behind including global information, i.e., cluster and node-level features besides just the VM information, is to aid the agent in making more ``coordinated" scaling decisions across VMs, by also taking into account availability of resources in the host(s), oversubscription levels, etc. For instance, when the side information shows that a node's resources are highly overcommitted, the agent might decide not to increase the resources of its VMs. Or when it detects sinusoidal usage patters in VMs, it may decide to augment and decrease their resources depending on the part of the cycle it is in, and so on.

\paragraph{Actions}
We use a special case of the general contextual bandit framework introduced before, in which the action set $\A_t$ remains unchanged for every round $t$. In particular, we define a total of three actions per resource type (\emph{scale up}, \emph{scale down} or \emph{noop}). For example, the agent can choose to scale up memory and scale down vCPUs, scale down both, neither, etc.
Actions result in resource allocations updates to VMs, and in turn, VMs respond to the new allocations by exhibiting an updated set of utilization and performance metrics, which the agent then uses to update its model.

\paragraph{Reward}

The final step in setting up the bandits formulation is to define the reward function. The primary objective in defining the reward function is to steer the cluster configurations towards states that correspond to minimal VM-level resource allocations without compromising VM performance. Our framework is agnostic to the way the reward function is defined; the only constraint it imposes is that the reward must be a function of the various metrics gathered by the Sensing Service. 

We give a reward of 1 when, irrespective of the action, we move from a ``bad" state to a ``good" one, e.g., from a context with swapping and/or CPU overload to a context without. We also give a payoff of 1 if we make ``good" actions, e.g., if we scale down to increase the usage, but the VMs do not end up incurring in swapping or CPU overload, or if we scale up to try to escape from a state with swapping or high CPU load. 
On the other hand, we penalize (zero reward) actions that lead to bad states, e.g., if we are not swapping and after scaling down we start doing so. 
Finally, we also penalize scaling up/down recommendations of PS if the domain knowledge encoded in DS heuristics (i.e., hard bounds) don't allow them.

We note that there are likely many formulations of the reward function that achieve the desired objective of maximizing system efficiency without hurting VM performance.  We plan to provide the cluster operator with the ability to configure the reward function by incorporating additional information from application-level performance metrics, as that would allow for more precise reward valuations and faster convergence to optimal configurations.

\subsubsection{Safe Allocations and Faster Training: \emph{Sim2Real}}
\label{sec:sim2real}
    
Another challenge of applying bandit-based approaches in our setting is that we need to ensure reasonable performance and respect ``safety'' constraints during the learning process. We need to be extra cautious not to mess up with VMs while exploring different actions but, at the same time, we want to make the right decisions as soon as possible. Incorporating ``prior knowledge'' before the agent is deployed might help to speed up learning and reduce the amount of interactions with the real VMs, which may be limited and costly~\cite{DBLP:journals/corr/abs-1708-05866, DBLP:conf/icml/KanskySMELLDSPG17}.

Inspired by the robotics community, as well as prior work on the systems space~\cite{Ferguson:2012:JGJ:2168836.2168847}, herein, we build a cluster simulator to pre-train our agent. The idea is to then transfer the knowledge gained while training on this (cost-less) simulator to \emph{bootstrap} our agent before it is deployed in real clusters.
We start the section by stating what we need from the simulator, the challenges its construction presents, and how we address those challenges in our work.

\paragraph{Requirements}

The simulator should provide an easy mechanism to emulate, to some extent, the dynamics of a cluster. We are interested in modeling what happens to VM performance metrics once we perform configuration changes. In other words, we need (simplistic) analytical models of the environment that our Sensing Service can query to obtain the contextual information (or features) and rewards necessary to train our agent.

\paragraph{Challenges}

Although we brought robotics into the picture, building a simulator of a robot is a completely different endeavor. Therein, the well-defined rules of physics (e.g., gravity) aid in the otherwise even harder process. Herein, we don't have those; the large number of components and connections (e.g., VMs, nodes, storage devices, queues), the intricate component dependencies (e.g., hypervisors multiplexing shared resources), and the irregular interactions and resource needs (e.g., different workloads changing over time) complicate our ability to create a simulator that faithfully represents a real cluster. Nevertheless, from a machine learning standpoint, we don't need an entirely ``accurate" simulator, we need a reasonable initialization of what we believe the dynamics are, and then we can keep updating those beliefs as we keep on training in the real cluster. By incorporating (incomplete) initial knowledge, the agent would be exposed to the relevant regions of the context and action spaces from the earliest steps of the learning process, thereby eliminating the time needed in random exploration for the discovery of these regions, as in safe reinforcement learning~\cite{Garcia:2015:CSS:2789272.2886795}.

\paragraph{Data-driven approach}

Following the ``reasonable" premise above, we use a data-driven strawman approach to build our cluster simulator. We run a set of controlled experiments on synthetic workloads that aim to mimic the ones we observe in real clusters, and we perform different changes to VM configurations and record their impact. For example, we change the amount of vCPUs assigned to VMs and observe how those changes affect their CPU usage. Further, we run different I/O benchmarks using Vdbench~\cite{vdbench} to profile IOPS and latencies for different representative workloads (e.g., 8k random reads, 8k random writes, 1M sequential writes, 8k 50\% random reads and 50\% random writes, burst, sequential) at different rates, and with different outstanding I/O per node. This profile data gives us an idea of the rates at which our system can (roughly) serve the different types of I/O. Given that we have an estimate of the service rates, and as we know the amount of outstanding I/O in a node, we then resort to queueing theory (single server model or M/M/1) to compute arrival rates per node, and then derive approximate latencies (or wait times) in the system. Finally, we also create multi-queue multiprocessor schedulers with round robin per node, to roughly estimate CPU ready times among the VMs running in those nodes. 
Although some initial results on the fidelity of our simulator are in~\OldS\ref{sec:sim_fidelity}, we acknowledge that the addition of extra features to the simulator can (and probably will) get us better results on real clusters. We leave that to future work.

\subsection{Contextual Bandits meet \textsc{AdaRes}}

Having introduced the core constructs of \textsc{AdaRes}, and MAB with contextual information , in this section, we show how we use our system together with the latter framework to dynamically adjust VM resources. 

The core services described in \OldS\ref{sec:adares_overview} are orchestrated by a controller running in the cluster manager node. Listing~\ref{alg:controller_loop} shows a (simplified) example of the main controller loop, the heart of our agent.
The agent starts sensing the cluster state (cluster, node, and VM-level information). Note that in our setting we define contexts $\x_t \in \mathbb{R}^d$ per VM, thus here $X_t \in \mathbb{R}^{nxd}$, where $n$ is the number of VMs in the cluster, and $d$ the size of our feature vector. The $i^{th}$ row in matrix $X_t$ represents the context of the $i^{th}$ VM.
 
\begin{algorithm}
\floatname{algorithm}{Listing}
\begin{algorithmic}[1]
  \STATE{$X_t \gets$ ss.sense(cluster) (sense context)}
  \FOR{$t = 1, 2 ... $}
    \STATE{$X_t \gets$ fs.filter($X_t$) (filter VMs)}
    \STATE{$P_t \gets$ ps.predict($X_t$) (get prediction values)}
    \STATE{$A_t \gets$ ds.decide($P_t$) (explore/exploit + domain knwl)}
    \STATE{es.execute($A_t$) (execute actions)}
    \STATE{$X_{t+1} \gets$ ss.sense(cluster) (sense new context)}
    \STATE{$R_{t,A_t} \gets$ reward($X_{t}$, $A_t$, $X_{t+1}$) (compute rewards)}
    \STATE{ps.learn($X_t$, $A_t$, $R_{t,A_t}$) (online learning)}
    \STATE{$X_t \gets X_{t+1}$ (update context)}
  \ENDFOR
\end{algorithmic}
\caption{\textsc{AdaRes} Controller}
\label{alg:controller_loop}
\end{algorithm}

The agent then uses FS to select $b$ VMs eligible for allocation updates in the current round, where $b \leq n$, and contacts the Predictive Service to obtain the recommendations for those filtered VMs ($P_t \in \mathbb{R}^{bx|\A_t|}$, where $|\A_t| = 9$ is the number of possible actions) (Line 4). In this and the next step is where the bandits algorithm comes into play. After obtaining the predictions, the Decision Service uses an exploration/exploitation strategy together with its domain knowledge to decide which actions to take ($A_t \in \mathbb{R}^{bx1}$, i.e., only one action per VM). 
The set of actions are passed to ES for the actual execution (Line 6). After the actions are executed, the agent uses the Sensing Service to get a sense of the actions' impact on the VMs performance metrics. Note that $X_{t+1} \in \mathbb{R}^{nxd}$, i.e., we sense the whole cluster, not just the previous $b$ filtered VMs. We do this because we will use these new contexts in the next iteration (Line 10), and because the filtering step (Line 3) may select a different subset of VMs than in previous iterations. 
The agent computes the rewards \emph{only} for the $b$ filtered VMs of the current round. 
Finally, the agent learns the benefits/drawbacks of taking actions $A_t$ for contexts $X_t$ in Line 9.

\section{Evaluation}

We implemented \textsc{AdaRes} in about 7.8 kLOC of Python. Our current prototype is built in the context of the same Nutanix commercial virtualization product that we used to collect the cluster measurements. In this section we present the evaluation of our prototype, with experiments on a real cluster. 

\subsection{Evaluation Setup}
\label{sec:infra_setup}

\paragraph{Cluster}
We have full control over an experimental cluster. This mainly homogeneous cluster consists of a total of 48 cores, a CPU capacity of 115.2 GHz and 512 GiB of RAM, on which we run around 20-36 VMs. 



\paragraph{Virtualization Software}
We use VMware ESXi 5.5.0 as the hypervisor, and our Execution Service talks to vSphere to change the virtual hardware associated with the different VMs.
We generate VM images with CentOS Linux 7, kernel version 3.10.x, which supports hot add/removal of CPU and memory. We use VMware vSphere APIs to execute programs on the guests to perform the adaptations. Only VMware Tools software needs to be installed in the guest OS, as the resources addition/removal can be done with native Linux programs (echo and grep)~\cite{linux_cpu_hotplug, linux_mem_hotplug}.
Further, we clone the VMs in our experiments from the three instance types shown in Table~\ref{tab:instances_and_min_max_bounds}. None of the VMs can have less than 1 vCPUs and 2 GiB of RAM, but their maximums differ based on the type. Finally, we set the same tuning aggressiveness for all VMs,  $\pm~1$ for vCPUs and $\pm~512$ MiB for memory.

\begin{table}[th!]
\begin{center}
{\footnotesize
\begin{tabular}{ccccc} 
\toprule
  \multirow{3}{*}{\textbf{VM Instances}} &
  \multicolumn{4}{c}{\textbf{Resources}} 
\\
\cmidrule(lr){2-5}
 & \multicolumn{2}{c}{\textbf{Initial}} & \multicolumn{2}{c}{\textbf{Min-Max }}
\\
  & \textbf{vCPUs} & \textbf{Mem (GiB)} & \textbf{vCPUs} & \textbf{Mem (GiB)} \\
\midrule
\emph{large} & 2 & 3.75 & 1-4 & 2-7.5 \\
\emph{xlarge} & 4 & 7.5 & 1-8 & 2-15 \\
\emph{2xlarge} & 8 & 15 & 1-16 & 2-30 \\
\bottomrule
\end{tabular}
}
\end{center}
\vspace{-0.3cm}
\caption{VM Instance Types and Min-Max Ranges}
\label{tab:instances_and_min_max_bounds}
\end{table}

\paragraph{Workloads}
We simulate different workloads using a modified version of flexible I/O tester~\cite{fio}, where we can configure the VM CPU load, the workload active memory size, and the I/O operations per second. We attempt to mimic the real workloads we observe in our traces, some VDI-based workloads, other Server-like workloads (e.g., SQL server), etc. 
We mainly issue 8k block-sized I/O. Depending on the workload, we do random reads, random writes, and both random reads and writes (50\% each, 70-30\%, or 80-20\%). 



\paragraph{Methods}
We use the following methods in our experiments:
\squishlist
\item \emph{passive}, where no configurations adjustments are done to VMs. This is the baseline currently deployed in our clusters,
\item \emph{reactive}, where we sense information about the VMs and if their usages are above/below certain threshold(s), we perform the adaptations, 
\item \emph{proactive}, similar to \emph{reactive}, but uses a machine learning model to predict maximum usages sometime in the near future (e.g., 10 minutes). It performs changes if the predicted utilization levels deviate from the configured target threshold(s), and 
\item \emph{bandits}, our method, where we adjust resources using contextual bandits. 
\squishend

We use 75\% as the underprovisioned threshold for the \emph{reactive} and \emph{proactive} baselines; that is, if the current or predicted VM resource usage (either CPU or memory) is above 75\%, the system scales the resource(s) up. Similarly, we use a 25\%-threshold to indicate overprovisioning, i.e., if the current or predicted VM resource usage is below that threshold, we scale the resource(s) down. Our system makes decisions every 5 minutes.

\paragraph{Machine Learning}
Further, we use two linear models, one for each resource, to predict the max utilization of each resource in the next 10 minutes, in the \emph{proactive} baseline. We train the models using stochastic gradient descent~\cite{Bottou10large-scalemachine} with $l_2$ regularization and the squared loss, and we use the default hyperparameters of scikit-learn~\cite{scikit-learn}.
For our method, \emph{bandits}, we use LinUCB~\cite{Li:2010:CAP:1772690.1772758}, a popular Upper Confidence Bound (UCB)~\cite{ucb_algo} algorithm. UCB algorithms are based on the principle of \emph{optimism in the face of uncertainty}. On an incoming context, LinUCB computes the estimated reward and the uncertainty, and chooses the action with the highest score (estimated reward + uncertainty). We set the exploration constant to $0.5$ (higher means more exploration), and the regularization parameter of the ridge regression to $0.01$.

\subsection{Results}
\label{sec:results}

\subsubsection{Cluster Simulator Fidelity}
\label{sec:sim_fidelity}

We start off by evaluating the fidelity of our cluster simulator. Herein, we instantiate 26 \emph{xlarge} VMs in our cluster. We group them in four distinct groups, each with a different workload pattern and I/O intensity. We perform random configuration changes during a 8-hour period. We record all the actions done along the way, and we then replay those exact same actions in our simulator.

\begin{figure}[h!]
  \centering
  \begin{subfigure}{0.48\columnwidth}
    \includegraphics[width=1\linewidth]{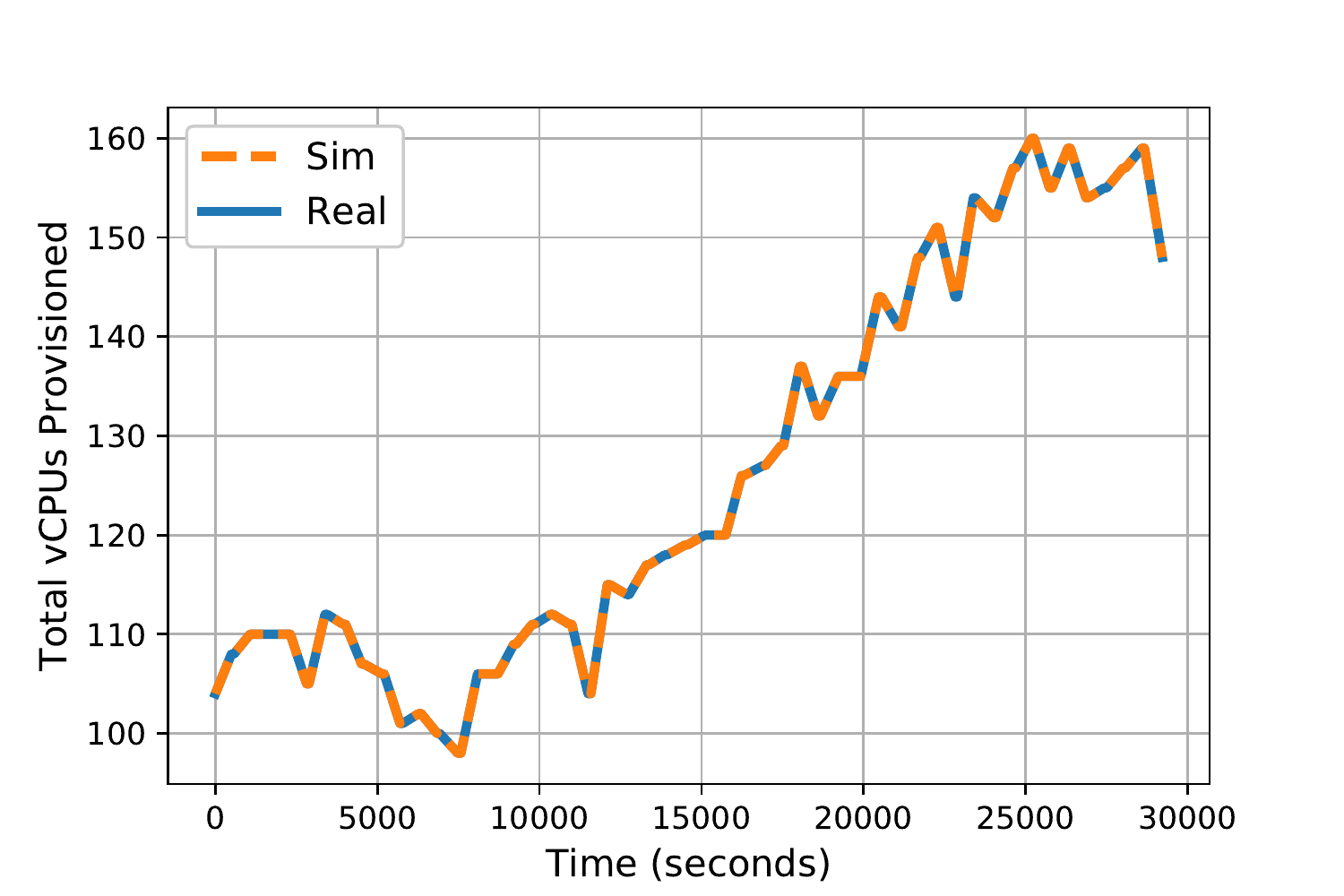}
    \caption{Provisioned vCPUs}
    \label{fig:sim_fidelity_vcpus_prov}
  \end{subfigure}
  \begin{subfigure}{0.48\columnwidth}
    \includegraphics[width=1\linewidth]{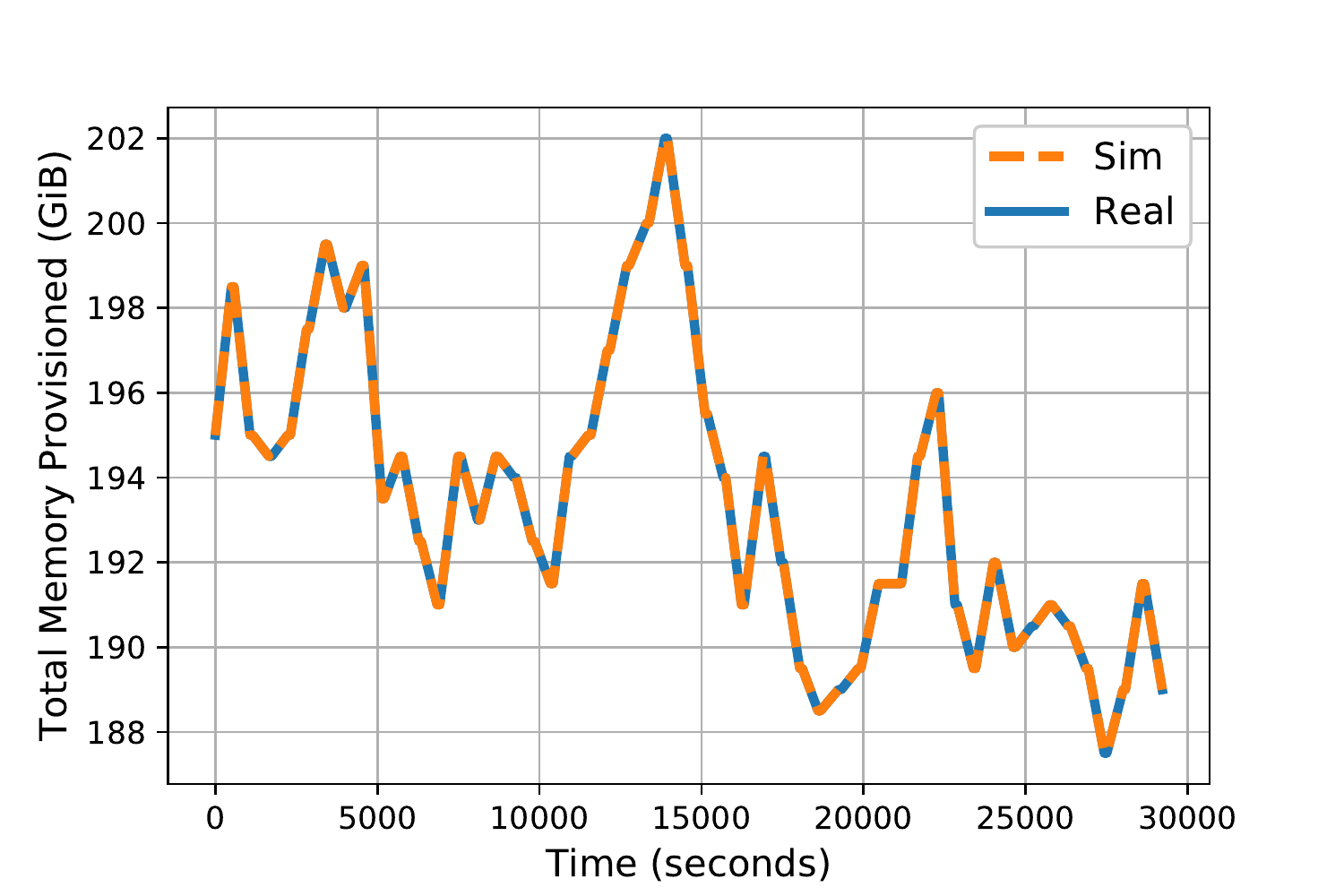}
    \caption{Provisioned Memory}
    \label{fig:sim_fidelity_mem_prov}
  \end{subfigure}
  \begin{subfigure}{0.48\columnwidth}
    \includegraphics[width=1\linewidth]{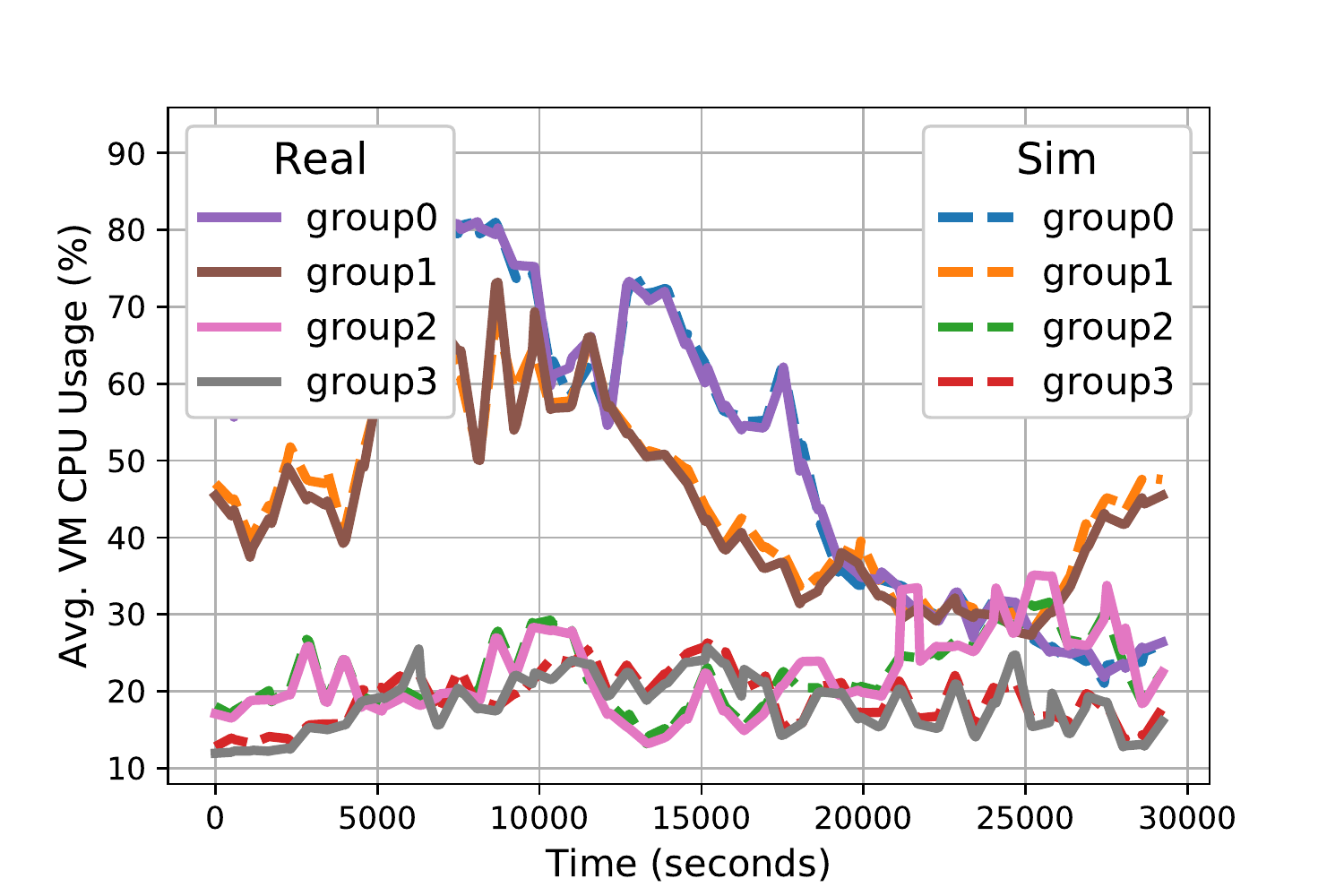}
    \caption{CPU Usage}
    \label{fig:sim_fidelity_cpu_usage}
  \end{subfigure}
  \begin{subfigure}{0.48\columnwidth}
    \includegraphics[width=1\linewidth]{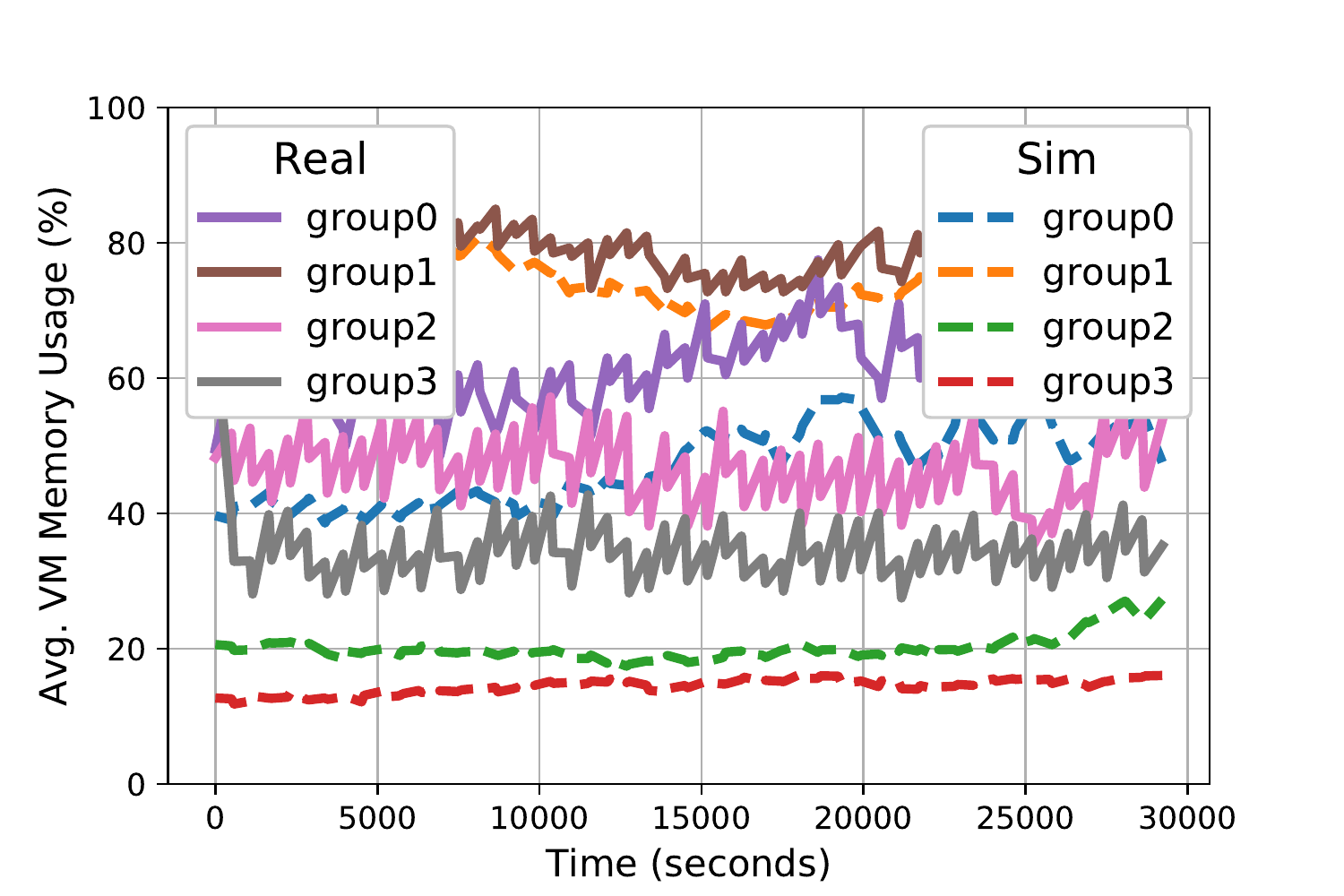}
    \caption{Memory Usage}
    \label{fig:sim_fidelity_mem_usage}
  \end{subfigure}  
  \caption{VM Resource Provisioning and Utilization}
  \label{fig:sim_fidelity_prov_utilization}
\end{figure}

Figures~\ref{fig:sim_fidelity_vcpus_prov} and~\ref{fig:sim_fidelity_mem_prov} show the total vCPUs and memory provisioned across the VMs over time. Both the simulator (Sim) and the real cluster (Real) lines overlap, as we are replaying the same actions in the simulator.
More interestingly, Figure~\ref{fig:sim_fidelity_cpu_usage} shows the average VM CPU usage across the four different VM groups. We observe that the simulator is doing a pretty good job in estimating the CPU usage of all the groups when we perform adaptations. Similarly, Figure~\ref{fig:sim_fidelity_mem_usage} illustrates the memory usage across groups. Herein, we note that our simulator mostly underestimates the usage, which is most notoriously for groups 2 and 3. However, it seems to follow the line trend (e.g., groups 0 and 1) but is off by some constant factor.

\begin{figure}[h!]
  \centering
    \includegraphics[width=0.48\linewidth]{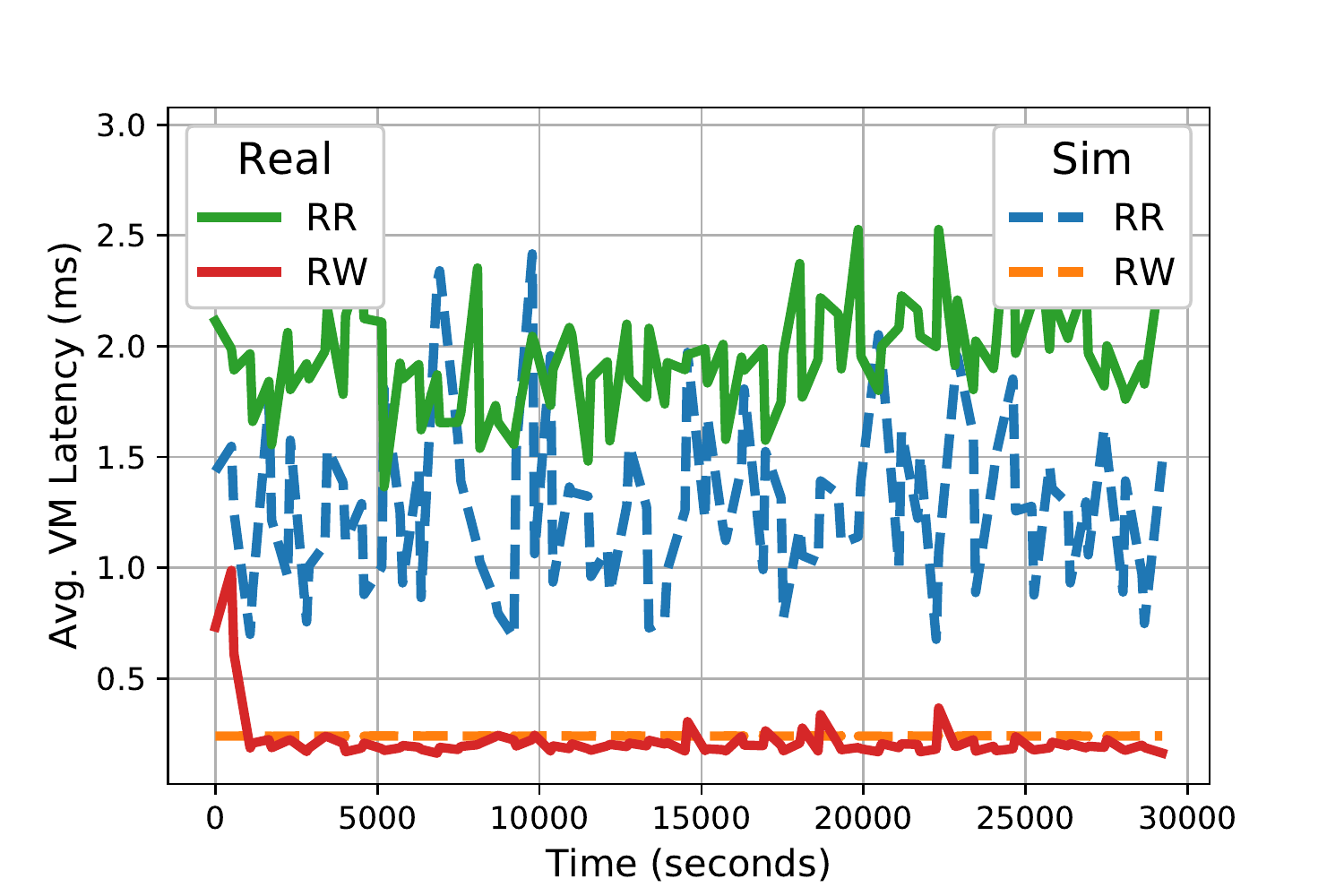}
  \caption{Latency}
  \label{fig:sim_fidelity_latency}
\end{figure}

Finally, Figure~\ref{fig:sim_fidelity_latency} shows the average VM latency decomposed in groups doing random reads (RR) and random writes (RW). We see that our simulator does a better job at estimating RWs operations, though it also does a decent job for random read I/O.

\subsubsection{Transfer Learning: \emph{Sim2Real}}
\label{sec:results_tl}

In this section we evaluate how transfer learning helps to speed up training in real clusters. We run different static workloads across a set of 36 VMs, 12 of each of the instance types described in Table~\ref{tab:instances_and_min_max_bounds}, for a period of $\sim$4 hours. Herein, we compare the two flavors of our bandit-based approach, with and without transfer learning. Note that we pre-train in our simulator using VMs that run other workloads in order to avoid overfitting. Still, if we were running the same workloads and overfitting, it would be an extra evidence of the reasonable performance of our simulator.

\begin{figure}[h!]
  \begin{subfigure}{0.48\columnwidth}
    \includegraphics[width=1\linewidth]{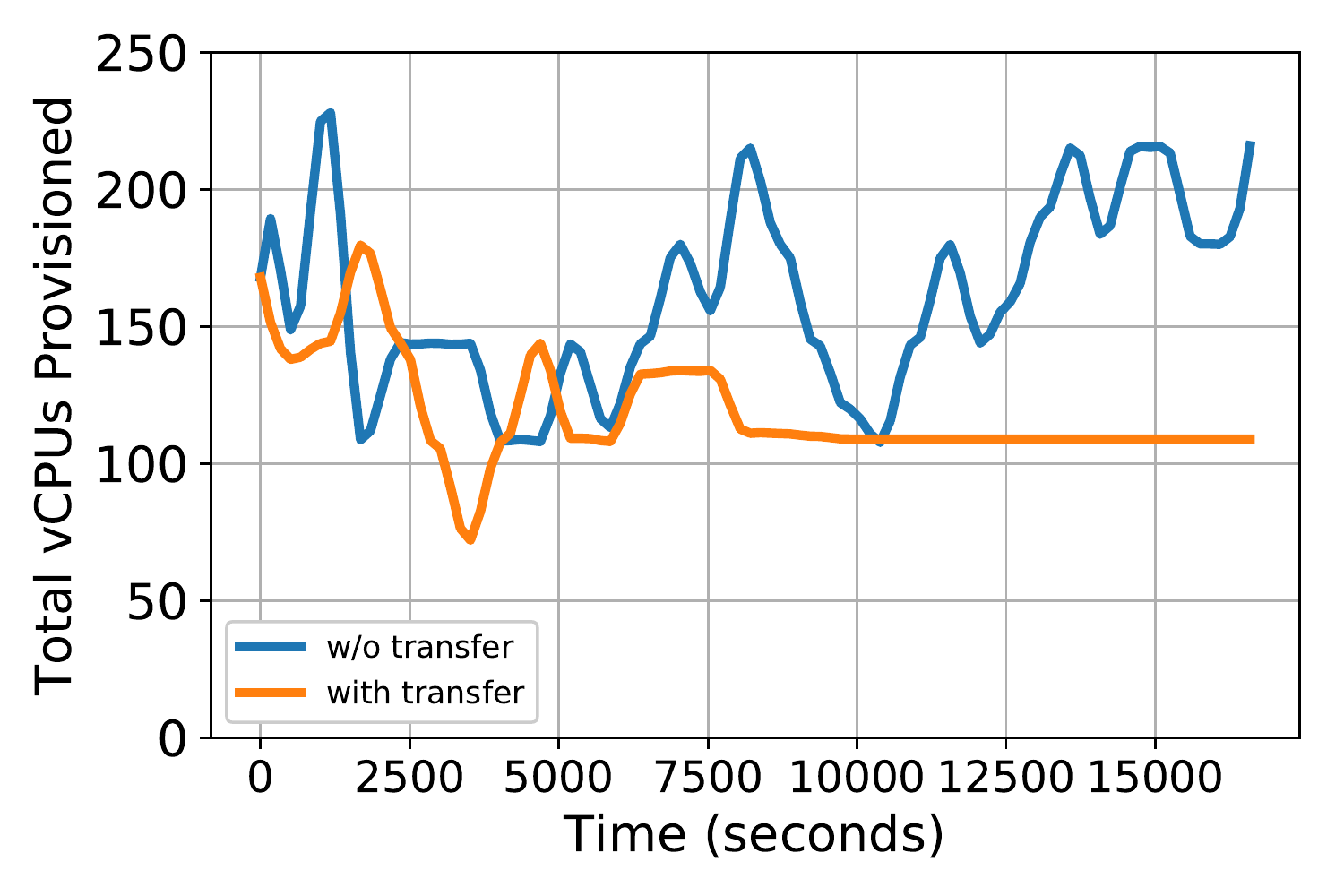}
    \caption{Provisioned vCPUs}
    \label{fig:sabine_transfer_no_transfer_vcpus_allocations}
  \end{subfigure}
  \begin{subfigure}{0.48\columnwidth}
    \includegraphics[width=1\linewidth]{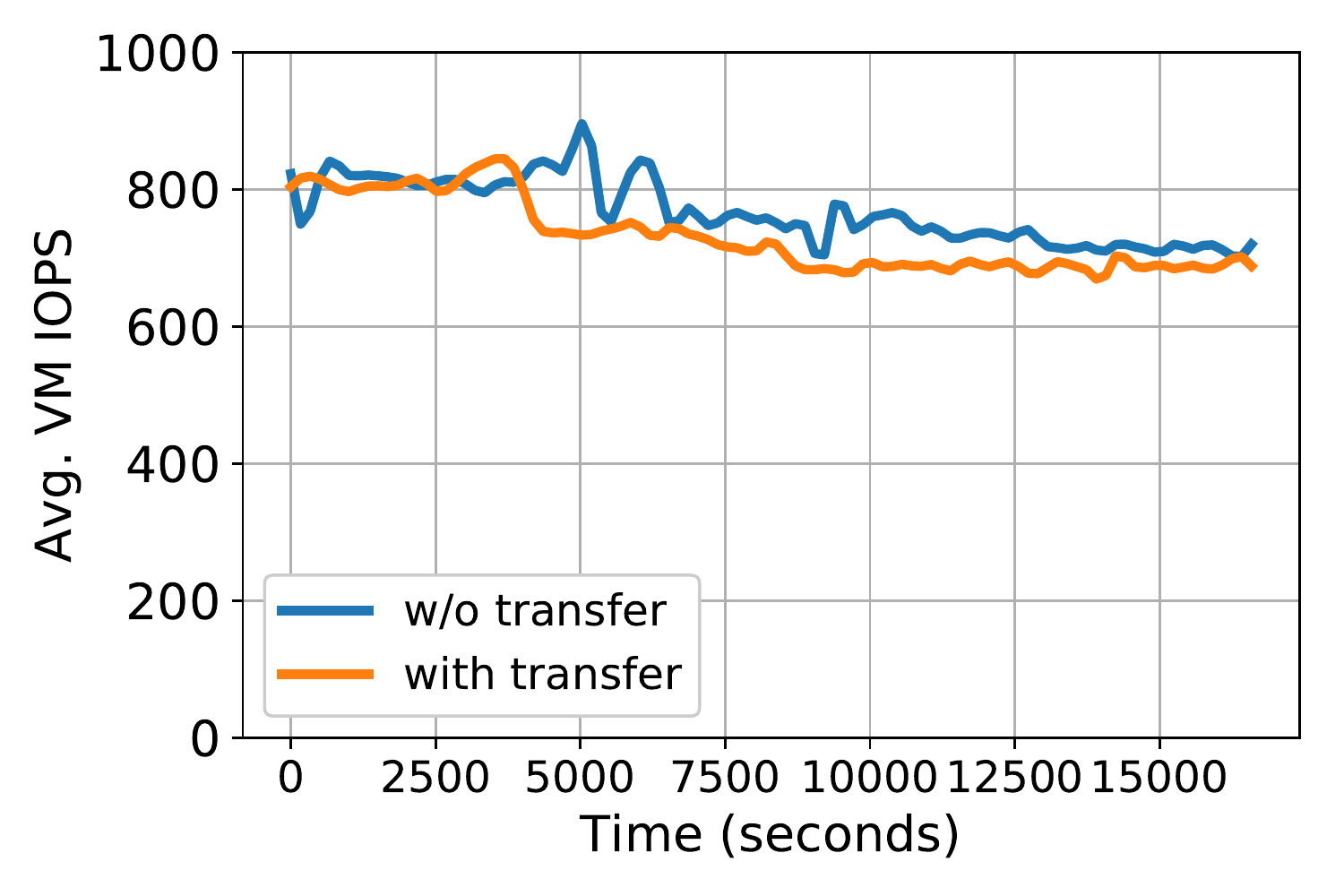}
    \caption{VM IOPS}
    \label{fig:sabine_transfer_no_transfer_vm_iops}
  \end{subfigure}
  \caption{vCPUS Allocations and VM IOPS}
  \label{fig:transfer_no_transfer_alloc_iops}
\end{figure}

Figure~\ref{fig:sabine_transfer_no_transfer_vcpus_allocations} shows the total vCPUs provisioned over time for both bandit-based approaches, with and without transfer learning. We observe that the allocations are much more stable when we pre-train. Transfer learning lead us to a 2$\times$ saving of vCPUs allocations for this workload (109 vCPUs as opposed to 216). Even more, without pre-training, the agent ends up allocating more vCPUs than the ones it started with. This latter statement highlights the importance of safe exploration while applying these type of methods. 
Figure~\ref{fig:sabine_transfer_no_transfer_vm_iops} shows the average I/O operations per second of the VMs in this workload. We observe that, even though we saved 2x vCPUs, we are still able to perform very close to the vanilla bandit version in terms of IOPS.

\begin{figure}[h!]
  \begin{subfigure}{0.48\columnwidth}
    \includegraphics[width=1\linewidth]{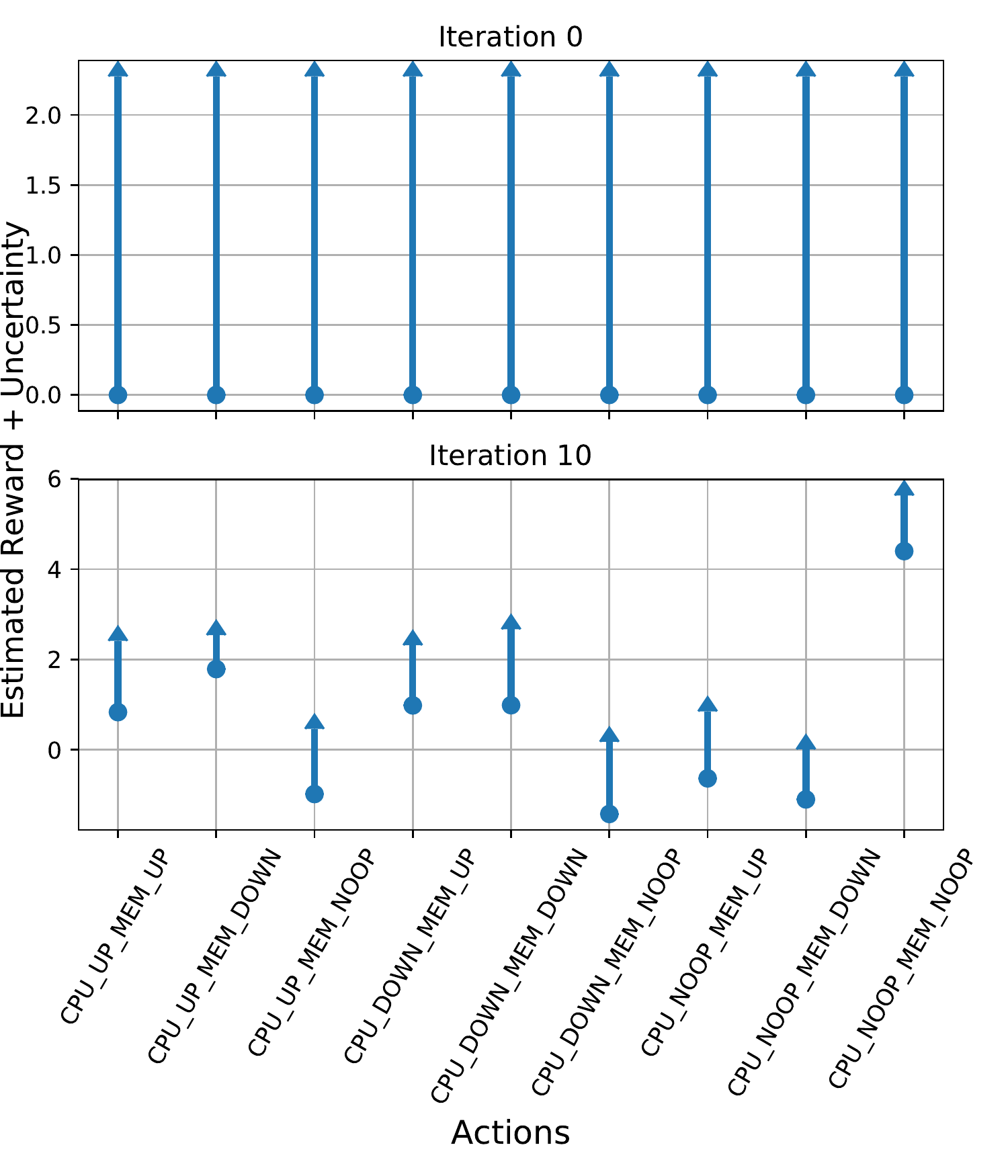}
    \caption{Without Transfer Learning}
    \label{fig:sabine_explanation_without_transfer}
  \end{subfigure}
  \begin{subfigure}{0.48\columnwidth}
    \includegraphics[width=1\linewidth]{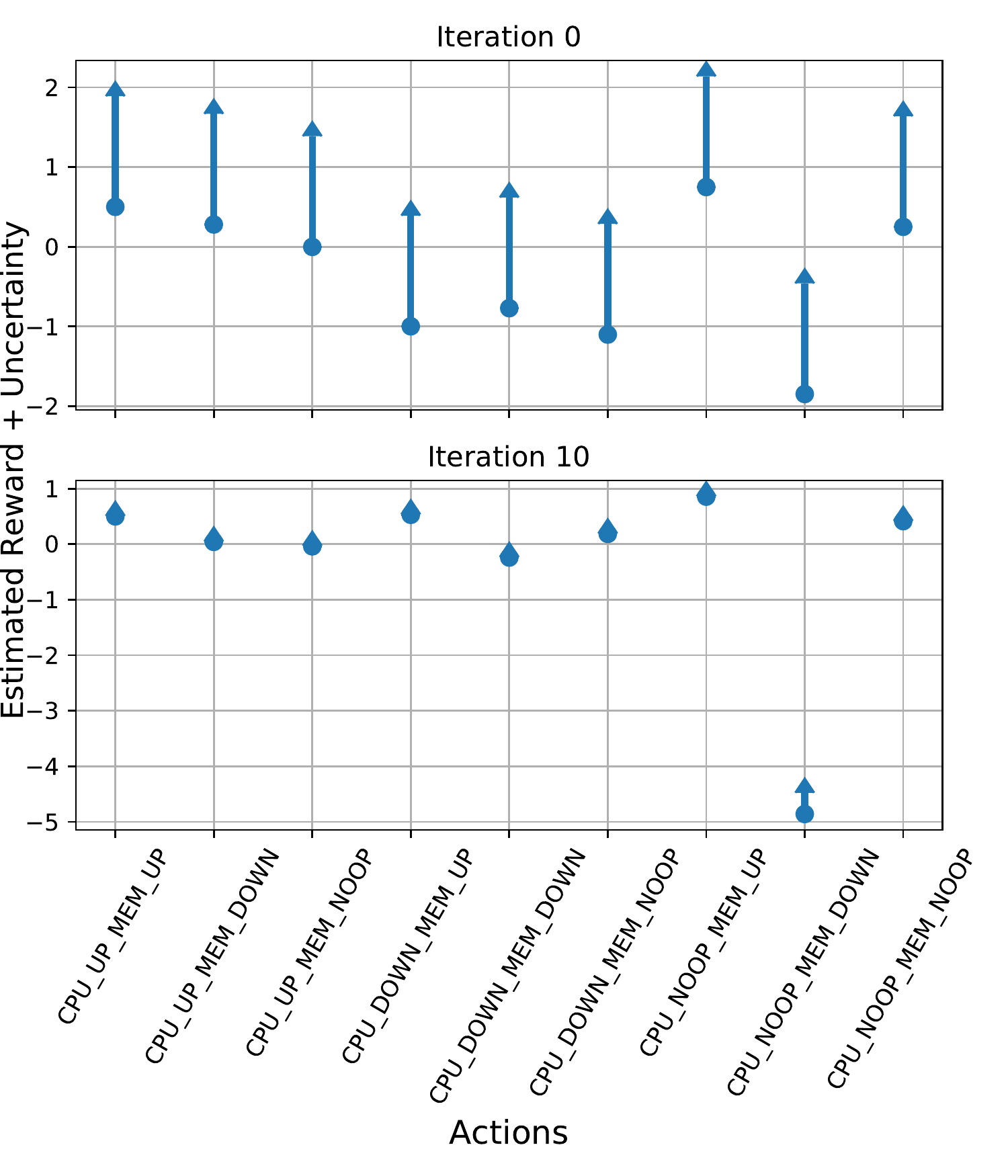}
    \caption{With Transfer Learning}
    \label{fig:sabine_explanation_with_transfer}
  \end{subfigure}
  \caption{LinUCB and Transfer Learning}
  \label{fig:transfer_learning_explanations}
  \vspace{+0.4cm}
\end{figure}

We now illustrate how transfer learning helps LinUCB to accelerate training. Figure~\ref{fig:transfer_learning_explanations} shows the estimated reward and uncertainty of the different actions for a random VM context that has memory underprovisioning. We observe that the estimated rewards start at zero (solid dots) and uniform uncertainty (long lines with caps), when we start training from scratch (top of Figure~\ref{fig:sabine_explanation_without_transfer}). As the agent learns, the confidence bounds shrink for that same context. However, the agent still recommends to do nothing CPU\_NOOP\_MEM\_NOOP, the action with highest score. On the other hand, Figure~\ref{fig:sabine_explanation_with_transfer} shows the benefits of ``bootstrapping" our model. At the top, we see non-uniform confidence bounds. Note that the agent is able to recommend the right action for this context (CPU\_NOOP\_MEM\_UP), from the beginning, due to the knowledge transfer. Few iterations later, the upper confidence bounds are close to the expected reward, and the leading actions are still the ones that involve scaling up memory. 

\subsubsection{Workloads}

\paragraph{Static} 

Herein, we evaluate static workloads, which are characterized by a somewhat flat utilization profile over time. 
To that end, we use the same setting as~\OldS\ref{sec:results_tl}, where we run workloads on a set of 36 VMs, 12 of each instance type, during 4 hours. We only report results on the \emph{bandits} version that uses transfer learning.

\begin{figure}[h!]
  \begin{subfigure}{0.48\columnwidth}
    \includegraphics[width=1\linewidth]{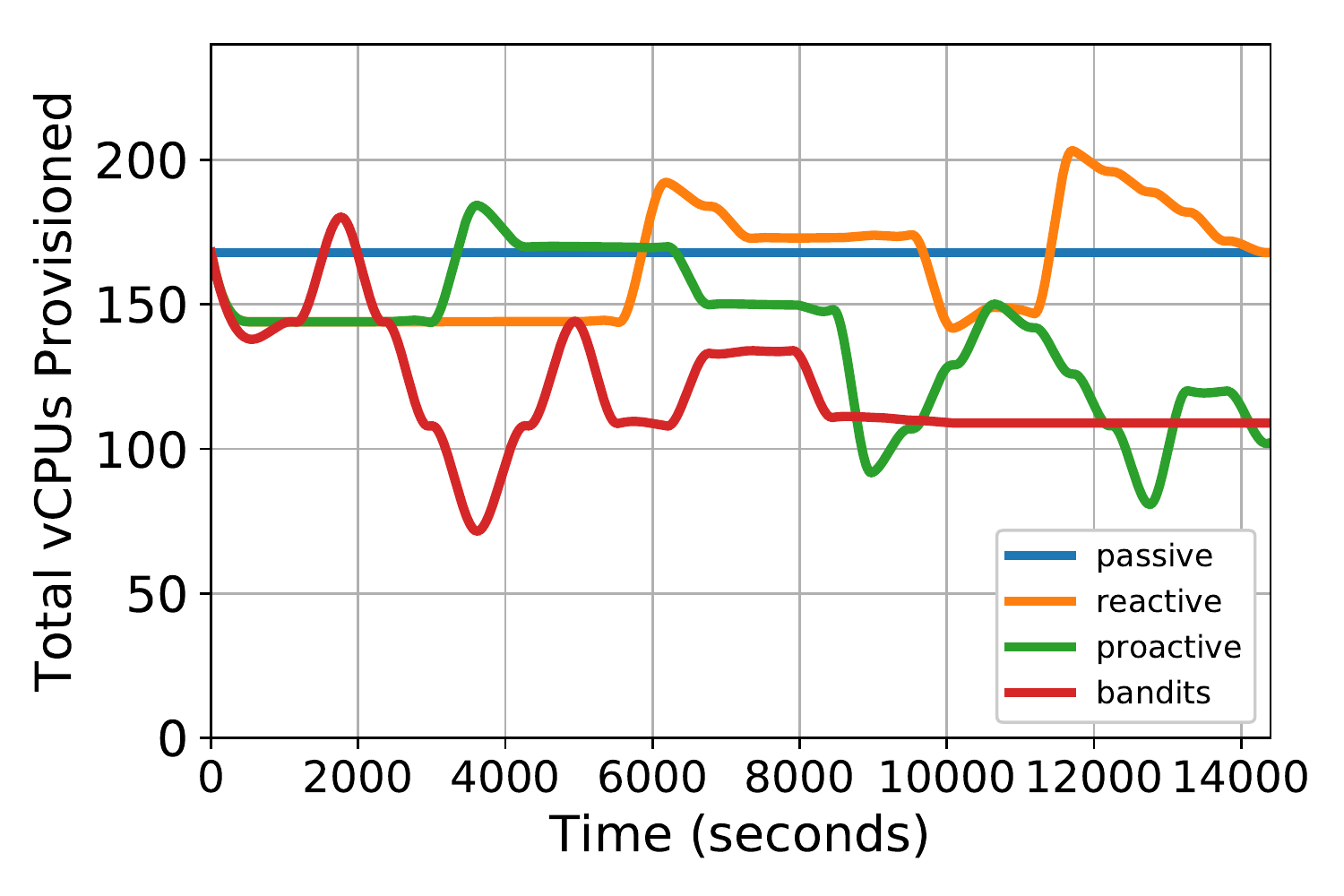}
    \caption{Provisioned vCPUs}
    \label{fig:sabine_provisioned_vcpus_static_wkl}
  \end{subfigure}
  \begin{subfigure}{0.48\columnwidth}
    \includegraphics[width=1\linewidth]{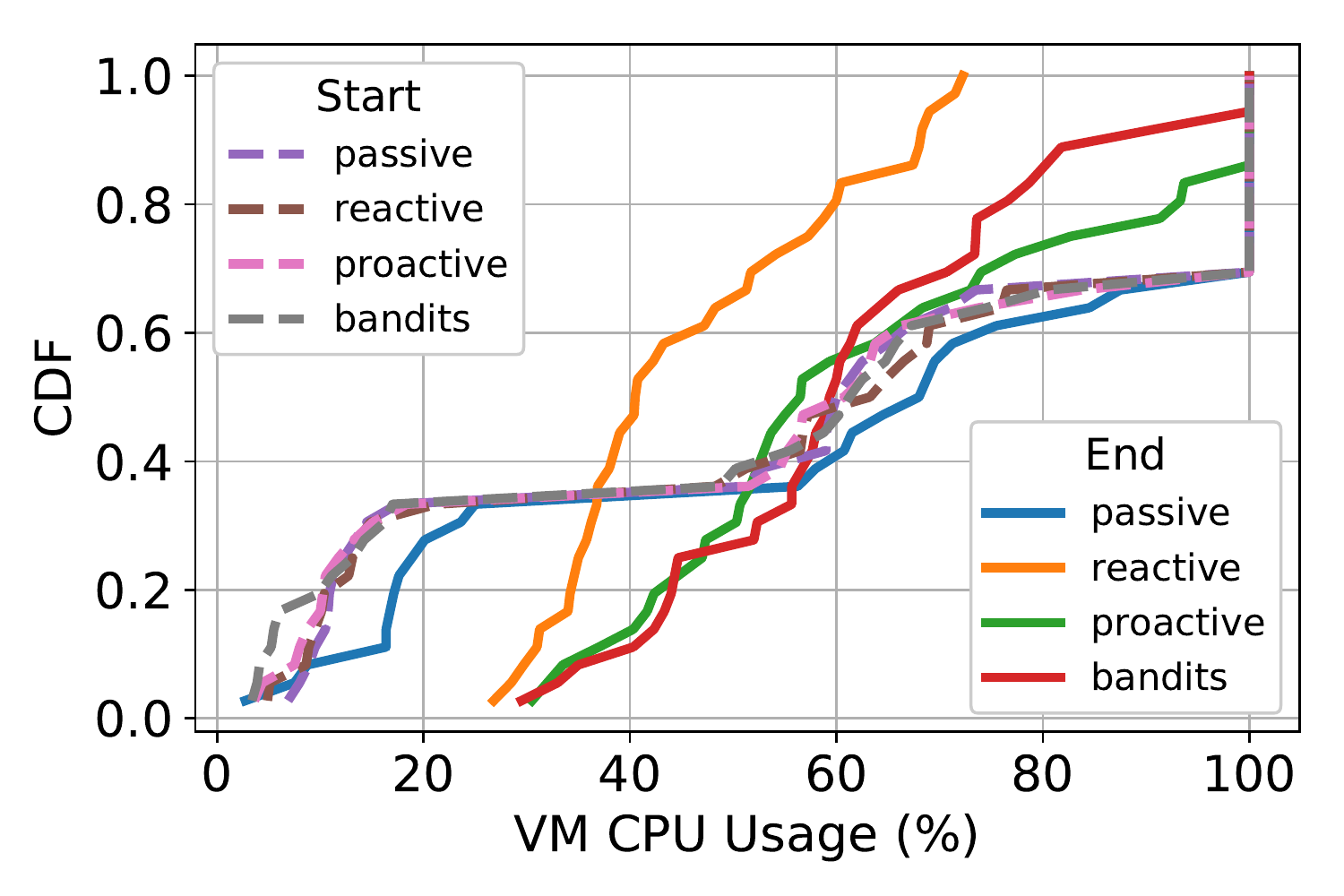}
    \caption{VM CPU Usage (Start-End)}
    \label{fig:sabine_vm_cpu_usage_methods_static_wkl}
  \end{subfigure}
  \caption{Static Workload}
  \label{fig:static_wkl}
\end{figure}

Figure~\ref{fig:sabine_provisioned_vcpus_static_wkl} shows the vCPUs allocations over time for the different methods. We see that both \emph{proactive} and \emph{bandits} result in the fewest allocations, although our method converges to a steady state sooner. 


Further, Figure~\ref{fig:sabine_vm_cpu_usage_methods_static_wkl} plots the CDF of CPU usages of VMs, both at the beginning and at the end of the runs. We observe that around 30\% of the VMs start with 100\% CPU usage, and $\sim$35\% are using less than 20\% of their computational resources. As expected, the initial curves have an almost perfect overlap, as every method runs the same workload. 
More interestingly, at the end of the runs, we can see how the adaptive methods increase the usages of overprovisioned VMs (by scaling them down), as well as decrease the usage of underprovisioned ones (by scaling up). For example, in the \emph{bandits} method, 35\% of the VMs have at most 55\% of CPU usage, and only less than 10\% of the VMs have 100\% CPU usage, as opposed to the initial 30\%.

Overall, we see a 35\% improvement, in terms of amount of vCPUs allocated, for the ML-based methods (\emph{bandits} and \emph{proactive}), when compared to static or threshold-based approaches. Further, at the end of the run, the standard deviation of the VMs CPU usage is 18\% and 22\% for \emph{bandits} and \emph{proactive} respectively, as opposed to 35\% of \emph{passive}, i.e., a 48\%-37\% improvement. Although the deviation of \emph{reactive} is lower (14\%), the average VM CPU utilization also is, 46\% as opposed to 62\% of \emph{bandits}.

\paragraph{Increasing}

\begin{figure}[h!]
  \begin{subfigure}{0.48\columnwidth}
    \includegraphics[width=1\linewidth]{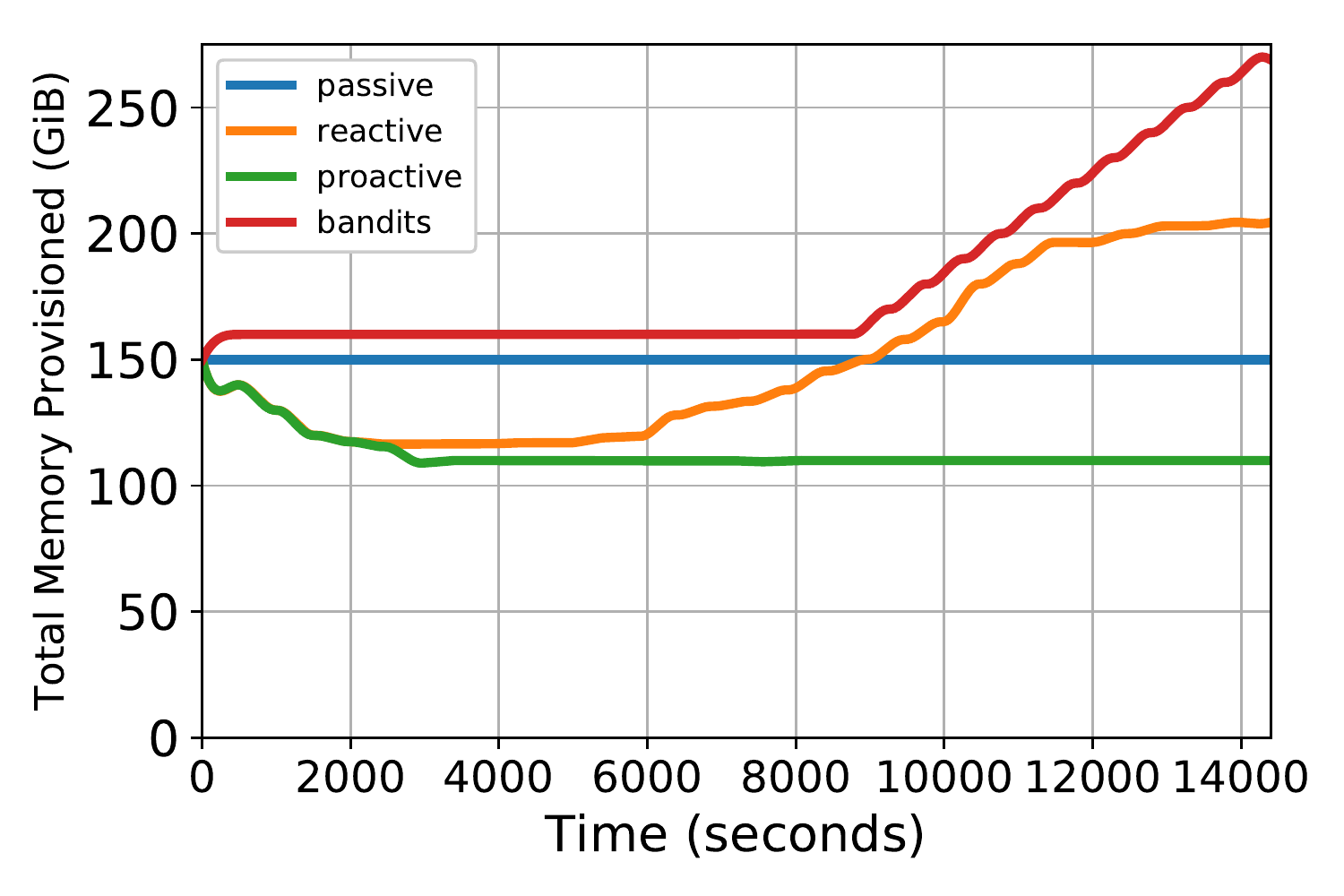}
    \caption{Provisioned Memory}
    \label{fig:sabine_provisioned_memory_increasing_wkl}
  \end{subfigure}
  \begin{subfigure}{0.48\columnwidth}
    \includegraphics[width=1\linewidth]{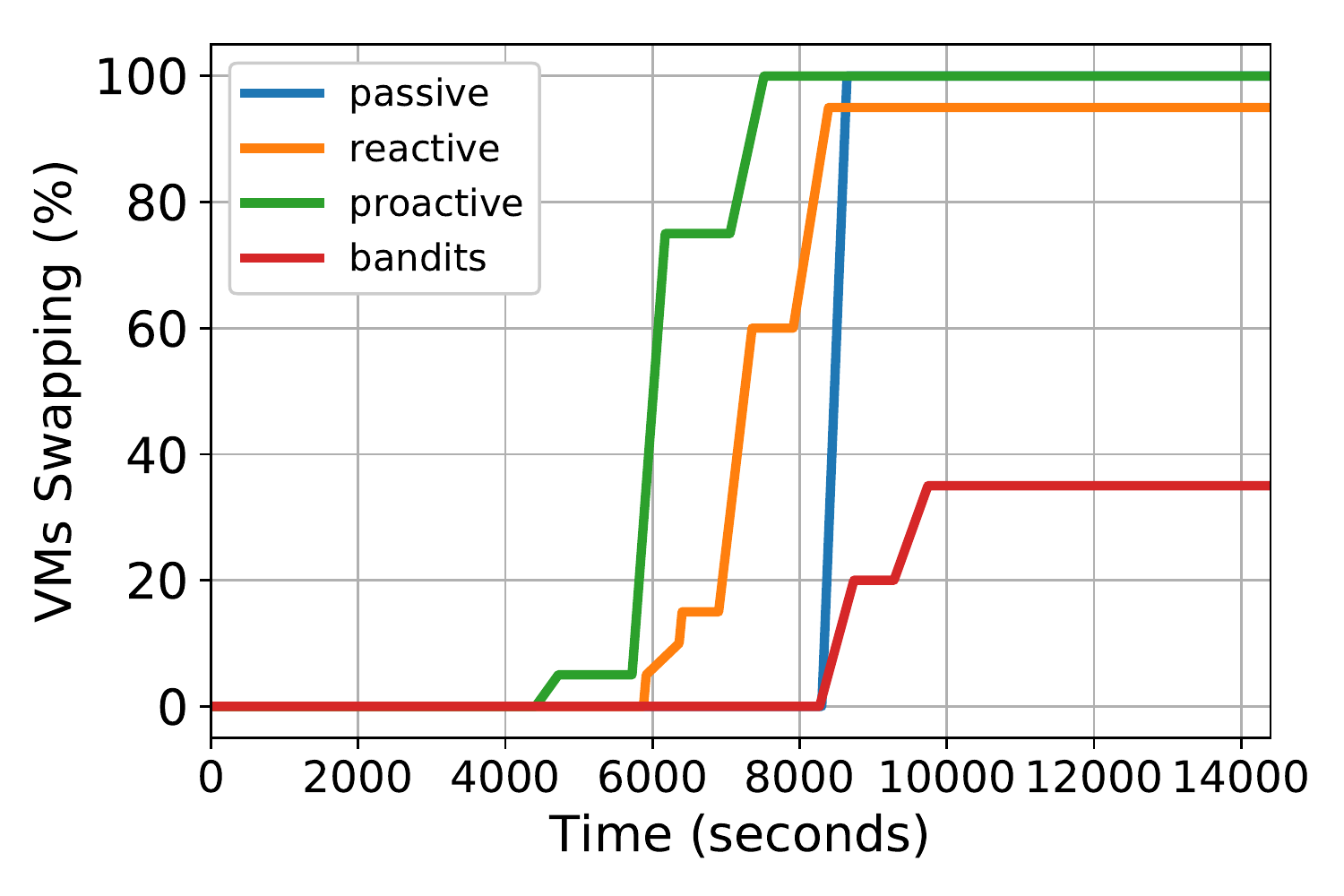}
    \caption{VMs doing Swapping}
    \label{fig:vms_swapping_increasing_wkl}
  \end{subfigure}
  \begin{subfigure}{0.48\columnwidth}
    \includegraphics[width=1\linewidth]{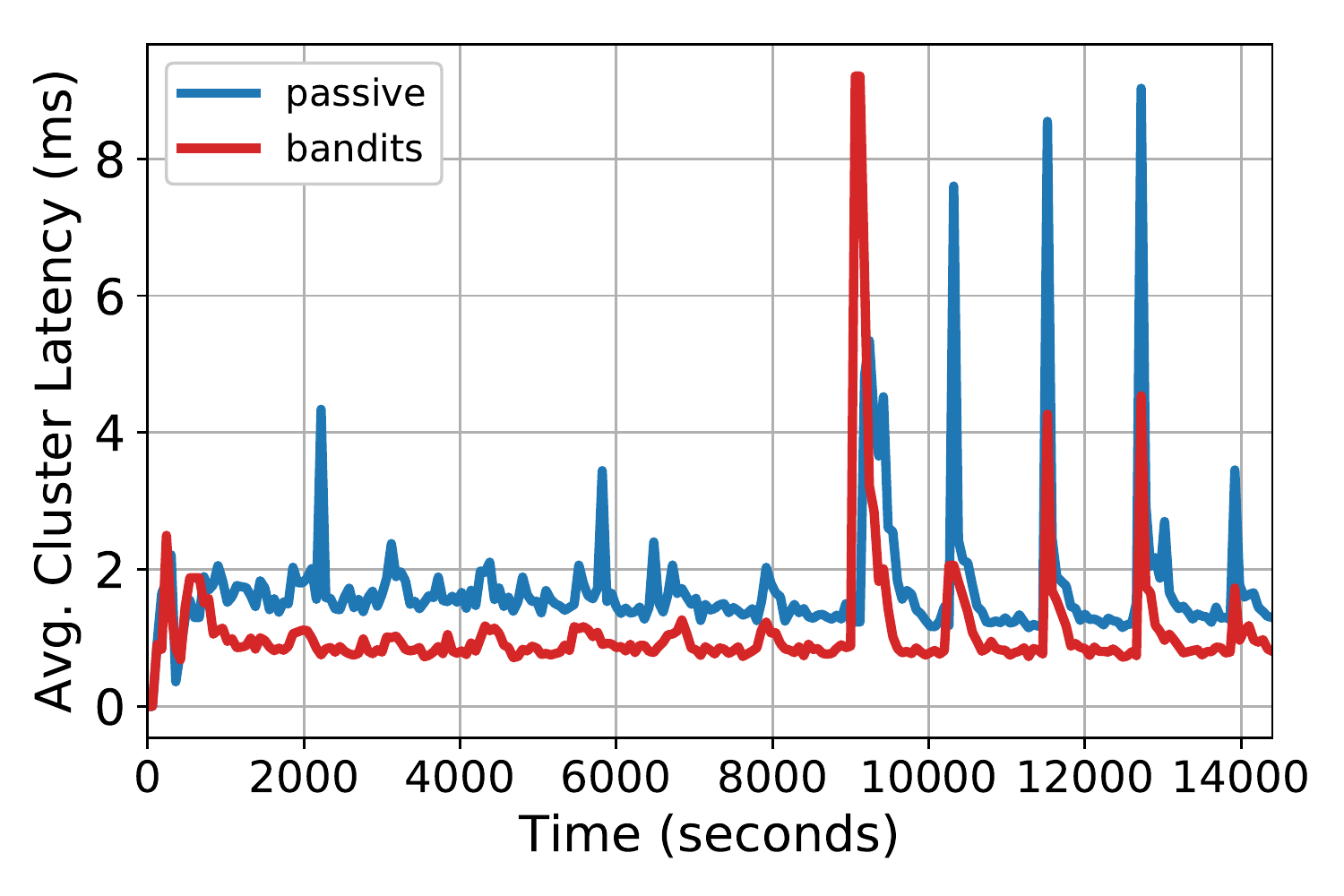}
    \caption{Cluster Latency}
    \label{fig:sabine_cluster_latency_inc_wkl}
  \end{subfigure}
  \begin{subfigure}{0.48\columnwidth}
    \includegraphics[width=1\linewidth]{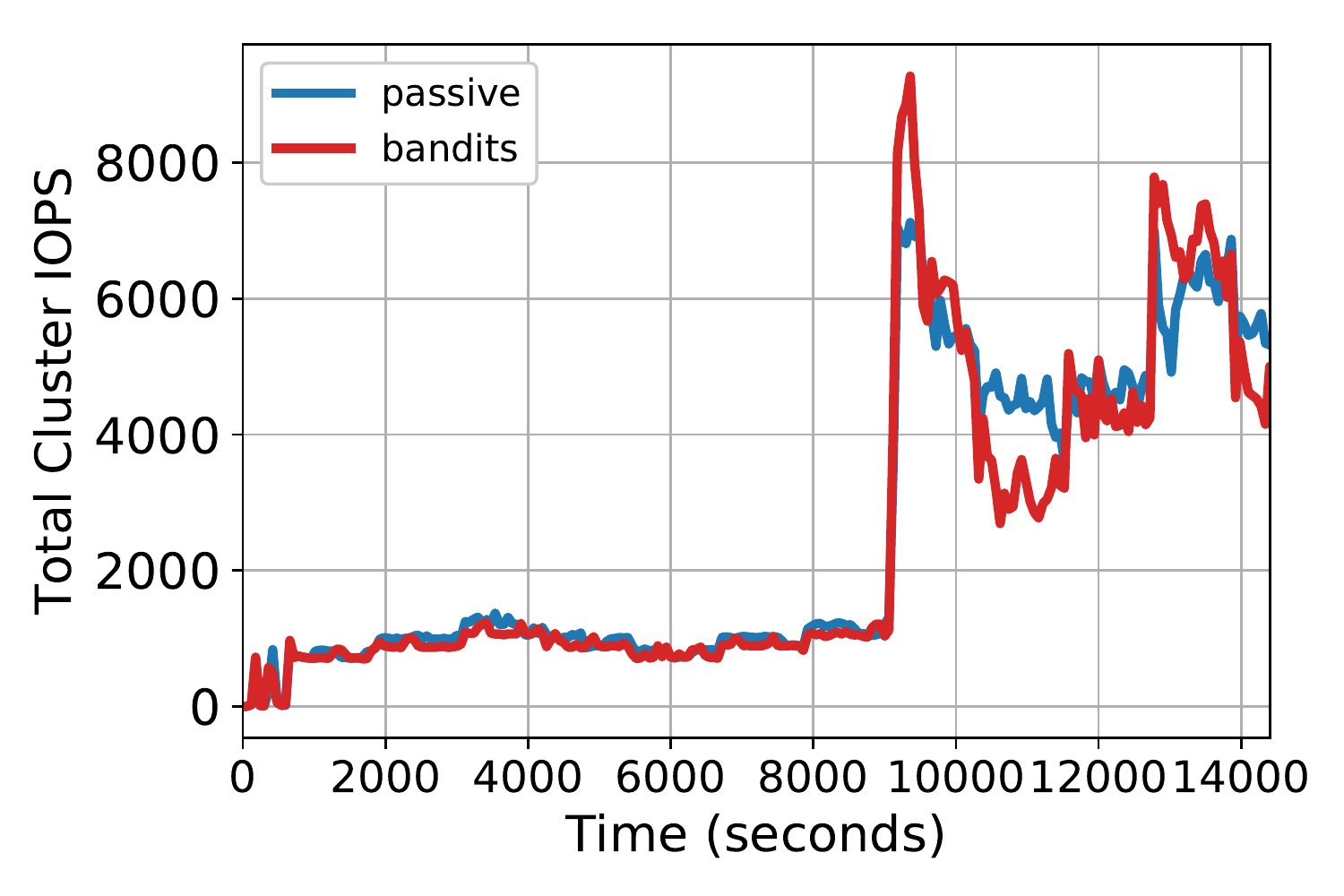}
    \caption{Cluster IOPS}
    \label{fig:sabine_cluster_iops_inc_wkl}
  \end{subfigure}  
  \caption{Increasing Memory Workload}
  \label{fig:increasing_wkl}
\end{figure}

Another example of workloads we observe in practice are those with increasing resource demands. In this case, we simulate a workload with increasing working set size (WSS). We augment the WSS every 20 minutes for a group of 20 \emph{xlarge} VMs running in our controlled cluster.
Figure~\ref{fig:increasing_wkl} shows the results of 4-hour runs. From~\ref{fig:sabine_provisioned_memory_increasing_wkl} we observe that both \emph{reactive} and \emph{proactive} begin by hot removing memory from VMs. Around 6k seconds, the sensed memory usage goes above 75\%, thus \emph{reactive} starts scaling up. The surprising fact is the \emph{proactive} allocations do not change. By looking at the predictions from this method, we observe that it always predicts a maximum memory utilization less than 75\%, therefore, it does not perform adaptations. We speculate the reason is that it has not enough information to start making ``accurate" predictions yet. Bootstrapping the method with our simulator using the same idea of transfer learning could have helped.

On the other hand, we can see that the \emph{bandits} method allocates slightly extra memory than \emph{passive} during the initial $\sim$130 minutes of the run. As the agent starts receiving punishments (or zero rewards) because of increasing swapping levels in the guest OSes, it starts scaling up (around 9k seconds). This phenomenon can be observed in Figure~\ref{fig:vms_swapping_increasing_wkl}, where we show the percentage of VMs that experience swapping over time. As expected, \emph{bandits} performs the best, as it is being trained to avoid such states (or contexts). Further, Figures~\ref{fig:sabine_cluster_latency_inc_wkl} and~\ref{fig:sabine_cluster_iops_inc_wkl} compares the average cluster latency and the total cluster IOPS of \emph{passive} and \emph{bandits} methods. We observe that our method shows lower I/O latency in general, and it can keep up with the workload IOPS.
Overall, if we consider the number of VMs that are experiencing swapping at the end of the run, we can see \emph{bandits} has a 63-65\% improvement over the other baselines. 


\paragraph{Periodic and Static}

\begin{figure}[h!]
  \begin{subfigure}{0.48\columnwidth}
    \includegraphics[width=1\linewidth]{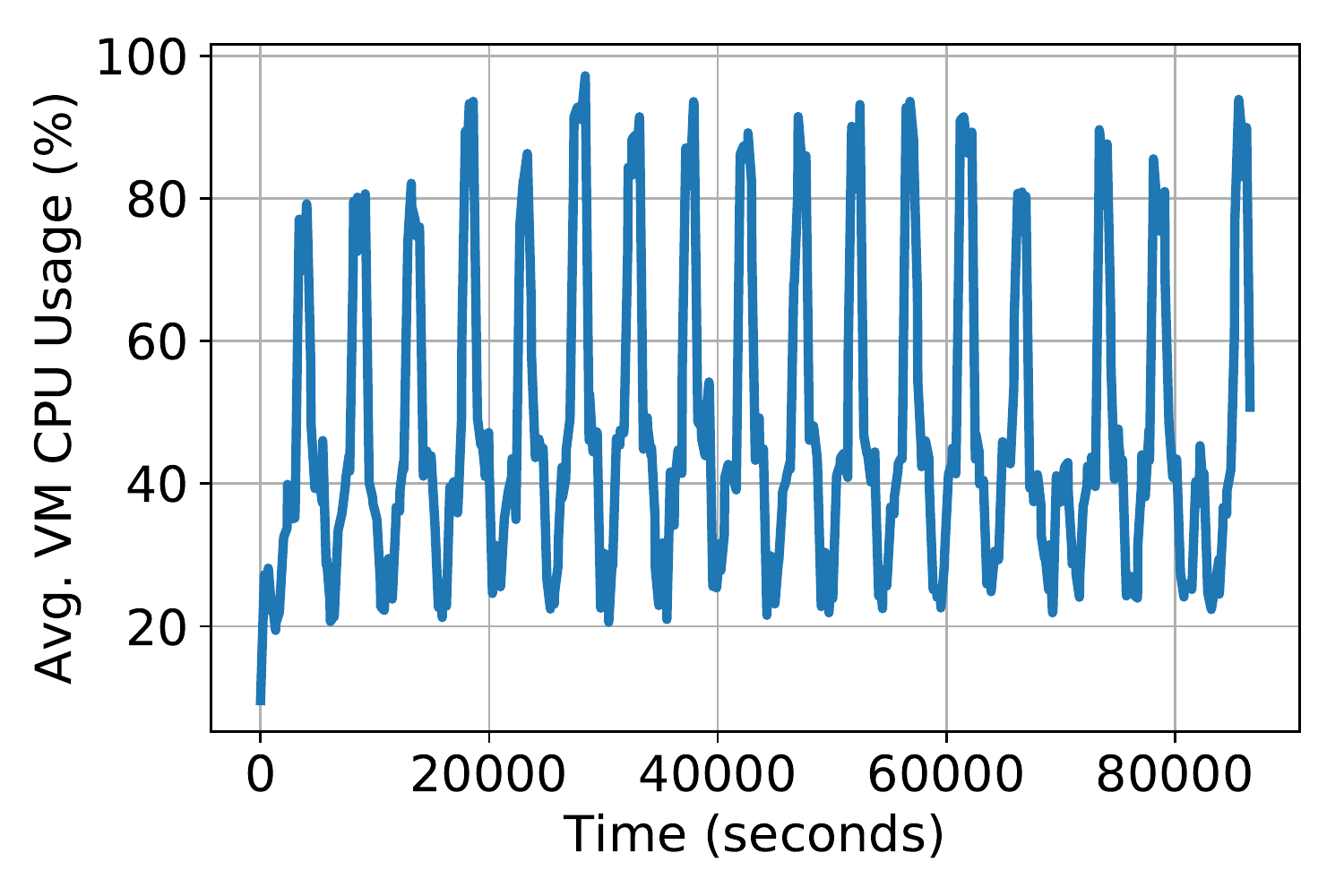}
    \caption{CPU Pattern}
    \label{fig:sabine_periodic_cpu_static_mem_wkl_cpu}
  \end{subfigure}
  \begin{subfigure}{0.48\columnwidth}
    \includegraphics[width=1\linewidth]{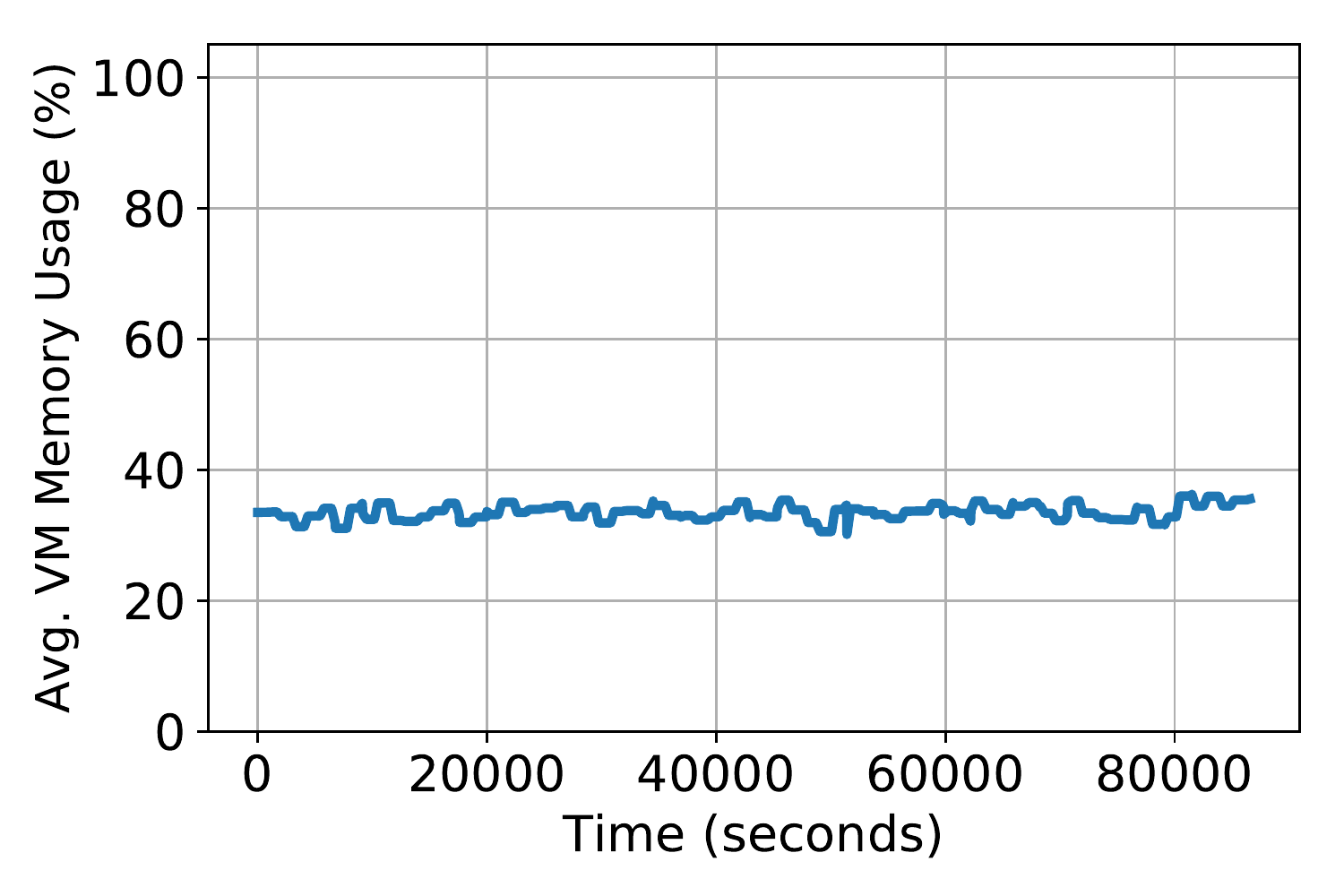}
    \caption{Memory Pattern}
    \label{fig:sabine_periodic_cpu_static_mem_wkl_mem}
  \end{subfigure}
  \begin{subfigure}{0.48\columnwidth}
    \includegraphics[width=1\linewidth]{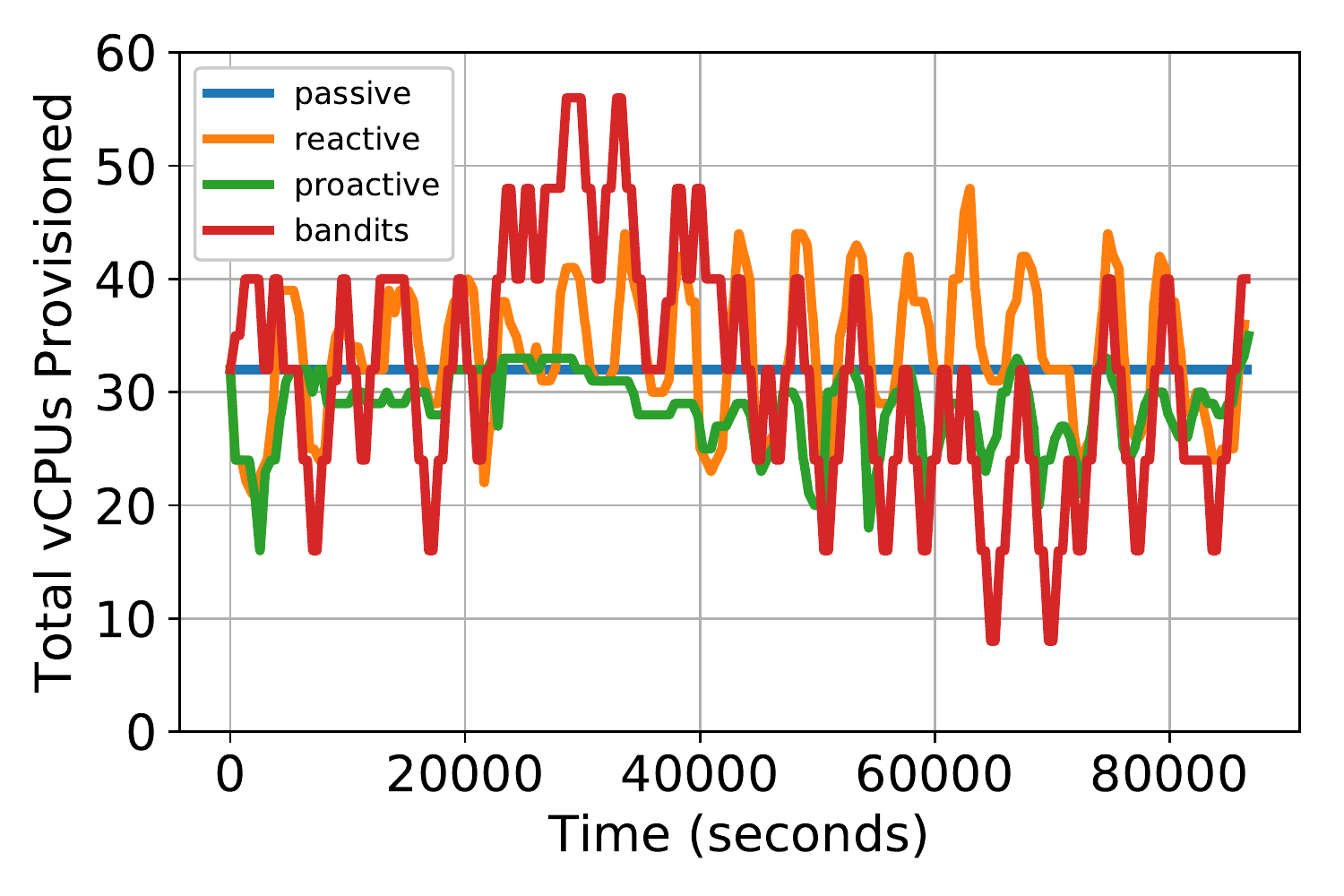}
    \caption{Provisioned vCPUs}
    \label{fig:sabine_provisioned_vcpus_periodic_cpu_static_mem_wkl}
  \end{subfigure}
  \begin{subfigure}{0.48\columnwidth}
    \includegraphics[width=1\linewidth]{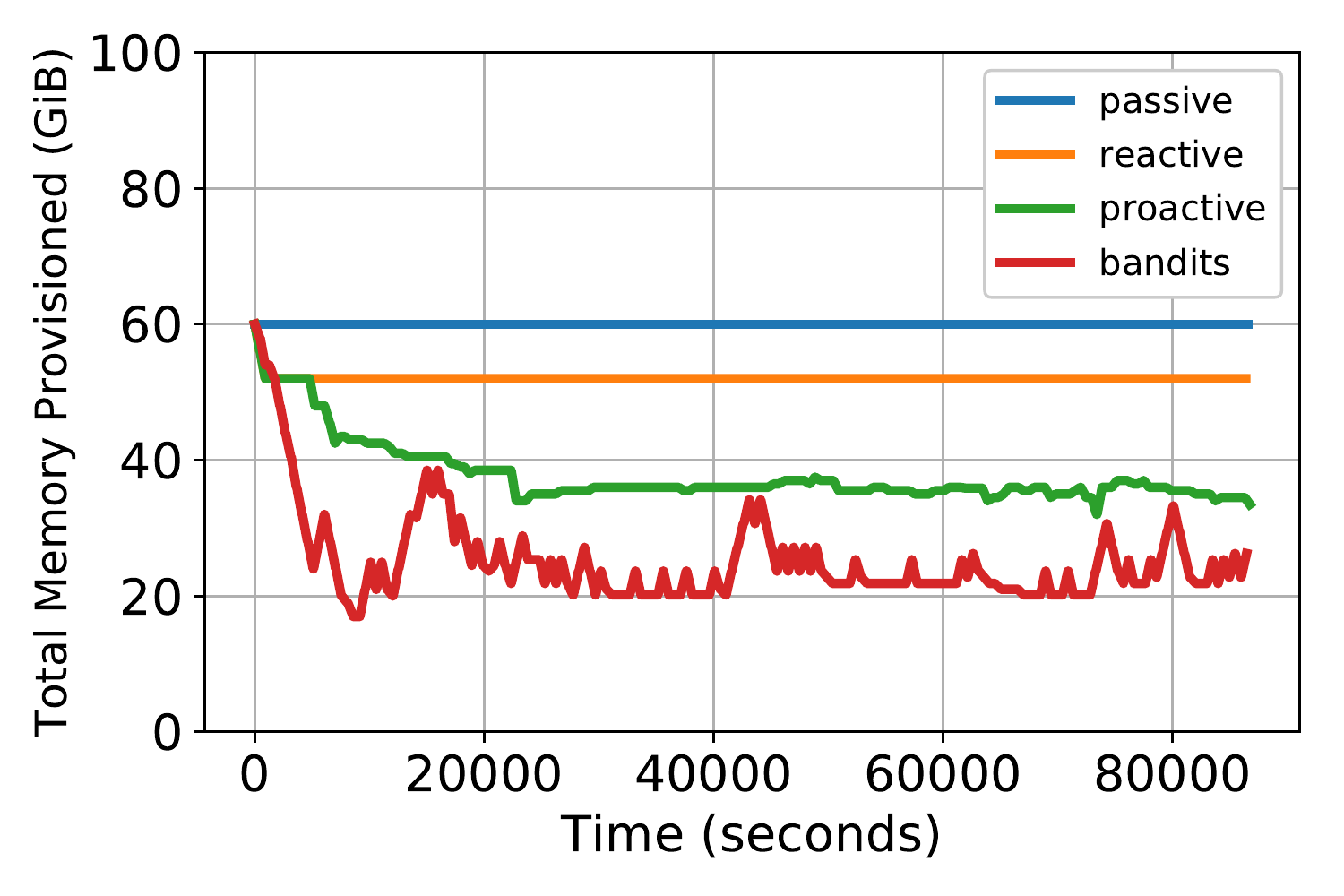}
    \caption{Provisioned Memory}
    \label{fig:sabine_provisioned_memory_periodic_cpu_static_mem_wkl}
  \end{subfigure}
  \begin{subfigure}{0.48\columnwidth}
    \includegraphics[width=1\linewidth]{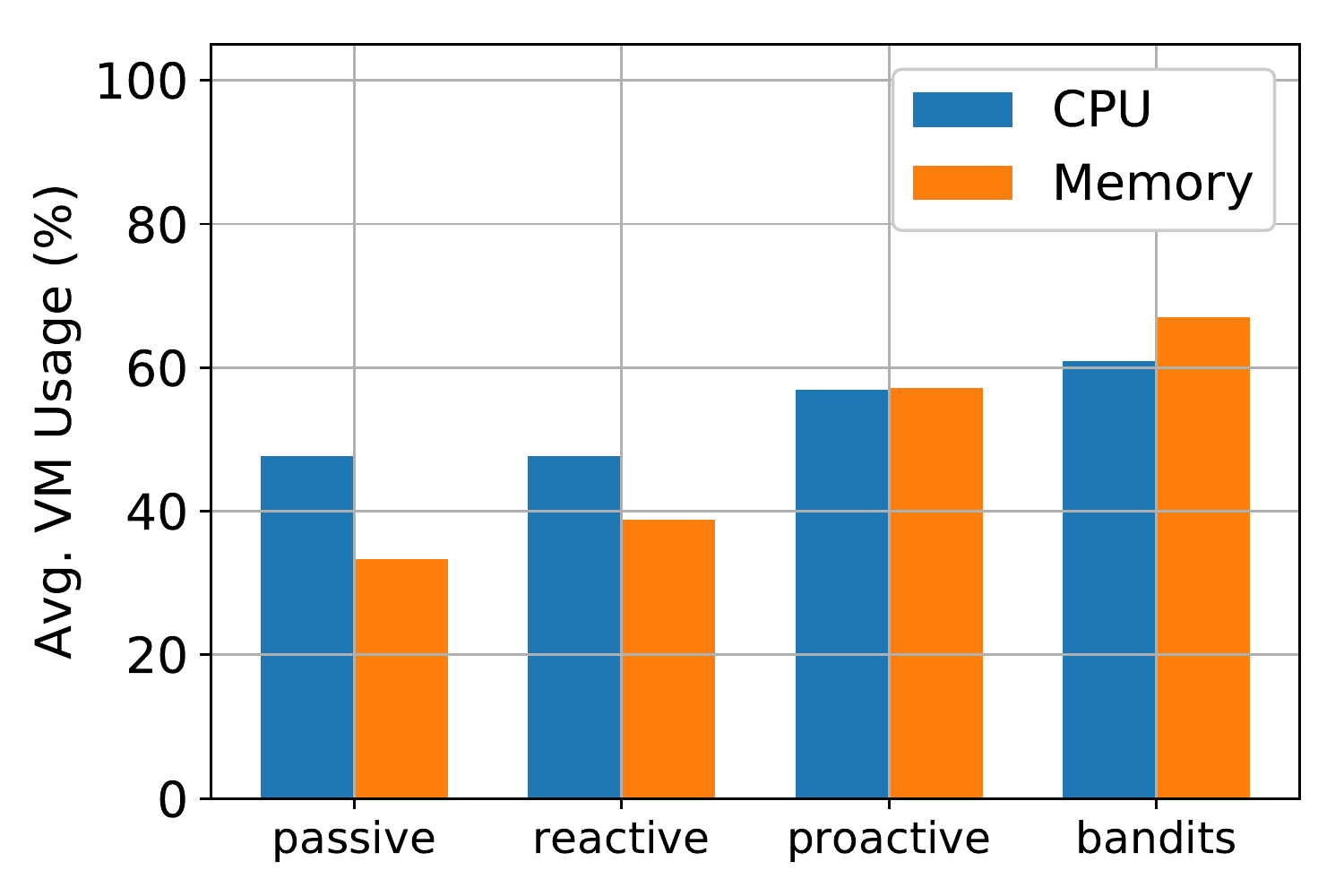}
    \caption{VM Usage}
    \label{fig:sabine_vm_avg_usage_periodic_cpu_static_mem_wkl}
  \end{subfigure}
  \begin{subfigure}{0.48\columnwidth}
    \includegraphics[width=1\linewidth]{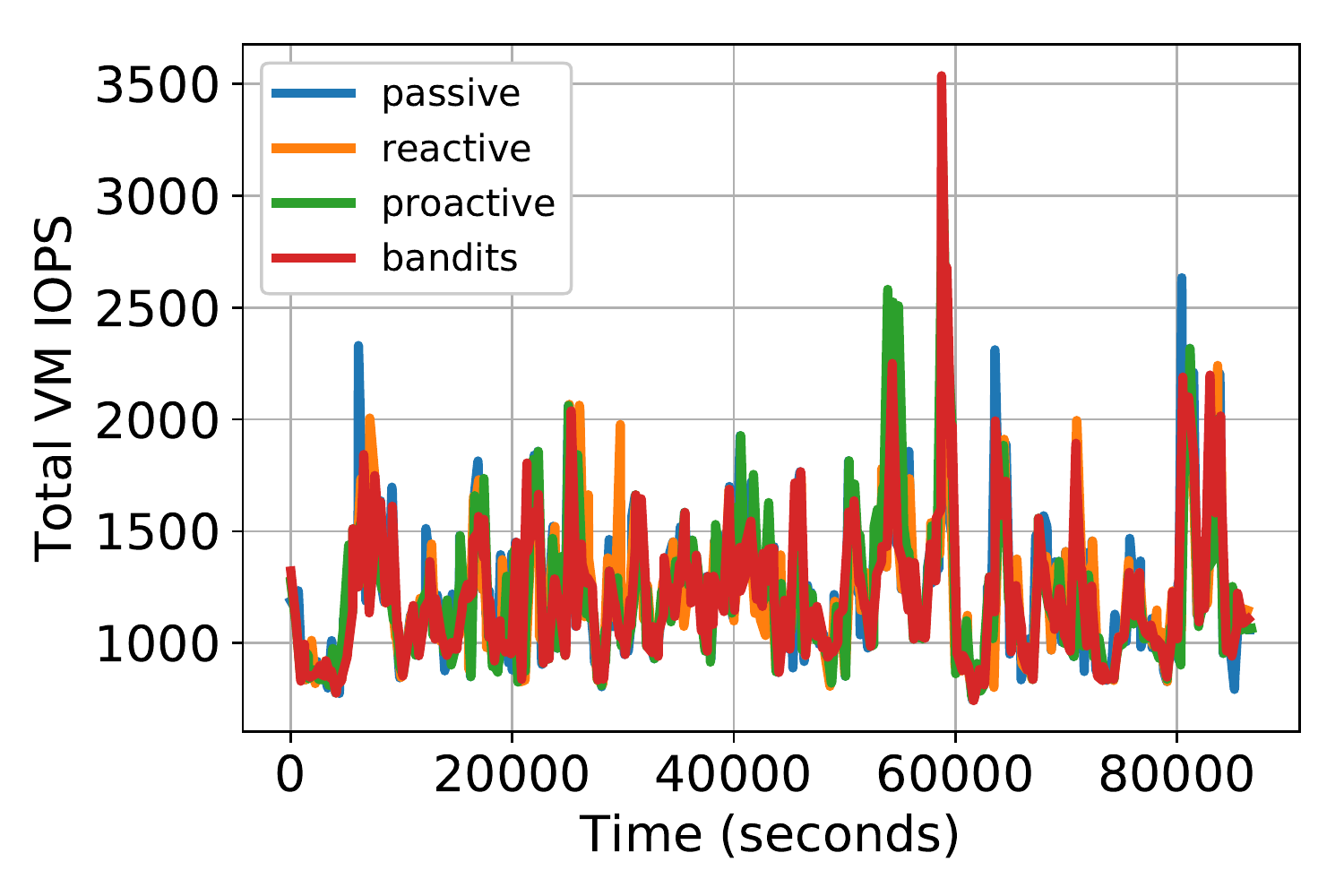}
    \caption{VM IOPS}
    \label{fig:sabine_vm_iops_periodic_cpu_static_mem_wkl}
  \end{subfigure}  
  \caption{Periodic and Static Workload}
  \label{fig:periodic_cpu_static_mem_wkl_II}
\end{figure}

We now focus on periodic and static workloads. In particular, we vary CPU utilization levels of VMs (Figure~\ref{fig:sabine_periodic_cpu_static_mem_wkl_cpu}), but keep constant the memory usage (Figure~\ref{fig:sabine_periodic_cpu_static_mem_wkl_mem}). We expect the adaptive methods to decommission CPU resources during non-peak times, and restore them back during high demand, and also, reduce the amount of provisioned memory to increase utilization. We run a 1-day long experiment, where we deploy four \textsc{AdaRes} agents and execute them in parallel, one for each method. Each agent controls 8 \emph{xlarge} VMs, which are evenly spread across the nodes in the cluster.

On average, we observe that the ML-based adaptive methods provision less vCPUs and memory than the other two (Figures~\ref{fig:sabine_provisioned_vcpus_periodic_cpu_static_mem_wkl} and~\ref{fig:sabine_provisioned_memory_periodic_cpu_static_mem_wkl}), which translates into higher resource utilization (Figure~\ref{fig:sabine_vm_avg_usage_periodic_cpu_static_mem_wkl}). For example, the average memory usage of \emph{bandits} almost doubles \emph{passive}'s usage. 

Further, even though \emph{bandits} uses less resources, it can still keep up with the IOPS of the workload (Figure~\ref{fig:sabine_vm_iops_periodic_cpu_static_mem_wkl}). Overall, \emph{bandits} ends up using around 25 GiB of RAM, almost a 60\% reduction over the static baseline, while at the same time keeps reducing the CPU overload during peak times. 

We acknowledge that different threshold settings can cause completely different behavior for the \emph{reactive} and \emph{proactive} approaches. Even for \emph{bandits}, hyperparameter tuning of exploration constants,  more advanced feature engineering, or non-linear models (both for \emph{bandits} and \emph{proactive}), could boost these numbers up. We leave that to future work.

\subsubsection{LinUCB in Practice}

\begin{figure}[h!]
  \centering
    \includegraphics[width=0.80\columnwidth]{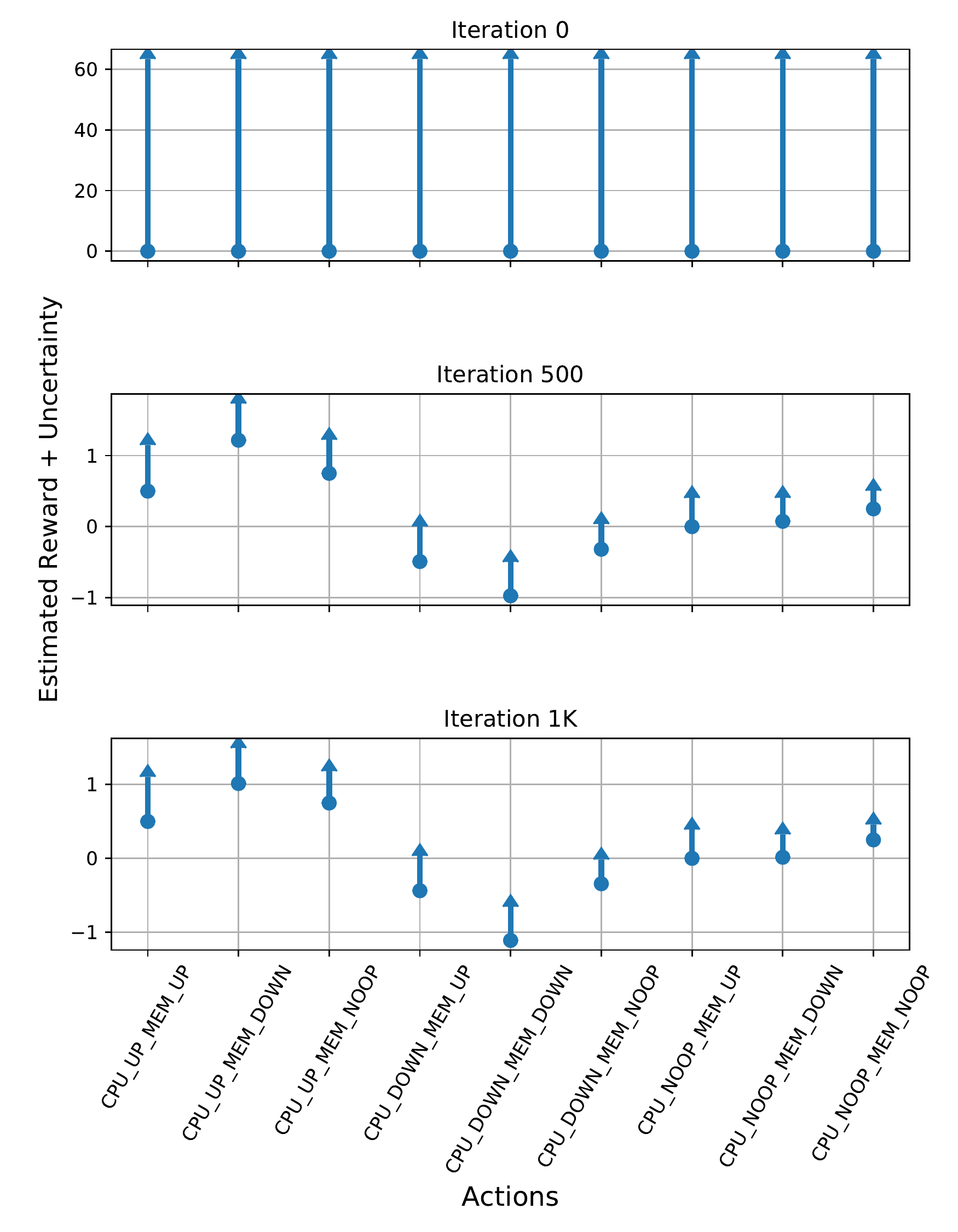}
    \caption{CPU Overload Context}
    \label{fig:explanation_cpu_overload}
\end{figure}

\begin{figure}[h!]
  \centering
    \includegraphics[width=0.80\columnwidth]{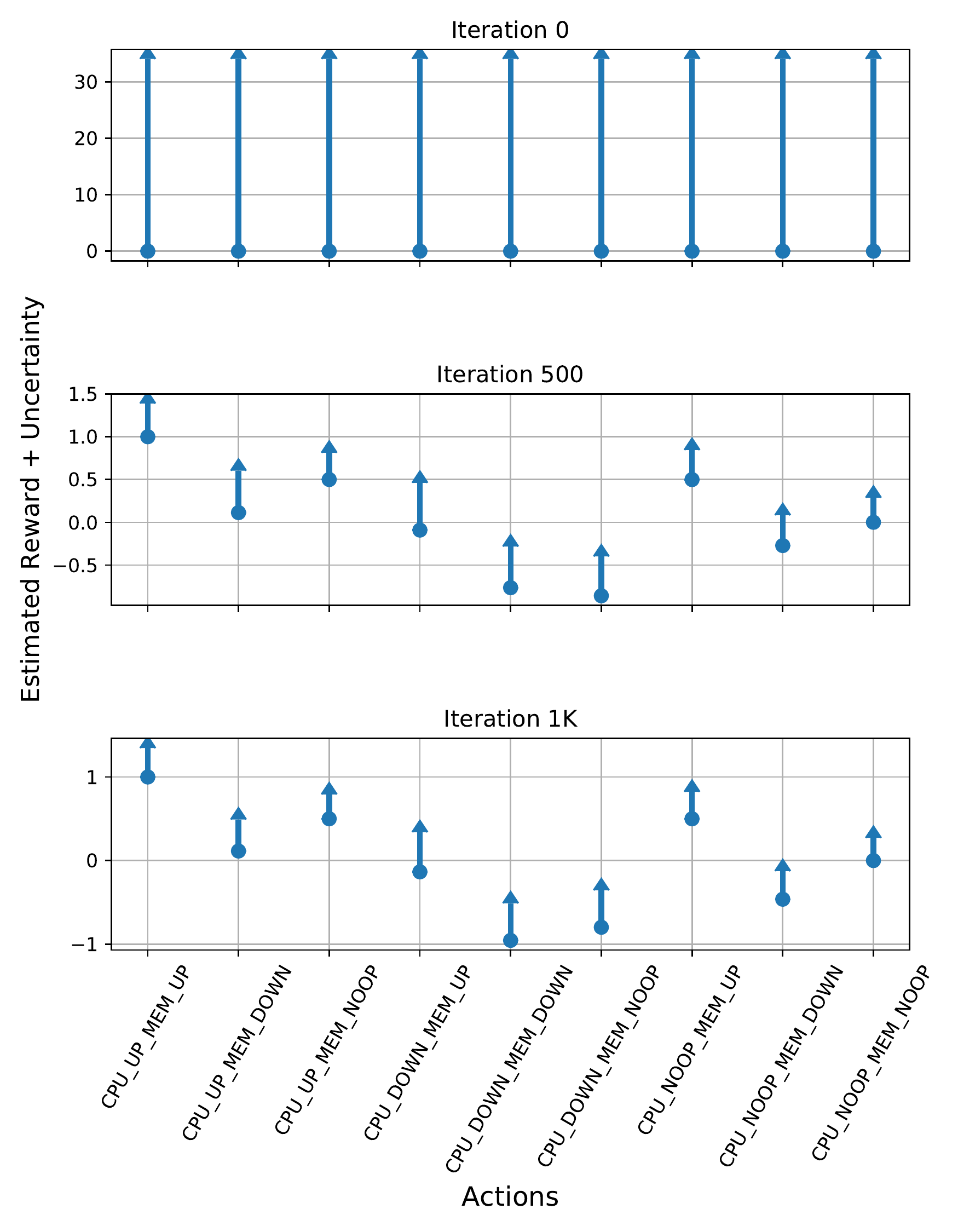}
    \caption{CPU Overload and Memory Swapping Context}
    \label{fig:explanation_cpu_overload_mem_swapping}
    \vspace{+0.4cm}    
\end{figure}

Finally, we illustrate few more examples of how LinUCB operates in practice. We use our cluster simulator to replicate the real cluster environment, and we run heterogeneous workloads across 36 VMs during 1k iterations. We checkpoint our model every 500 iterations to be able to track the progress. 
We randomly select a VM with CPU underprovisioning and one with both CPU and memory underprovisioning. We show the estimated reward and uncertainty of the examples in Figures~\ref{fig:explanation_cpu_overload} and~\ref{fig:explanation_cpu_overload_mem_swapping} respectively. 

In both cases, we observe that the estimated rewards start at zero and there is high uncertainty in every action. As the algorithm performs exploration, those intervals shrink and the estimated rewards get closer to the expected rewards for each action (Iteration 500). The algorithm then starts exploiting and choosing the actions with the highest expected reward. From Figure~\ref{fig:explanation_cpu_overload}, we see that the ``best" actions are the ones that involve scaling up vCPUs, as the VM experiences high CPU overload. Although the \emph{noop} action seems to be the most explored one, as its confidence interval shrinked the most, its estimated reward is still below the aforementioned actions. On a similar note, Figure~\ref{fig:explanation_cpu_overload_mem_swapping} illustrates that scaling up both vCPUs and memory is the clear winner for VM contexts with underprovisioning of both resources. 

\section{Related Work}
\label{sec:related}

We discuss work relevant to \textsc{AdaRes} in the areas of measurements and different approaches towards resource management (RM).

\paragraph{Measurements:}
Google traces \cite{clusterdata:Reiss2011,clusterdata:Wilkes2011} have enabled research on a broad set of topics, from workload characterizations~\cite{Mishra:2010:TCC:1773394.1773400} to
new algorithms for machine assignment~\cite{Reiss:2012:HDC:2391229.2391236}. However, they characterized a month-long trace of non-VM workloads. In this work, on the other hand, we focus on VM workloads running in enterprise clusters. There has been some recent work on VM workloads characterization~\cite{Cortez:2017, Kilcioglu:2017}, but mainly in the public cloud setting. Prior work on measurements of enterprise clusters~\cite{Srini:2016:CPC:2987550.2987584} do not quantify issues related to VM resource allocations. Other measurement studies mainly concentrate on network traffic and communication patterns within data center networks to reduce bandwidth utilization but do not focus per se on VM workload characterization~\cite{Bodik:2012:SFB:2342356.2342439}.

\paragraph{Profiling RM Approaches:}

The prior work on resource management based on profiling approaches has focused on empirically deriving application demands by online and offline profiles of real workloads~\cite{Urgaonkar:2002:ROA:844128.844151,Zheng:2009:JEM:1855807.1855825, Govindan:2009:SPT:1519065.1519099}.

PseudoApp~\cite{6572999} chooses the right VM size by creating a pseudo application to mimic the resource consumption of a real application; that is, it runs the same set of distributed components and executes the same sequence of system calls as those of the real application. 
CherryPick~\cite{Alipourfard:2017:CAU:3154630.3154669} leverages Bayesian Optimization to build performance models for various applications, and uses those models to identify the best (or close-to-the-best) configuration, but using extra profiling runs.

\paragraph{Model-Driven RM Approaches:}

In general, model-driven approaches focus on building models to estimate the result of different allocation strategies on the performance of applications. Oftentimes, they rely on historical resource demands to train statistical learning models to drive the allocation decisions~\cite{Stewart:2008:DCC:1404014.1404029, 4550838, Ganapathi:2009:PMM:1546683.1547490, Shivam:2006:AAL:1182635.11641734}.

PARIS~\cite{Yadwadkar:2017:SBV:3127479.3131614} leverages established machine learning techniques, such as random forests, to identify the best VM across multiple cloud providers. Ernest~\cite{Ernest:2016} is a system to efficiently run applications on shared infrastructure by choosing the right hardware configuration. Their insight is that a number of jobs have predictable structure in terms of computation and communication, thus they build performance models that can predict the running time (or other performance metric of interest) of jobs on specified hardware configurations. 
One key difference with our approach is that they do not adjust VM resources on-the-fly, rather, their work assumes fix-sized instance types (as is the case of the public cloud), and they aim to choose the optimal instance type (and optimal number of instances) to run a particular job.

Gmach et al.~\cite{Gmach:2007:WAD:1524302.1524818} propose a resource allocation system for datacenter applications that depends on predicting their behavior a priori based on the repetitive nature of their workloads. On a similar note, DejaVu~\cite{Vasic:2012:DAR:2150976.2151021} identifies a few workload categories and leverages them to reuse previous resource allocations so as to minimize re-allocation overheads.
In contrast, we assume our workload patterns can change over time, thus we propose a contextual bandits model to dynamically adapt to changes.

Soror et al.~\cite{Soror:2008:AVM:1376616.1376711} leverages cost models that are built into database query optimizers to recommend workload-specific VM configurations. However, their framework only works for SQL-like workloads, as opposed to ours, which is agnostic to the application. Finally, PRESS~\cite{press_2010} extracts dynamic patterns
in application resource demands and adjusts their resource allocations automatically using signal processing and statistical learning algorithms. They only tune VM CPU limits, although they mention their approach is extensible to other resources, such as memory and networking.

\paragraph{Adaptive RM Approaches:}

Some prior work investigate auto-scaling using adaptive control loops and reinforcement learning~\cite{Bodik:2009:AED:1555271.1555273, Dutreilh:2010:DCR:1844768.1845365, dutreilh:hal-01122123, Ferguson:2012:JGJ:2168836.2168847, Zhu:2009:IIA:1507559.1507586, Kalyvianaki:2009:SSC:1555228.1555261, Padala:2007:ACV:1272996.1273026,Rao:2009:VRL:1555228.1555263,6216363}, though none of the above use the contextual bandits framework.
Other adaptive auto-scaling systems, such as the ones offered in Google Cloud Platform~\cite{gcp-autoscaler:2017} or Amazon Web Services~\cite{aws-auto-scaling:2017}, allow users to maintain application availability by dynamically scaling their resources according to conditions they define. For example, users can set target utilization metrics (e.g., average CPU utilization, requests per second) and the system will then automatically adjust the number of instances as needed to maintain those targets (similar to \emph{reactive}). Such systems mainly focus on horizontal scaling, whereas our work targets vertical one. In general, these threshold-based systems (either horizontal/vertical) are simple to implement and use, however their performance depends on the quality of the thresholds~\cite{Arabnejad:2017}.

Perhaps the most prominent work on the VM resource allocation problem has been done by Delimitrou et al.~\cite{Delimitrou:2014:QRQ:2541940.2541941, Delimitrou:2016:HRP:2872362.2872365}. They mainly use collaborative filtering techniques to classify workloads using four different classification tasks (scale up/out, heterogeneity, and interference), and they rely on (small) online workload profiling as well as monitoring tasks for allocation re-adjustment. 

\paragraph{Scheduling/Migration Approaches}

A great deal of previous research into resource management has focused on VM/task scheduling and migration~\cite{Novakovic:2013:DTI:2535461.2535489, conf/im/BobroffKB07, Yang:2013:BPO:2485922.2485974, Delimitrou:2013:PQS:2451116.2451125, Ruprecht:2018:VLM}. They are somewhat orthogonal to our work, as we focus on the problem of maximizing the resource usage efficiency of VMs, which should result in easier scheduling, i.e., packing of smaller VMs~\cite{Hermenier:2013:BFC:2554123.2554126}.

\section{Discussion and Future Work}
\label{sec:discussion}

Although we have proposed an initial framework for adjusting vCPUs and memory of VMs on-the-fly using ML techniques, some natural extensions of this work come to mind, both from a systems as well as an ML perspective.
On the systems front, besides improving our simulator and adding support for more application-level metrics (e.g., SQL transactions per second), we are also planning on being able to tune other type of resources, such as networking and storage, as well as managing other entities, such as containers. Further, we plan to include sensitivity analyses of the different threshold choices (e.g., 25\%, 75\%), as well as augment the experiments with real workloads.
Regarding ML, apart from experimenting with more complex models, an interesting step to take would be to enable smarter filtering policies in FS. By borrowing ideas from active learning literature~\cite{settles2009active}, we could potentially filter the VMs that would provide the most useful information to our agent. For example, we could pick the instances in a greedy fashion, according to some informativeness measure used to evaluate all the instances in the cluster, or select the most ``diverse" instances using submodularity~\cite{submodularity}, which would allow the agent to have a better coverage of the state space, thus improving generalization and speeding up training.

\section{Conclusions}
\label{sec:conclusions}

Virtual execution environments enable a more efficient use of server's resources by consolidating multiple applications onto the same physical hardware. However, provisioning a VM with more (or less) resources than it requires can drastically impact its performance as well as that of other VMs in the cluster.
As part of this work, we first provided a characterization of resource allocation and utilization of virtual machines from thousands of enterprise clusters running production workloads. Given that we observed a high degree of overprovisioning and underprovisioning, mainly due to inaccurate user guesses, as well as significant variability in load demands over time, we proposed \textsc{AdaRes}, an adaptive system that dynamically tunes resources of VMs.
\textsc{AdaRes} uses the contextual bandits framework together with transfer learning to optimize configurations of VMs in a cluster, and exploits cluster, node and VM-level information to promote efficient resource utilization across VMs. 

\section*{Acknowledgments}
{
Special thanks to Kevin Jamieson and Lalit Jain for their valuable and constructive feedback. 
We are also grateful to Robert Marver for helping with the experiments.
One of the authors was supported by the Argentinean government program BEC.AR.
}

{\footnotesize \bibliographystyle{acm}
\bibliography{biblio}}




\end{document}